\documentclass[preprint,12pt]{elsarticle}

\usepackage{lineno,hyperref}
\usepackage{natbib}
\usepackage{geometry}
\usepackage{fleqn}
\usepackage{graphicx}
\usepackage{newtxtext,newtxmath}
\usepackage{hyperref}
\usepackage{booktabs}
\usepackage{bm}
\usepackage{amsmath}
\usepackage{enumitem}
\usepackage[justification=centering]{caption}

\usepackage{array}
\newcommand{\PreserveBackslash}[1]{\let\temp=\\#1\let\\=\temp}
\newcolumntype{C}[1]{>{\PreserveBackslash\centering}p{#1}}

\modulolinenumbers[5]

\usepackage{framed} 
\usepackage{nomencl} 
\makenomenclature

\graphicspath{{./figures/}}
\journal{Nuclear Engineering and Design}

\makenomenclature

\begin{document}

\begin{frontmatter}

\title{\large{A Comprehensive Survey of Inverse Uncertainty Quantification of Physical Model Parameters in Nuclear System Thermal-Hydraulics Codes}}

\author[NCSU]{Xu Wu\corref{mycorrespondingauthor}}
\cortext[mycorrespondingauthor]{Corresponding author}
\ead{xwu27@ncsu.edu}

\author[NCSU]{Ziyu Xie}

\author[NCSU]{Farah Alsafadi}

\author[UIUC]{Tomasz Kozlowski}

\address[NCSU]{Department of Nuclear Engineering, North Carolina State University    \\ 
	Burlington Laboratory, 2500 Stinson Drive, Raleigh, NC 27695, USA \\}
	
\address[UIUC]{Department of Nuclear, Plasma \& Radiological Engineering, University of Illinois at Urbana-Champaign    \\  	Talbot Laboratory, 104 South Wright Street, Urbana, IL 61801, USA \\}

\begin{abstract}
		
Uncertainty Quantification (UQ) is an essential step in computational model validation because assessment of the model accuracy requires a concrete, quantifiable measure of uncertainty in the model predictions. The concept of UQ in the nuclear community generally means forward UQ (FUQ), in which the information flow is from the inputs to the outputs. Inverse UQ (IUQ), in which the information flow is from the model outputs and experimental data to the inputs, is an equally important component of UQ but has been significantly underrated until recently. FUQ requires knowledge in the input uncertainties which has been specified by expert opinion or user self-evaluation. IUQ is defined as the process to inversely quantify the input uncertainties based on experimental data.

This review paper aims to provide a comprehensive and comparative discussion of the major aspects of the IUQ methodologies that have been used on the physical models in system thermal-hydraulics codes. IUQ methods can be categorized by three main groups: \textit{frequentist (deterministic)}, \textit{Bayesian (probabilistic)}, and \textit{empirical (design-of-experiments)}. We used eight metrics to evaluate an IUQ method, including \textit{solidity}, \textit{complexity}, \textit{accessibility}, \textit{independence}, \textit{flexibility}, \textit{comprehensiveness}, \textit{transparency}, and \textit{tractability}. Twelve IUQ methods are reviewed, compared, and evaluated based on these eight metrics. Such comparative evaluation will provide a good guidance for users to select a proper IUQ method based on the IUQ problem under investigation.
	
\end{abstract}


\end{frontmatter}
\newpage



\nomenclature{AA}{Average Amplitude}
\nomenclature{ASM}{Adjoint Sensitivity Method}
\nomenclature{BEMUSE}{Best-Estimate Methods Uncertainty and Sensitivity Evaluation}
\nomenclature{BEPU}{Best Estimate plus Uncertainty}
\nomenclature{BFBT}{BWR Full-Size Fine Mesh Bundle Test}
\nomenclature{CDF}{Cumulative Distribution Function}
\nomenclature{CET}{Combined Effect Tests}
\nomenclature{CIRC{\'E}}{Calcul des Incertitudes Relatives aux Corr{\'e}lations {\'E}lementaires}
\nomenclature{CSAU}{Code Scaling Applicability and Uncertainty}
\nomenclature{DA}{Data Assimilation}
\nomenclature{DFT}{Discrete Fourier Transform}
\nomenclature{DoE}{Design-of-experiments}
\nomenclature{E-M}{Expectation-Maximization}
\nomenclature{EMDAP}{Evaluation Model Development and Assessment Process}
\nomenclature{FBA}{Full Bayesian Approach}
\nomenclature{FEBA}{Flooding Experiments with Blocked Arrays}
\nomenclature{FFT}{Fast Fourier Transform}
\nomenclature{FFTBM}{Fast Fourier Transformation Based Method}
\nomenclature{FOM}{Figure-of-merit}
\nomenclature{FUQ}{Forward Uncertainty Quantification}
\nomenclature{GP}{Gaussian Processes}
\nomenclature{GPMSA}{Gaussian Process Models for Simulation Analysis}
\nomenclature{HTC}{Heat Transfer Coefficient}
\nomenclature{IPREM}{Input Parameter Range Evaluation Methodology}
\nomenclature{IUQ}{Inverse Uncertainty Quantification}
\nomenclature{MAP}{Maximum a Posteriori}
\nomenclature{MBA}{Modular Bayesian Approach}
\nomenclature{MCMC}{Markov Chain Monte Carlo}
\nomenclature{MLE}{Maximum Likelihood Estimation}
\nomenclature{M\&S}{Modeling \& Simulation}
\nomenclature{MSLB}{Main Steam Line Break}
\nomenclature{NEA}{Nuclear Energy Agency}
\nomenclature{NUPEC}{Nuclear Power Engineering Corporation}
\nomenclature{OECD}{Organisation for Economic Co-operation and Development}
\nomenclature{PCA}{Principal Component Analysis}
\nomenclature{PDF}{Probability Density Function}
\nomenclature{PMP}{Physical Model Parameter}
\nomenclature{PREMIUM}{Post-BEMUSE Reflood Models Input Uncertainty Methods}
\nomenclature{PSBT}{PWR Sub-channel and Bundle Test}
\nomenclature{QoI}{Quantity-of-Interest}
\nomenclature{SAPIUM}{Systematic APproach for Input Uncertainty quantification Methodology}
\nomenclature{SET}{Separate Effects Test}
\nomenclature{TH}{Thermal-Hydraulics}
\nomenclature{UAM}{Uncertainty Analysis in Modeling}
\nomenclature{UQ}{Uncertainty Quantification}
\nomenclature{VVUQ}{Verification, Validation and Uncertainty Quantification}

\printnomenclature[5em]

\section{Introduction}

Historically in nuclear system design and licensing practices, with the conservative approach, computer codes tried to model the physical phenomena at the worst-case scenario with deliberate pessimistic and unphysical assumptions. Extreme or unfavorable values of input parameters were used to produce conservative predictions of the outputs. Consequently, the evaluated reactor designs had considerable margins to assure their safety by over-predicting safety-related outputs, such as the peak cladding temperature. Conservetism led to considerable inaccuracies in modeling \& simulation (M\&S) and damaged the economic performance of nuclear energy.

In the late 1980s, best-estimate (BE) safety analysis strategy started to be embedded in the Code Scaling Applicability and Uncertainty (CSAU) \cite{boyack1990quantifying} \cite{young1998application}, and Evaluation Models Development and Assessment Procedure (EMDAP) \cite{martin2012evaluation} \cite{kaizer2018the} methodologies, which were accepted by the U.S. Nuclear Regulatory Commission. This strategy is commonly referred to as Best Estimate plus Uncertainty (BEPU) \cite{d2012best} \cite{wilson2013historical} \cite{rohatgi2020historical}. The goal of BEPU aims to capture the physical phenomena as realistically as possible by implementing a wide range of modeling options and accurate calculation methods to capture physical phenomena at a greater fidelity. According to the BEPU methodology, uncertainties must be quantified in order to prove that the investigated design stays within acceptance criteria.

Uncertainty Quantification (UQ) \cite{smith2013uncertainty} is the process to quantify the uncertainties in Quantity-of-Interest (QoIs) by propagating the uncertainties in input parameters through the computer model. QoIs are also frequently referred to as output, responses, or system response quantity in the open literature. UQ is an essential step in computational model validation because assessment of the model accuracy requires a concrete, quantifiable measure of uncertainty in the model predictions. In the nuclear community, the significance of UQ has been widely recognized and numerous publications have been devoted to UQ methods and applications in response to the BEPU methodology. UQ play a more significant role in nuclear engineering compared to other fields to reduce conservatism while dealing with high-consequence systems. The design decision-making process, development of public policy and preparation of safety procedures all rely on reliable computer codes that have undergone extensive Verification, Validation and Uncertainty Quantification (VVUQ) process \cite{oberkampf2002verification} \cite{oberkampf2010verification}. UQ is also critical for the development of advanced nuclear reactors to make confident, risk-informed decisions when considering alternative designs and operations impacting economics and nuclear safety.

\begin{figure}[ht]
	\centering
	\includegraphics[width=0.75\textwidth]{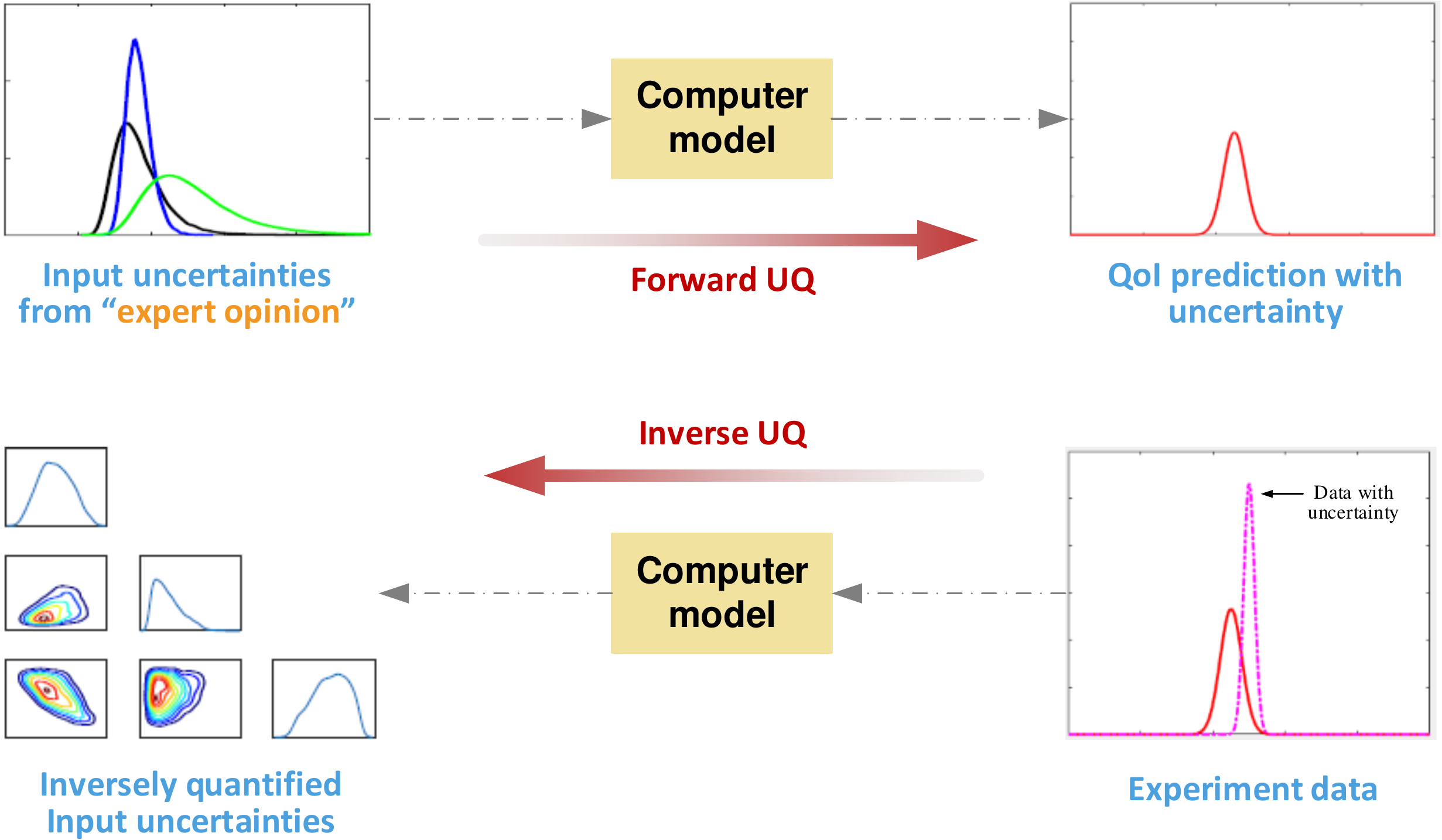}
	\caption[]{Comparison of the information flow in FUQ and IUQ.}
	\label{figure:Intro1-Forward-vs-Inverse-UQ}
\end{figure}

The concept of UQ in the nuclear community generally means forward UQ (FUQ), in which the information flow is from the inputs to the outputs. However, there is another equally important component of UQ - inverse UQ (IUQ), that has been significantly underrated until recently. With IUQ, the information flow is from the model outputs and experimental data to the inputs. FUQ requires knowledge in the computer model input uncertainties, such as the statistical moments (e.g., mean and variance), probability density functions (PDFs), upper and lower bounds, which are not always available. Historically, expert opinion or user self-evaluation have been predominantly used to specify such information in VVUQ studies. Such ad-hoc specifications are subjective, lack mathematical rigor, and can sometimes lead to inconsistencies. \textit{IUQ is defined as the process to inversely quantify the input uncertainties based on experimental data \cite{wu2018inversePart1}. It seeks statistical descriptions of the uncertain input parameters that are consistent with the observation data.} Figure \ref{figure:Intro1-Forward-vs-Inverse-UQ} illustrates the differences between the FUQ and IUQ processes.

An early appearance of the term ``inverse UQ'' can be found in \cite{oberkampf2002verification}, in which it was also called ``backward problem''. Other researchers have called it ``inverse uncertainty propagation'' \cite{unal2011improved}. According to Oberkampf and Trucano \cite{oberkampf2002verification}, ``The backward problem asks whether we can reduce the output uncertainty by updating the statistical model using comparisons between computations and experiments''. \textit{This review paper aims to provide a comprehensive survey of the recently developed IUQ methodologies, with a focus on the physical models in system thermal-hydraulics (TH) codes}.

IUQ of the physical model parameters (PMPs) in system TH codes offers a representative case study. In FUQ of system TH codes, a significant uncertainty source comes from physical models, which are closure laws (also known as correlations or constitutive relationships) that are used to describe the transfer terms in the balance equations. These physical models govern the mass, momentum and energy exchange between the fluid phases and surrounding medium. When the closure models were originally developed, their accuracy were studied with a particular experiment, called separate effect tests (SETs). Once they are implemented in a system TH code as empirical correlations and used for prediction at different experimental conditions, the accuracy and uncertainty characteristics of these correlations are no longer known to the code user. IUQ of the physical model uncertainty is therefore essential due to the wide application of system TH codes.

Estimating the uncertainties in physical models of system TH codes is a difficult problem because these models are not directly measurable in the majority of the cases. It is widely accepted that IUQ of system TH physical models should rely on SETs that exhibit relevant phenomena. SETs are usually simple tests that involve one physical phenomenon that can be described by one physical model. Direct comparison of code simulations with experimental data for the selected phenomena can be used for IUQ. When no SETs exist for a certain physical model, combined effect tests (CETs, also called intermediate experiments) can be used. In this more frequent case, several physical models must be considered together for IUQ based on CETs.

\subsection{Overview of International Benchmarks Related to BEPU and UQ}

There have been multiple international activities relevant to BEPU and UQ \cite{martin2021progress}. The Organisation for Economic Co-operation and Development/Nuclear Energy Agency (OECD/NEA) proposed and completed several projects targeting at the assessment of uncertainties associated with BE simulations, including UMS (Uncertainty Methods Study) \cite{wickett1998report}, BEMUSE (Best-Estimate Methods Uncertainty and Sensitivity Evaluation) \cite{glaeser2011bemuse} \cite{perez2011uncertainty}, UAM (Uncertainty Analysis in Modeling) \cite{ivanov2013benchmarks} \cite{bratton2014oecd} \cite{avramova2015summary}, etc. But international projects focusing on IUQ is relatively limited. The PREMIUM (Post-BEMUSE Reflood Models Input Uncertainty Methods) project \cite{reventos2016premium} \cite{mendizabal2017post} \cite{skorek2019quantification} was launched in 2012 with the main concern of comparing different methodologies to quantify the uncertainty of the physical models in system TH codes. Another notable project is the European Commission-funded NURESAFE project \cite{chanaron2017overview}, which also had a partial focus on IUQ of the closure laws.

The PREMIUM benchmark was completed in 2015 and is by far the most comprehensive international activity that deals with IUQ of phenomenon modeling for which no SETs are available. It brought together participants from sixteen institutions at eleven countries. Its scope includes a review of the existing IUQ methods, identification of influential PMPs, quantification of their uncertainties, and verification/validation of the IUQ results. PREMIUM focused on a concrete case: core reflood that takes place at the end of a large break loss-of-coolant accident. Two reflood experimental test facilities were selected, with the German FEBA (Flooding Experiments with Blocked Arrays) tests \cite{ihle1984feba1} \cite{ihle1984feba2} used for quantification and verification, and the French PERICLES tests \cite{deruaz1985study} as blind tests for the validation of the IUQ results. The verification and validation of IUQ results were performed based on the so-called ``envelop calculations'', i.e., whether FUQ with the physical model uncertainties obtained from IUQ can fully cover/envelop the observation data. A great variety of system TH codes were involved in PREMIUM, including TRACE, RELAP5, CATHARE, ATHLET, COBRA-TF, MARS-KS, KORSAR and APROS.

Besides FEBA and PERICLES, there are several international benchmarks that can provide data for different physical phenomena in various types of nuclear reactors. A few notable examples are: ACHILLES \cite{denham1989achilles}, Finnish VEERA \cite{puustinen1994veera} reflooding experiments, OECD/NEA MSLB (Main Steam Line Break) \cite{ivanov1999pressurised}, OECD/NRC NUPEC BFBT (BWR Full-size Fine-mesh Bundle Test) \cite{neykov2006nupec}, OECD/NRC PSBT (PWR Sub-channel and Bundle Test) \cite{rubin2012oecd}, etc. Among these experiments, BFBT and PSBF tests have also been widely used for IUQ of system TH codes. However, more extensive research employing the other experiments are needed in the nuclear community.

The PREMIUM benchmark was a first step toward the development and application of IUQ methods. However, the IUQ results were widely dispersed among the codes and methods used by the participants \cite{mendizabal2017post} \cite{skorek2019quantification}. The main reason was a lack of common consensus and practices in the followed process and method. OECD/NEA proposed another project in 2017, called SAPIUM (Systematic APproach for Input Uncertainty quantification Methodology) \cite{baccou2019development-NED}, to develop a systematic approach for quantification and validation of the uncertainty of the physical models in system TH codes. The main outcome of the project \cite{baccou2020sapium} is a first good-practices document that can be exploited for safety study in order to reach consensus among experts on recommended practices in IUQ.

\subsection{Overview of Available IUQ Methods for System TH Codes}
\label{section:Overview-IUQ-Methods}

Even though the nuclear community didn't witness a wide interest on IUQ research until the early 2010s, there have been a lot of IUQ methods available today, mainly owing to the PREMIUM project. Some of these methods were introduced from other disciplines but have been adapted/improved for applications in system TH codes. The majority of these IUQ methods are based on statistical analysis, and they can be categorized by three main groups: \textit{frequentist}, \textit{Bayesian}, and \textit{empirical} \cite{baccou2020sapium}. They are sometimes referred to as \textit{deterministic (optimization-based)}, \textit{probabilistic (sampling-based)}, and \textit{design-of-experiments (DoE, forward propagation-based)}, respectively.

The \textit{frequentist/deterministic IUQ methods} consider that the PMPs have fixed but unknown values. Consequently, IUQ is formulated as an optimization problem, more specifically, maximization of the likelihood to find the ``best-fit'' values. Note that this does not necessarily mean the results of frequentist IUQ are always point estimates. It is also common for the frequentist IUQ methods to treat the PMPs to follow normal or log-normal distributions, and to evaluate the ``best-fit'' distributional parameters (e.g., mean and variance for a normal distribution) instead of the PMPs themselves.

The \textit{Bayesian/probabilistic IUQ methods} also assume that the PMPs have true but unknown values, but always use probabilistic treatment of these parameters with uncertain distributions, because it is impossible to quantify the exact values given limited available information. The Bayesian IUQ methods are built upon the Bayes' rule as a procedure to update information after observing experimental data \cite{gelman2013bayesian}. Knowledge about the physical model uncertainties is first characterized as prior distributions, which will be updated to posterior distributions based on a comparison of model and data.

Both frequentist and Bayesian IUQ methods are built on rigorous mathematical frameworks. The \textit{empirical/DoE IUQ methods}, as the name suggests, are based on adjusting the parameters in a trial-and-error manner without a robust mathematical basis. The general idea is to first generate random samples of the PMPs according to a prescribed uncertainties, propagate the uncertainties to QoIs by FUQ,  compare with measurement data and try to fulfil certain requirement, such as coverage of the physical data by the simulations. Because several iterations may be needed to adjust the parameter uncertainties before a desired coverage rate is satisfied, the empirical IUQ methods can be computationally expensive. Efficient DoE procedures can be exploited to reduce the cost by reducing the number of samples. The empirical IUQ methods still involve a considerable amount of engineering judgment. It is worth noting that Bayesian and empirical IUQ methods rely on completely different types of sampling. Empirical IUQ uses random Monte Carlo sampling for forward propagation of parameter uncertainties, while Bayesian IUQ uses Markov Chain Monte Carlo (MCMC) sampling to explore non-standard posterior PDFs.

\begin{table}[!ht]
	\scriptsize
	\centering
	\captionsetup{justification=centering}
	\caption{List of available IUQ methods that have been applied to physical models of system TH codes.}
	\label{table:Intro-List-of-IUQ-Methods}
	\begin{tabular}{p{0.4\linewidth} | p{0.2\linewidth} | p{0.15\linewidth} | p{0.1\linewidth}}
		\toprule
		IUQ methods  &  Category  &  References  &  PREMIUM?  \\ 
		\midrule
		CIRC{\'E} (CEA, France)                   &  Frequentist  &  \cite{de2001determination} \cite{de2004quantification} \cite{de2012circe}  &  yes   \\
		
		IPREM (UNIPI, Italy)                      &  Empirical    &  \cite{kovtonyuk2012procedure} \cite{kovtonyuk2014development} \cite{kovtonyuk2017development}  &  yes   \\
		
		CET-based Sample Adjusting (GRS, Germany) &  Empirical    &  \cite{skorek2017input}  &  yes   \\
		
		DIPE (IRSN, France)                       &  Empirical    &  \cite{joucla2008dipe}  &  yes   \\
		
		MCDA (KAERI, South Korea)                 &  Frequentist/Bayesian  &  \cite{heo2014implementation}  &  yes   \\
		
		Sampling-based IUQ (Tractebel, Belgium)   &  Empirical    &  \cite{zhang2019development}  &  yes   \\
		
		Modular Bayesian Approach (UIUC, USA)     &  Bayesian     &  \cite{wu2018inversePart1} \cite{wu2018inversePart2} \cite{wu2018Kriging} \cite{wu2019demonstration}  &  no   \\
		
		Bayesian CIRC{\'E} (CEA Saclay, France)   &  Bayesian     &  \cite{damblin2018bayesian} \cite{damblin2020bayesian}  &  no   \\
		
		Non-parametric Clustering (PSI, Switzerland)  &  Empirical  &  \cite{vinai2007statistical}  &  no   \\
		    
		Data Adjustment and Assimilation (KIT, Germany)  &  Frequentist  &  \cite{cacuci2010best} \cite{petruzzi2010best} \cite{cacuci2010sensitivity}  &  no   \\
		    
		CASUALIDAD (NINE, Italy)                  &  Frequentist  &  \cite{petruzzi2008development} \cite{petruzzi2014uncertainties} \cite{petruzzi2019casualidad}  &  no   \\
		
		MLE and MAP (UIUC, USA)                   &  Frequentist/Bayesian  &  \cite{shrestha2016inverse} \cite{hu2016inverse} \cite{saleem2019effect} \cite{saleem2019estimation}  &  no   \\
		\bottomrule
	\end{tabular}
\end{table}

All these three groups of IUQ methods depend on a comparison between code simulations and physical observations, though in different manners. Frequentist IUQ tries to identify most likely PMP values, with which the TH model can reproduce the experimental data. Bayesian IUQ targets at reducing the disagreement between simulation and data, while finding parameter uncertainties that can explain the disagreement. Empirical IUQ seeks parameter ranges with which the model predictions can envelop the measurement data to a desired level. Because of these different mechanisms, these three types of IUQ methods have very different assumptions, application scenarios, treatment of various sources of uncertainties, etc. This review paper aims at providing a comprehensive and comparative discussion of the major aspects of these IUQ methods.

Table \ref{table:Intro-List-of-IUQ-Methods} presents a list of IUQ methods that have been used in the nuclear community, as well as their categories and original/representative references. We have specifically focused on IUQ methods that have been applied for IUQ of PMPs in system TH codes. Table \ref{table:Intro-List-of-IUQ-Methods} also indicates whether an IUQ method was used in the PREMIUM benchmark or not. Due to the limitation in the authors' knowledge, there may be other IUQ methods that are not included in Table \ref{table:Intro-List-of-IUQ-Methods}.

A good IUQ method should: (1) be able to capture the input uncertain distributions, instead of finding the best-fit point estimates, (2) have a comprehensive formulation, which means it can simultaneously consider all available sources of uncertainties, (3) be able to avoid over-fitting, since IUQ is inherently a calibration process, which makes it easy to be over-fitted to the selected data, (4) be easily understandable and applicable, etc. \textit{In this paper, we used eight metrics to evaluate an IUQ method, including solidity, complexity, accessibility, independence, flexibility, comprehensiveness, transparency, and tractability. Twelve IUQ methods are reviewed, compared, and evaluated based on these eight metrics. Such comparative evaluation will provide a good guidance for users to select a proper IUQ method based on their IUQ problem under investigation.}

\subsection{IUQ vs. Calibration}

There are some differences between IUQ and calibration. IUQ by definition is similar to several applications in other areas, including \textit{inverse problems} \cite{tarantola2005inverse} \cite{kaipio2006statistical} \cite{stuart2010inverse} \cite{kirsch2011introduction}, \textit{Bayesian calibration} \cite{kennedy2001bayesian} \cite{wilkinson2010bayesian}, \textit{parameter estimation} \cite{aster2018parameter}, etc. Calibration can be classified as deterministic and statistical \cite{campbell2006statistical}. \textit{Deterministic calibration}, or \textit{parameter tuning}, determines the point estimates of best-fit input parameters such that the discrepancies between code and data can be minimized. \textit{Statistical calibration}, also called \textit{Bayesian calibration} \cite{kennedy2001bayesian}, \textit{probabilistic inversion} \cite{van2005bayesian}, or \textit{calibration under uncertainty} \cite{trucano2006calibration}, is very similar to Bayesian IUQ because Bayesian inference \cite{gelman2013bayesian} and MCMC sampling \cite{brooks2011handbook} are employed.

\begin{figure}[ht]
	\centering
	\includegraphics[width=0.75\textwidth]{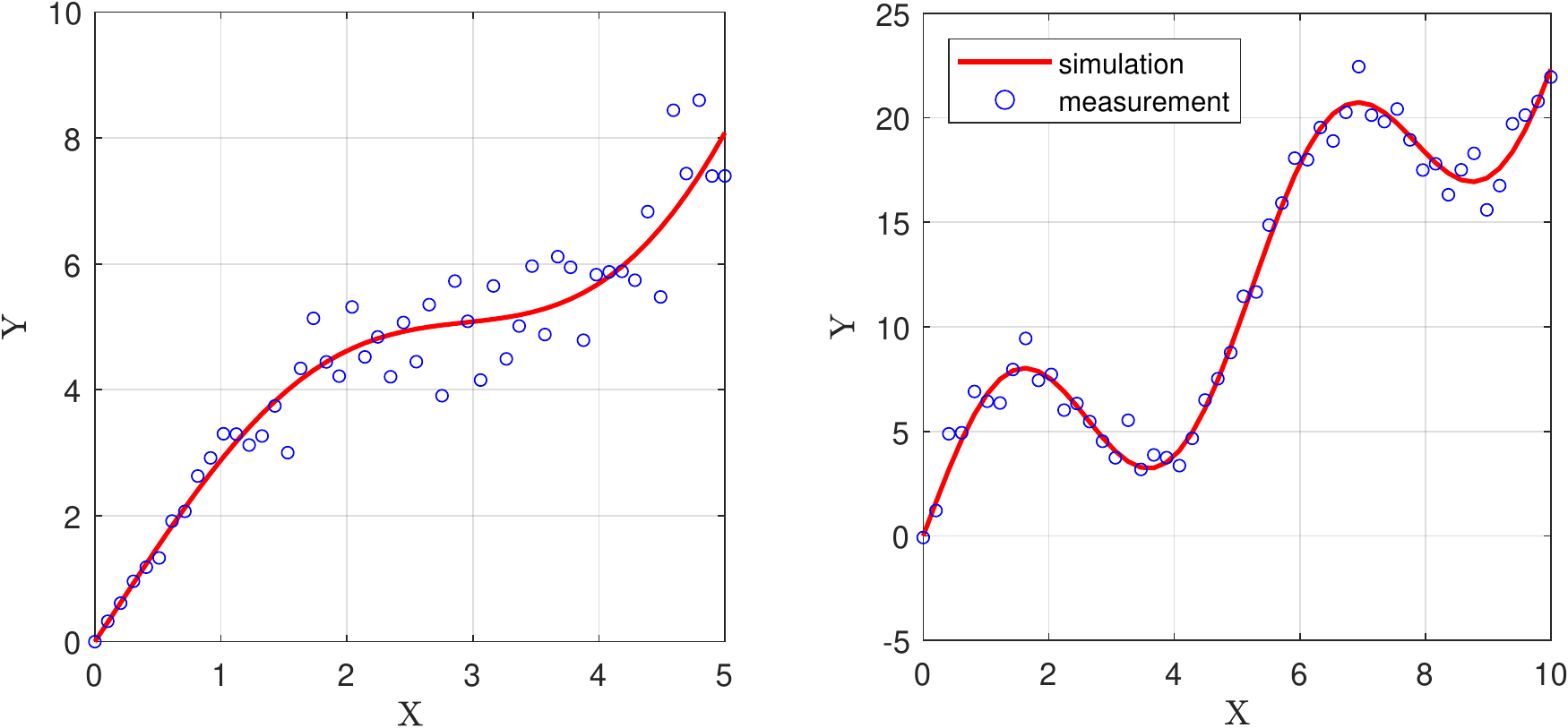}
	\caption[]{Two simple cases when calibration will not improve the agreement between model and data.}
	\label{figure:Intro2-Calibration-vs-IUQ}
\end{figure}

Despite the similarity in statistical calibration and Bayesian IUQ, IUQ and calibration should be treated as different activities because they have different goals: calibration aims at reducing the differences between model and data, while IUQ aims at inferring the uncertainties in the calibration parameters. When the model simulation agrees very well with experimental data, no calibration deemed necessary. However, IUQ may still be needed because the uncertainties in model input parameters have to be quantified. Figure \ref{figure:Intro2-Calibration-vs-IUQ} illustrates two simple cases, in which the differences between model and data approximately follow zero-mean Gaussian noise. In these cases, calibration won't improve the agreement between model and data, but IUQ is still useful. Moreover, there are many non-Bayesian IUQ methods available, as shown in Table \ref{table:Intro-List-of-IUQ-Methods}.

The advantage of IUQ over deterministic calibration is more apparent. Firstly, information from physical observations is usually limited and never sufficiently accurate to allow inference of the ``true'' values of the unknown input parameters. Therefore, the ignorance in the parameters can be reflected by an uncertainty characterization with distributions. Secondly, it is impossible for deterministic calibration to quantify correlations between different input parameters. Correlations are usually calculated based on samples but deterministic calibration only estimates best-fit values. Thirdly, inverse problems are usually ill-posed due to the fact that multiple combinations of input parameters can yield simulations that have similar agreement with the data. Deterministic calibration, which relies on optimization techniques to select best-fit values, may end up with getting only one of a set of equally well-fitting values. This is especially true for over-parameterized models given limited data \cite{van2005bayesian}. Finally, the observed data usually contains certain degree of uncertainty, which can be considered in many IUQ methods but not easily in deterministic calibration.

This review paper is organized as follows. Section 1 introduces the concept of FUQ and IUQ and compares their differences. Brief overviews of international benchmarks and available IUQ methods are also provided. Section 2 provides a mathematical formulation of the IUQ problem. Section 3 introduces the Bayesian IUQ method that originates from the ``model updating equation'' and employs surrogate-based MCMC sampling. A comparison of full and modular Bayesian approaches will also be included. Section 4 contains a brief but self-contained descriptions of all the IUQ methods used in the PREMIUM benchmark. Section 5 summarizes the major aspects of the SAPIUM project. Section 6 reviews a few other IUQ methods and their applications in the nuclear community. Section 7 discusses the advantages and disadvantages of all the 12 reviewed IUQ methods, and evaluates them based on 8 metrics. Section 8 presents a list of IUQ challenges and research needs for future development. Section 9 concludes this review paper.

\section{Definition of the IUQ Problem}
\label{section:Definition-of-IUQ}

Consider a general computer model $\mathbf{y}^{\text{M}} = \mathbf{y}^{\text{M}} ( \mathbf{x}, \bm{\theta} )$ where $\mathbf{y}^{\text{M}}$ is the model output (also called response or QoI) which can be either a scalar or vector that corresponds to multi-dimensional outputs. The computer model has two types of inputs. $\mathbf{x}$ denotes the \textit{design variables}, while $\bm{\theta}$ represents the \textit{calibration parameters}. Table \ref{table:Bayesian-IUQ1-X-vs-Theta} presents a detailed comparison of $\mathbf{x}$ and $\bm{\theta}$. The distinction between $\bm{x}$ and $\bm{\theta}$ is not important for many purposes like FUQ and sensitivity analysis. But in IUQ, calibration parameters are the inputs to be estimated. It does not make physical sense to calibrate the values or quantify the uncertainties in $\bm{x}$ because they change from one experiment to another. Information learned about $\bm{x}$ from one experiment cannot be applied to another experiment. On the other hand, $\bm{\theta}$ are inputs associated with the computer model only. They have unknown true values that are invariant for different experiments.

Some PMPs, such as nuclear fuel thermal conductivities or heat transfer coefficients (HTCs), are classified as calibration parameters in Table \ref{table:Bayesian-IUQ1-X-vs-Theta}, but they are dependent on factors such as temperature, burnup, etc. As a result, they will change in different experiments. In this case, the calibration parameters are not the fuel thermal conductivities or HTCs themselves, but the coefficients in their empirical formulas that express the dependence on other factors (e.g., temperature, burnup).

\begingroup
\renewcommand{\arraystretch}{1.2}
\begin{table}[ht]
	\footnotesize
	\captionsetup{justification=centering}
	\caption{Comparison of the design variables $\mathbf{x}$ and the calibration parameters $\bm{\theta}$.}
	\label{table:Bayesian-IUQ1-X-vs-Theta}
	\centering
	\begin{tabular}{p{0.12\textwidth} p{0.4\textwidth}  p{0.4\textwidth}}
		\toprule
		&  Design variables $\mathbf{x}$  &  Calibration parameters $\bm{\theta}$ \\ 
		\midrule
		Needed by $\mathbf{y}^{\text{M}}$ or $\mathbf{y}^{\text{E}}$
		&
		$\mathbf{x}$ are needed by both the computer model $\mathbf{y}^{\text{M}} ( \mathbf{x}, \bm{\theta} )$ and experiment $\mathbf{y}^{\text{E}} (\mathbf{x})$.
		& 
		$\bm{\theta}$ are only needed by the computer model $\mathbf{y}^{\text{M}} ( \mathbf{x}, \bm{\theta} )$.
		\\ 
		\midrule
		Physical meaning
		&
		$\mathbf{x}$ usually have clear and unambiguous physical meaning.
		& 
		$\bm{\theta}$ may have a physical meaning in nature or be purely numerical.
		\\ 
		\midrule
		Relation to experiments.
		&
		$\mathbf{x}$ are used to describe the conditions or scenarios under which the experiments have been performed.
		& 
		$\bm{\theta}$ have inherent values that remain unchanged under different scenarios or experimental conditions.
		\\ 
		\midrule
		Value and uncertainty
		&
		$\mathbf{x}$ are usually assumed to be known or at least observable during experimentation. These variables may also be subject to uncertainties due to known ``variability'' that may be reported along with the benchmark.
		& 
		$\bm{\theta}$ are usually subjected to unknown uncertainties (aleatory, epistemic or mixed).
		\\ 
		\midrule
		Other names
		&
		system inputs, experimental conditions, controllable/observable variables
		& 
		ancillary variables, fitting/model parameters
		\\ 
		\midrule
		Examples
		&
		Initial conditions (e.g., initial temperature), boundary conditions (e.g., pressure, power, mass flow), geometries of the physical system.
		& 
		PMPs (e.g., heat transfer coefficients), tuning parameters (multiplicative and additive factors), context-specific constants (e.g., switch between different scenarios)
		\\
		\bottomrule
	\end{tabular}
\end{table}
\endgroup

Note that some different classifications on model inputs have been used in the literature. In the work of Nouy and De Cr{\'e}cy \cite{nouy2017quantification}, the authors grouped the model inputs as: (1) \textit{input global parameters}, which are inputs associated with a physical model, (2) \textit{input basic parameters}, like boundary or initial conditions, geometrical or material property parameters or discretization parameters, and (3) \textit{input coefficient parameters}, which are single coefficients inside correlations. Such categorization is not straightforward for the purpose of IUQ, because all groups contain parameters that can be calibrated. In another work by Bachoc et al. \cite{bachoc2014calibration}, the vector $\mathbf{x}$ was called \textit{experimental conditions}, which was further categorized as control variables and environment variables. \textit{Control variables} define the physical system, independently of the environment, such as the geometric parameters of the system that remain fixed regardless of what happens to the system. \textit{Environment variables} are the inputs of the physical system such as initial/boundary conditions. This work referred to $\bm{\theta}$ as \textit{fitting/model parameters}. In the work of Campbell \cite{campbell2006statistical}, ``inputs'' and ``parameters'' are used to represent $\mathbf{x}$ and $\bm{\theta}$ respectively. However, this can easily cause confusion for the readers. Therefore, we assume no difference between ``input", ``variable'' and ``parameter'', and will use ``design'' and ``calibration'' in front of these terms to explicitly refer to $\mathbf{x}$ and $\bm{\theta}$. Similarly, we use ``experiment'', ``observation'' and ``measurement'' interchangeably in this work, assuming no difference between them, and also use ``physical'' or ``field'' in front of them in order to be consistent with the open literature.

\textit{The goal of IUQ can be stated as: given experimental data $\mathbf{y}^{\text{E}} (\mathbf{x})$, inversely quantify the parameter uncertainties in $\bm{\theta}$ such that $\mathbf{y}^{\text{M}} ( \mathbf{x}, \bm{\theta} )$ is consistent with $\mathbf{y}^{\text{E}} (\mathbf{x})$}. Inverse problems are usually ill-posed because the solution is generally non-unique. The error between model simulation and measurement data is due to several sources of uncertainties, while there can be a myriad of combinations of these sources that lead to the same total error. In this paper, we classify the quantifiable uncertainties in M\&S by four different sources:
\begin{enumerate}[label=(\arabic*)]
	\setlength{\itemsep}{0.1pt}
	\item \textbf{Parameter uncertainty}, originated from ignorance in the exact values of input parameters (\textit{epistemic uncertainty}) or randomness (\textit{aleatory uncertainty}). They are associated with $\mathbf{x}$ and $\bm{\theta}$. We assume that uncertainties in $\mathbf{x}$ are known from the benchmark. Inferring the parameter uncertainties in $\bm{\theta}$ is the goal of IUQ.
	
	\item \textbf{Model uncertainty}, due to inaccurate and/or incomplete underlying physics incorporated in the computer models, as well as numerical approximation errors. Even though the verification process tries to minimize the coding and numerical errors, it can never eliminate them. Model uncertainties can also be learned during the IUQ process. Note that sometimes numerical approximation errors are referred to as a separate type called \textbf{numerical uncertainty}. But generally it is difficult to separate it from model uncertainty, and they can be treated in a similar way, so we consider numerical uncertainty as a part of model uncertainty.
	
	\item \textbf{Experiment uncertainty}, caused by measurement error. Generally, it is considered to be reported along with the experimental data.
	
	\item \textbf{Code uncertainty}, owing to the emulation of computationally prohibitive codes using surrogate models (also called metamodels). This source become zero when the original computational model is used instead of a surrogate model.
\end{enumerate}

A comprehensive IUQ method should be able to simultaneously consider all these sources of uncertainties. However, in practice it is usually difficult to do so. We will provide a more detailed discussion of the so-called non-identifiability issue in Section \ref{section:Identifiability}, which is a direct result of the ill-posedness of IUQ. Compare to the most classical example of calibration: least squares, which consists of finding the best-fit estimate of $\bm{\theta}$ that minimizes the quadratic misfit between data and model, the most notable feature of IUQ is the quantification of the uncertainties in $\bm{\theta}$.

\section{The Full and Modular Bayesian Approaches}

One of the most successful approaches for IUQ is based on Bayesian inference \cite{gelman2013bayesian}. In this review we will refer to it as Bayesian IUQ or Bayesian calibration. Bayesian IUQ seeks the posterior distributions of the uncertain input parameters, which are updated from the prior knowledge given experimental data. Most previous work on Bayesian IUQ followed the seminal work of Kennedy and O'Hagan \cite{kennedy2001bayesian}. In this section, we will first briefly introduce the fundamentals of Bayesian inference, followed by the Bayesian IUQ formulation. Two different solution processes, the Full Bayesian Approach (FBA) and Modular Bayesian Approach (MBA), will be introduced and compared. Finally, we will review the previous applications of Bayesian IUQ and its variations for nuclear TH codes.

\subsection{Fundamentals of Bayesian Inference}

\begin{figure}[ht]
	\centering
	\includegraphics[width=0.75\textwidth]{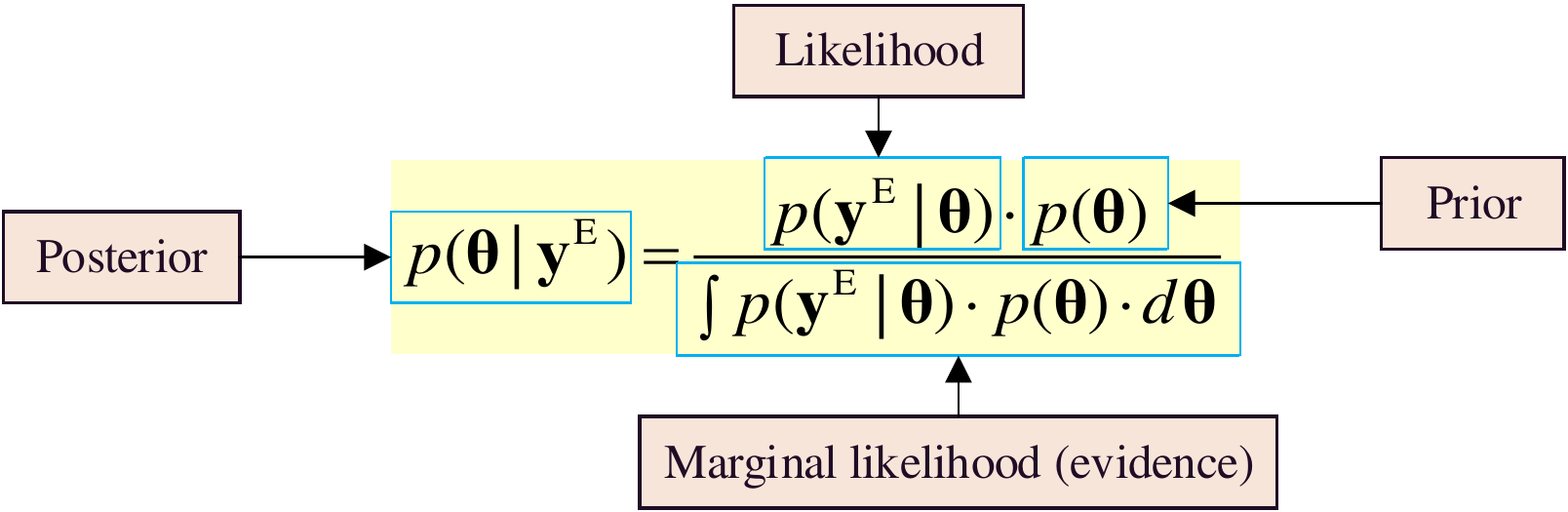}
	\caption[]{The Bayes' rule.}
	\label{figure:Bayesian-IUQ1-Bayes-Rule}
\end{figure}

Figure \ref{figure:Bayesian-IUQ1-Bayes-Rule} shows the Bayes rule. The \textit{prior} $p (\bm{\theta})$ is the knowledge about $\bm{\theta}$ \textit{before} observing the data $\mathbf{y}^{\text{E}}$. The \textit{posterior} $p ( \bm{\theta} | \mathbf{y}^{\text{E}} )$ is the updated knowledge about $\bm{\theta}$ \textit{after} observing $\mathbf{y}^{\text{E}}$. The probability to observe $\mathbf{y}^{\text{E}}$ given certain values of $\bm{\theta}$ is the \textit{likelihood} $p ( \mathbf{y}^{\text{E}} | \bm{\theta} )$. Prior and posterior represent degrees of belief about possible values of $\bm{\theta}$, before and after observing the data $\mathbf{y}^{\text{E}}$, which enters the formulation through the likelihood. The normalizing constant $p(\mathbf{y}^{\text{E}}) = \int p ( \mathbf{y}^{\text{E}} | \bm{\theta} ) p(\bm{\theta}) d \bm{\theta}$ is also called the \textit{evidence}, and it does not contain $\bm{\theta}$. Posterior is the solution to IUQ. Prior is chosen by the user, and it can be based on expert opinion or user self-evaluation. Bayesian IUQ needs a likelihood function to proceed.
\begin{equation}
	p ( \bm{\theta} | \mathbf{y}^{\text{E}} ) \propto  p ( \mathbf{y}^{\text{E}} | \bm{\theta} ) \cdot p (\bm{\theta})
\end{equation}

Define $\bm{\theta}^{*}$ as the true but unknown values for $\bm{\theta}$, with which the model produces the most accurate prediction. $\bm{\theta}^{*}$ is the ``best'' value only in the sense of most accurately representing the measurement data. Due to model bias, model prediction may not agree well with reality even when the model runs at $\bm{\theta}^{*}$. However, since the true value can never be learned, by convention the best value and true value are treated as the same.

\subsection{Bayesian IUQ}

\begin{figure}[htbp]
	\centering
	\includegraphics[width=0.9\textwidth]{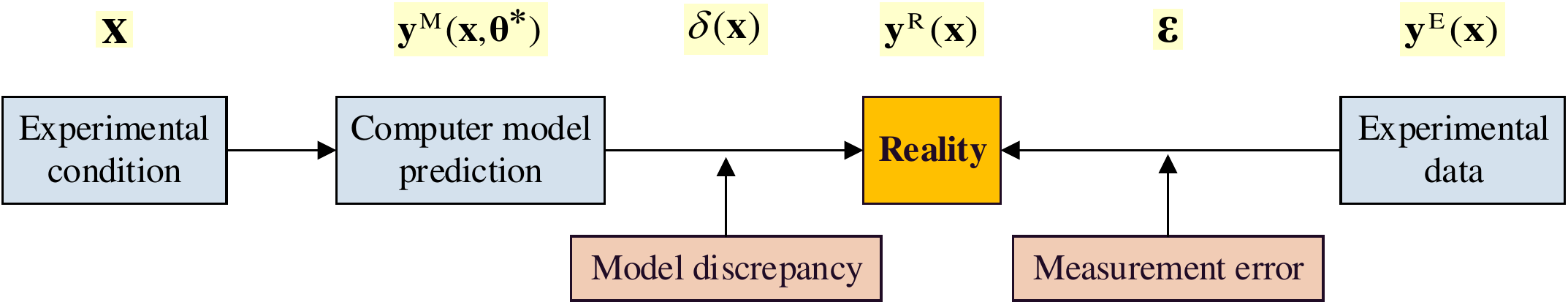}
	\caption[]{The connections between computer model prediction, reality and experimental data.}
	\label{figure:Bayesian-IUQ2-Model-Update-Equation}
\end{figure}

Table \ref{table:Bayesian-IUQ2-Symbols} presents the symbols and definitions used for Bayesian IUQ. Define the unknown reality or true value of the output as $\mathbf{y}^{\text{R}} (\mathbf{x})$. As illustrated in Figure \ref{figure:Bayesian-IUQ2-Model-Update-Equation}, given an experimental condition characterized by design variables $\mathbf{x}$, the reality $\mathbf{y}^{\text{R}} (\mathbf{x})$ can be predicted by computer model simulation. The computer model is only an approximation of the reality:
\begin{equation}	 \label{equation:Bayesian-IUQ1-Bias}
	\mathbf{y}^{\text{R}} (\mathbf{x}) = \mathbf{y}^{\text{M}} \left( \mathbf{x}, \bm{\theta}^{*} \right) + \delta(\mathbf{x})
\end{equation}

\begingroup
\renewcommand{\arraystretch}{1.3}
\begin{table}[!ht]
	\footnotesize
	\centering
	\captionsetup{justification=centering}
	\caption{Definitions of symbols used for Bayesian IUQ.}
	\label{table:Bayesian-IUQ2-Symbols}
	\begin{tabular}{p{0.08\linewidth} | p{0.32\linewidth} | p{0.08\linewidth} | p{0.32\linewidth}}
		\toprule
		Symbol  &  Description  &  Symbol  &  Description  \\ 
		\midrule
		$\mathbf{x}$                 &  design variables   &  $\mathbf{y}$  &  QoIs  \\
		
		$\mathbf{x}^{\text{IUQ}}$    &  IUQ (calibration) domain  &  $\mathbf{y}^{\text{M}}$  &  QoIs from model simulation  \\
		
		$\mathbf{x}^{\text{VAL}}$    &  validation domain  &  $\mathbf{y}^{\text{E}}$  &  QoIs from experiment  \\
		
		$\mathbf{x}^{\text{PRED}}$   &  prediction domain  &  $\mathbf{y}^{\text{R}}$  &  QoIs' unknown real values  \\
		
		$\bm{\theta}$                &  calibration parameters  &  $\bm{\Sigma}$  &  total uncertainty of the likelihood  \\
		
		$\bm{\theta}^{*}$            &  true but unknown values of $\bm{\theta}$  &  $\bm{\Sigma}_{\text{exp}}$  &  experimental uncertainty  \\
		
		$\delta(\mathbf{x})$         &  model bias  &  $\bm{\Sigma}_{\text{bias}}$  &  model uncertainty  \\
		
		$\bm{\epsilon}$              &  measurement noise  &  $\bm{\Sigma}_{\text{code}}$  &  code uncertainty  \\
		$\bm{\Sigma}_{\bm{\epsilon}}$   &  covariance matrix of $\bm{\epsilon}$  &  &  \\
		\bottomrule
	\end{tabular}
\end{table}
\endgroup

The learning of $\bm{\theta}^{*}$ is the goal of IUQ. $\delta(\mathbf{x})$ is the \textit{model bias}, also called \textit{model uncertainty}, \textit{model discrepancy}, \textit{model inadequacy} or \textit{model error} \cite{kennedy2001bayesian}. The model bias is due to incomplete or inaccurate underlying physics, numerical approximation errors, and/or other inaccuracies that would exist even if $\bm{\theta}^{*}$ was known \cite{arendt2012quantification}. To learn the reality $\mathbf{y}^{\text{R}} (\mathbf{x})$, we may also perform experiments to get observation $\mathbf{y}^{\text{E}} (\mathbf{x})$. The experimentation process will inevitably introduce measurement noise:
\begin{equation}	 \label{equation:Bayesian-IUQ2-Data-Noise}
	\mathbf{y}^{\text{E}} (\mathbf{x}) = \mathbf{y}^{\text{R}} (\mathbf{x}) + \bm{\epsilon}
\end{equation}
where $\bm{\epsilon} \sim \mathcal{N} ( \bm{\mu}, \bm{\Sigma}_{\bm{\epsilon}}  )$ represents the \textit{measurement error}. There can be multiple measurements that have homoscedastic experimental errors $\bm{\Sigma}_{\bm{\epsilon}} = \sigma_{\epsilon}^{2} \mathbf{I}$. Also, $\bm{\mu} = \mathbf{0}$ is frequently used, assuming that the instrumentation has no systematic bias and the mean value of the measurement is same with reality. Combining Equations (\ref{equation:Bayesian-IUQ1-Bias}) and (\ref{equation:Bayesian-IUQ2-Data-Noise}):
\begin{equation}        \label{equation:Bayesian-IUQ3-Model-Update-Eqn}
	\mathbf{y}^{\text{E}} (\mathbf{x}) = \mathbf{y}^{\text{M}} ( \mathbf{x}, \bm{\theta}^{*} ) + \delta(\mathbf{x}) + \bm{\epsilon}
\end{equation}

Equation (\ref{equation:Bayesian-IUQ3-Model-Update-Eqn}) is referred to as the \textit{model updating formulation} \cite{kennedy2001bayesian} \cite{arendt2012quantification}, which serves as the starting point of Bayesian IUQ. The measurement error $\bm{\epsilon}$ is usually assumed to be i.i.d. zero-mean Gaussian, whose variance is expected to be reported along with measurement data since the error rates for most instrumentation are known. In other words, $$\bm{\epsilon} = \mathbf{y}^{\text{E}} (\mathbf{x}) - \mathbf{y}^{\text{M}} ( \mathbf{x}, \bm{\theta}^{*} ) - \delta(\mathbf{x})$$ follows a multi-dimensional Gaussian distribution that has mean zero and covariance matrix $\bm{\Sigma}_{\bm{\epsilon}}$. The likelihood function $p ( \mathbf{y}^{\text{E}}, \mathbf{y}^{\text{M}} | \bm{\theta}^{*} )$ can be written as:
\begin{equation}        \label{equation:Bayesian-IUQ4-Likelihood}
	p \left( \mathbf{y}^{\text{E}}, \mathbf{y}^{\text{M}} | \bm{\theta}^{*} \right)
	\propto  \frac{ 1 }{ \sqrt{| \bm{\Sigma} |} }  \text{exp} \left[  - \frac{1}{2} \left( \mathbf{y}^{\text{E}} - \mathbf{y}^{\text{M}} - \delta \right)^\top \bm{\Sigma}^{-1} \left( \mathbf{y}^{\text{E}} - \mathbf{y}^{\text{M}} - \delta \right) \right]
\end{equation}

It denotes the probability of observing $\mathbf{y}^{\text{E}}$ given the model input parameters $\bm{\theta}^{*}$. Note that the variance of the likelihood is $\bm{\Sigma}$ instead of $\bm{\Sigma}_{\text{exp}}$ because the inclusion of extra uncertainties. The variance $\bm{\Sigma}$ will be explained later. The posterior $p \left( \bm{\theta}^{*} | \mathbf{y}^{\text{E}}, \mathbf{y}^{\text{M}}\right)$ becomes:
\begin{multline}        \label{equation:Bayesian-IUQ5-Posterior}
	p \left( \bm{\theta}^{*} | \mathbf{y}^{\text{E}}, \mathbf{y}^{\text{M}}\right)  \propto  p \left( \bm{\theta}^{*} \right) \cdot p \left( \mathbf{y}^{\text{E}}, \mathbf{y}^{\text{M}} | \bm{\theta}^{*} \right) 	\\
	\propto  p \left( \bm{\theta}^{*} \right) \cdot \frac{1}{\sqrt{|\bm{\Sigma}|}}  \cdot \text{exp} \left[  - \frac{1}{2} \left( \mathbf{y}^{\text{E}} - \mathbf{y}^{\text{M}} - \delta \right)^\top \bm{\Sigma}^{-1} \left( \mathbf{y}^{\text{E}} - \mathbf{y}^{\text{M}} - \delta \right) \right]
\end{multline}
where $p \left( \bm{\theta}^{*} \right)$ is the prior. In the following we will use $\bm{\theta}$ instead of $\bm{\theta}^{*}$ to represent the target of IUQ. In Bayesian analysis, all unknowns are considered random. Therefore, we drop the superscript for notational simplicity.

The covariance of the likelihood $\bm{\Sigma}$ consists of three parts:
\begin{equation}        \label{equation:Bayesian-IUQ6-Posterior-Variance}
	\bm{\Sigma} = \bm{\Sigma}_{\text{exp}} + \bm{\Sigma}_{\text{bias}} + \bm{\Sigma}_{\text{code}}
\end{equation}

The first part $\bm{\Sigma}_{\text{exp}}$ is the \textit{experimental uncertainty} due to measurement error. The second part $\bm{\Sigma}_{\text{bias}}$ represents the \textit{model uncertainty}. The third term $\bm{\Sigma}_{\text{code}}$ is called \textit{code uncertainty}, or \textit{interpolation uncertainty} \cite{kennedy2001bayesian}, because we do not know the computer code outputs at every input, especially when the code is computationally prohibitive. In this case, we might choose to use some kind of metamodels. Note that $\bm{\Sigma}_{\text{code}} = \mathbf{0}$ if the computer model is used instead of its surrogates. The formulation shown in Equation (\ref{equation:Bayesian-IUQ5-Posterior}) includes all the four major sources of the quantifiable uncertainties M\&S as discussed in Section \ref{section:Definition-of-IUQ}, i.e., parameter uncertainties as the target, experimental/model/code uncertainties in the covariance matrix, making it a very comprehensive formulation.

\begin{figure}[htbp]
	\centering
	\includegraphics[width=0.8\textwidth]{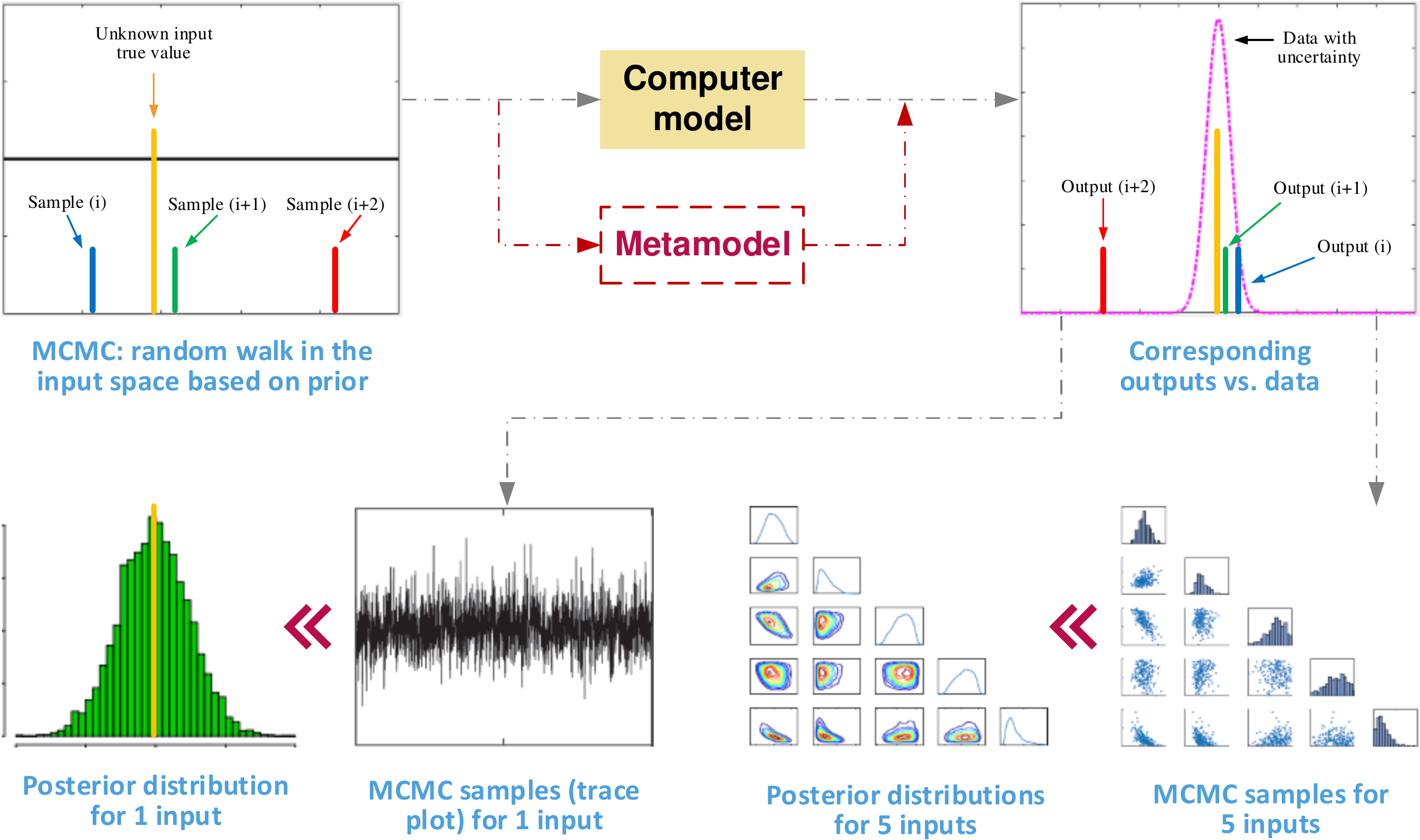}
	\caption[]{Illustration of MCMC sampling.}
	\label{figure:Bayesian-IUQ3-MCMC}
\end{figure}

The posterior PDF in Equation (\ref{equation:Bayesian-IUQ5-Posterior}) is non-standard and not normalized, needs numerical sampling to explore it. MCMC generates samples following a probability density which is proportional to the posterior PDF without knowing the normalizing constant. Figure \ref{figure:Bayesian-IUQ3-MCMC} illustrates the workflow of MCMC sampling. MCMC is widely used to sample from a complicated distribution without explicitly knowing the normalizing constant. The most popular MCMC algorithm is Metropolis-Hastings (MH) \cite{brooks2011handbook}, which defines a family of possible transitions from one Markov chain state to the next from a proposal distribution. Other widely used algorithms include Gibbs sampling, Adaptive Metropolis (AM) \cite{andrieu2008tutorial}, Delayed Rejection Adaptive Metropolis (DRAM) \cite{haario2006dram}, DiffeRential Evolution Adaptive Metropolis (DREAM) \cite{vrugt2008accelerating}, etc.

The most significant challenge of using MCMC to explore the posterior distributions is that a large number of model simulations are required. Typically MCMC requires over 10,000 samples to reach a good mixing. This can be infeasible when the computer code is very expensive to run. For MCMC sampling, surrogate models can be used to reduce the computational cost. Surrogate models, also called metamodels, response surfaces or emulators, are approximations of the input/output relation of the original computer model. They are built from a limited number of full model runs (training set) and a learning algorithm. Metamodels usually take much less computational time than the full model while maintaining the input/output relation to a desirable accuracy. Once validated, metamodels can be used in uncertainty, sensitivity, validation, and optimization studies, for which the original computer model can incur an excessive computational burden as hundreds or thousands of computer model simulations are needed.

\begin{figure}[htbp]
	\centering
	\includegraphics[width=0.9\textwidth]{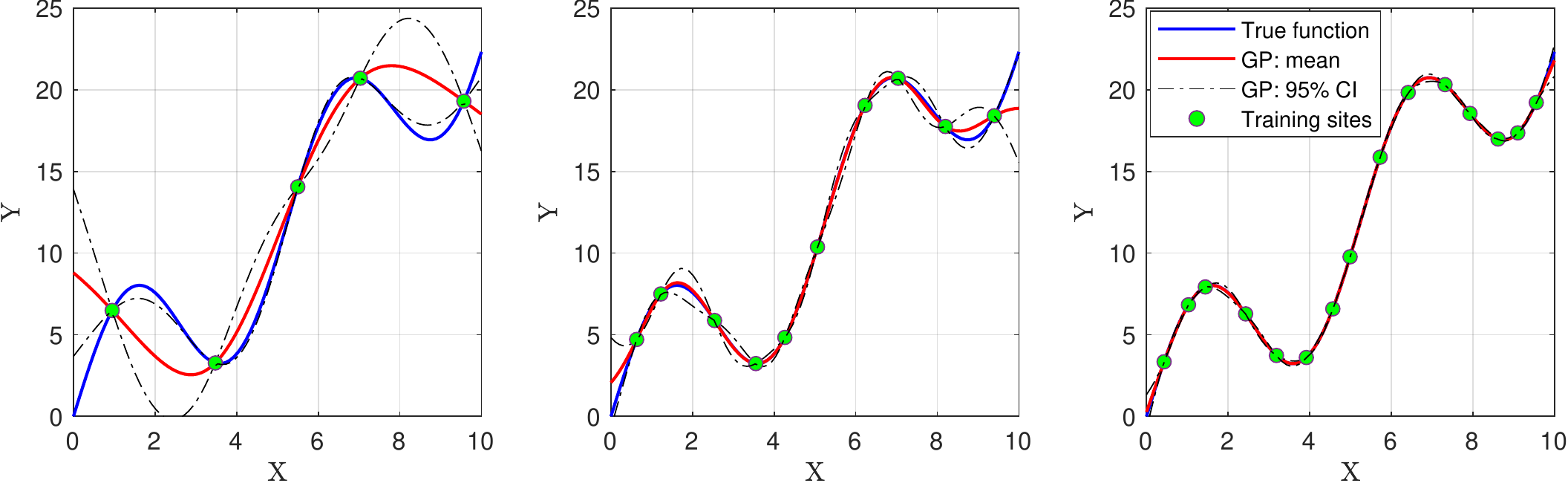}
	\caption[]{Demonstration of the capability of GP to approximate a function based on 5, 10 and 15 samples.}
	\label{figure:Bayesian-IUQ4-GP}
\end{figure}

Gaussian Process (GP, also called Kriging) \cite{jones1998efficient} \cite{santner2003design} is one of the most popular methods for surrogate modeling in Bayesian IUQ. Figure \ref{figure:Bayesian-IUQ4-GP} illustrates the way GP approximates a test function based on a limited number of training samples. At every input setting, GP estimation follows a Gaussian distribution with a mean prediction and a variance. GP model always interpolates the training points (there are exceptions for the interpolation property but it is outside of the scope of this paper). The variance of the prediction decreases as the untried point gets closer to training points. When the number of training sites is increased, the approximation accuracy improves quickly. The popularity of using GP as surrogate models is due to the fact that it provides the variance that can enter $\bm{\Sigma}_{\text{code}}$. Furthermore, GP has also been widely used to model the computer model bias $\delta(\mathbf{x})$ and provide information for $\bm{\Sigma}_{\text{bias}}$.

\subsection{Full vs. Modular Bayesian Approaches for Bayesian IUQ}

The inclusion of the model bias term $\delta(\mathbf{x})$ greatly complicates the Bayesian IUQ process. The solutions process based on the formulation introduced above can be either full Bayesian or modular Bayesian. To provide a self-contained introduction in this paper, we will briefly present the essential features of these two approaches. Interested readers can refer to \cite{wu2018inversePart2} \cite{wu2017metamodel} for a more detailed introduction and comparison of these two approaches.

In brief, both FBA and MBA use a GP metamodel to replace the computer code during MCMC sampling, and a second GP model to represent $\delta(\mathbf{x})$. Both GP models have unknown hyperparameters, $\bm{\Psi}^{\text{M}} = \{ \bm{\beta}^{\text{M}}, \sigma_{\text{M}}^2, \bm{\omega}^{\text{M}}, \mathbf{p}^{\text{M}} \}$ for the computer model $\mathbf{y}^{\text{M}}$, and $\bm{\Psi}^{\delta} = \{ \bm{\beta}^{\delta}, \sigma_{\delta}^2, \bm{\omega}^{\delta}, \mathbf{p}^{\delta} \}$ for model bias $\delta(\mathbf{x})$, where $\bm{\beta}$ is the vector of basis functions, $\sigma^2$ is the process variance, $\bm{\omega}$ is the vector of the characteristic length-scales and $\mathbf{p}$ is the vector of the roughness parameters. FBA and MBA differ in their treatment of $\bm{\Psi}^{\text{M}}$ and $\bm{\Psi}^{\delta}$. In FBA \cite{higdon2004combining}, both $\bm{\Psi}^{\text{M}}$ and $\bm{\Psi}^{\delta}$ are treated in a similar way as the calibration parameters $\bm{\theta}$. They are assigned priors which also enter the likelihood function. Joint posteriors of $\{ \bm{\theta}, \bm{\Psi}^{\text{M}}, \bm{\Psi}^{\delta} \}$ are solved together. Then $\bm{\Psi}^{\text{M}}$ and $\bm{\Psi}^{\delta}$ need to be integrated out from the joint posterior to get marginal distributions of $\bm{\theta}$. However, in MBA \cite{arendt2012quantification} \cite{liu2009modularization}, the estimation of $\bm{\theta}$, $\bm{\Psi}^{\text{M}}$ and $\bm{\Psi}^{\delta}$ are all separated in different modules. MBA uses plausible estimates of $\bm{\Psi}^{\text{M}}$ and $\bm{\Psi}^{\delta}$ evaluated by methods like Maximum Likelihood Estimation (MLE) and treat them as if they were the true values of $\bm{\Psi}^{\text{M}}$ and $\bm{\Psi}^{\delta}$. Finding the MLEs of $\bm{\Psi}^{\text{M}}$ and $\bm{\Psi}^{\delta}$ is done during the training process of the GP models. The major characteristics of FBA and MBA are summarized and compared in Table \ref{table:Bayesian-IUQ3-GP-FBA-vs-MBA}.

\begingroup
\renewcommand{\arraystretch}{1.2}
\begin{table}[ht]
	\footnotesize
	\captionsetup{justification=centering}
	\caption{Comparison of the main characteristics of FBA and MBA.}
	\label{table:Bayesian-IUQ3-GP-FBA-vs-MBA}
	\centering
	\begin{tabular}{p{0.47\linewidth}  p{0.47\linewidth}}
		\toprule
		FBA & MBA \\ 
		\midrule
		It treats unknown GP hyperparameters $\left\{ \bm{\Psi}^{\text{M}}, \bm{\Psi}^{\delta}  \right\}$ in a similar way with $\bm{\theta}$. 
		&  It uses modularization to separate various modules in IUQ.  
		\\ \midrule
		It requires that the user have reasonably good priors for $\left\{ \bm{\Psi}^{\text{M}}, \bm{\Psi}^{\delta} \right\}$, which is very difficult in practice especially for the model uncertainty term. 
		&  It only requires prior for $\bm{\theta}$, and uses MLE to evaluate $\left\{ \bm{\Psi}^{\text{M}}, \bm{\Psi}^{\delta} \right\}$.  
		\\ \midrule
		The joint posterior for all the unknowns $\left\{ \bm{\theta}, \bm{\Psi}^{\text{M}}, \bm{\Psi}^{\delta}  \right\}$ can be extremely complicated and have very high dimension, posing challenges for MCMC sampling. 
		&  The posterior function only contains calibration parameters $\bm{\theta}$.  
		\\ \midrule
		After MCMC sampling, $\left\{ \bm{\Psi}^{\text{M}}, \bm{\Psi}^{\delta}  \right\}$ need to be integrated out from the joint posterior to get marginal distributions of $\bm{\theta}$. 
		&  $\bm{\theta}^{\text{Posterior}}$ is conditioned on the MLEs of $\{ \bm{\Psi}^{\text{M}}, \bm{\Psi}^{\delta} \}$. Marginalization is not needed  
		\\ \midrule
		It is theoretically superior, but computationally intractable. 
		&  It is much easier to use, but does not consider uncertainties in $\{ \bm{\Psi}^{\text{M}}, \bm{\Psi}^{\delta} \}$. This is why such a method is only ``empirical'' or ``partial'' Bayesian.  \\
		\bottomrule
	\end{tabular}
\end{table}
\endgroup

\subsection{Applications and Variations of MBA}

MBA \cite{wu2018inversePart1} \cite{wu2017metamodel} significantly simplifies FBA by introduction modularization. However, it is still relatively difficult to understand and apply for engineers. Wu et al. \cite{wu2018inversePart1} proposed an improved MBA method that is more understandable. A sequential test source allocation algorithm was proposed \cite{wu2018inversePart2} to separate the given data for IUQ and training of $\delta(\mathbf{x})$. It was applied to the IUQ of TRACE uncertain PMPs \cite{wu2018inversePart2} based on the BFBT void fraction data \cite{neykov2006nupec}. It was demonstrated that the inclusion of model bias is capable of avoiding ``over-fitting'' for IUQ. The resulting distributions can effectively represent the input uncertainties that are consistent with data.

IUQ of TRACE PMPs using PSBT void fraction data \cite{wang2018surrogate} \cite{wang2019gaussian} was performed by Wang and Kozlowski, using the MBA method. Another application is \cite{borowiec2018modeling}, in which Borowiec and Kozlowski studied the effect of using biased measurement data for IUQ. It was shown that IUQ with biased data can lead to posterior distributions that cannot be extrapolated to different set of experimental condition. In a more recent work, Lu et al. \cite{lu2020enhancing} used MBA to an one-dimensional thermal stratification model for pool-type sodium-cooled fast reactors. The FUQ results showed that the quantified parameter uncertainties effectively improved the predictive capability of the model. In a series of papers \cite{liu2019uncertainty} \cite{liu2019validation-NSE} \cite{liu2019validation-NED}, Liu and his colleagues applied the MBA method to calibrate a two-fluid model-based multiphase computational fluid dynamics with high-resolution experimental data. Roma et al. \cite{roma2021bayesian} performed IUQ of RELAP5-3D model for the reliability analysis of passive safety systems, using a simplified version of the MBA method by ignoring the model bias term.

Earlier version of the MBA method explored different surrogate models. Wu and Kozlowski \cite{wu2017inversePCE} used generalized Polynomial Chaos Expansion (gPCE) to construct surrogate models for IUQ of a point kinetics coupled with lumped parameter TH feedback model. The developed approach was demonstrated to be capable of identifying the (pre-specified) true values of calibration parameters. In another application \cite{wu2017inverseSGSC}, the authors used a sparse gird stochastic collocation (SGSC) surrogate model. SGSC can also greatly reduce the computational cost. However, unlike GP, gPCE and SGSC cannot be provide their code uncertainties directly, also it is more difficult for them to represent the model bias.

It was pointed out in \cite{skorek2017input} that quantification of the model uncertainties on the basis of only one experiment is clearly not a good practice. IUQ for system TH physical models should be performed on the basis of multiple larger, better-defined experimental databases. Wang and co-authors \cite{wang2019inverse} \cite{wang2020hierarchical} studied a hierarchical Bayesian inference approach by extending the MBA method to account for the variability of TRACE physical model uncertainties in multiple experiments. The hierarchical model provides a more stable framework so that outliers will not have significant influence on the IUQ results. The hierarchical model introduces more parameters in the sampling process so a gradient-based sampling methods called No-U-Turn Sampler is used for the high-dimensional MCMC sampling. The results demonstrated that hierarchical model has better performance than traditional non-hierarchical model regarding the robustness to outliers.

\subsection{Applications and Variations of General Bayesian IUQ}

There are multiple works that employed Bayesian calibration/IUQ that is not MBA-based. For example, Bui and co-workers \cite{bui2013statistical} \cite{bui2013two} \cite{bui2014advanced} proposed the concept of ``total model-data integration'' which is based on the theory of Bayesian calibration and a mechanism to accommodate multiple data-streams and models. Such concept allows assimilation of heterogeneous multivariate data in comprehensive calibration and validation of computer models. This approach was demonstrated on the calibration/validation of subcooled boiling flow. The authors used the Gaussian Process Models for Simulation Analysis (GPMSA) toolbox \cite{gattiker2008gaussian} to constructed GP surrogate models on the basis of multiple GP kernels, while performed Principal Component Analysis (PCA) of time-dependent measurement data to reduce the dimensionality of the QoIs. The data uncertainty is artificially introduced because it is missing in public source. The authors represented the model bias with a 1-D function as a linear combination of seven normal kernels placed at equidistant locations along the pipe. A notable feature of this application is a systematic treatment of the data characterization and homogenization, for data obtained at different physical scales and having different qualities (a function of relevancy, scalability and uncertainty).

Some Bayesian calibration work considered the model bias term. Bachoc et al. \cite{bachoc2013parametric} \cite{bachoc2014calibration} applied the Bayesian calibration approach to the TH code FLICA4 in a single-phase friction model. The model bias was modeled with a GP after making a linear assumption in the relation between the QoIs and PMPs. Beyond the quantification of the parameter uncertainties, this work made inference of the model bias for each new potential experimental point, extrapolated from what had been learned from the available experimental data. The computer code predictive capability was reported to be improved based on the tested case. Wicaksono \cite{wicaksono2016bayesian} \cite{wicaksono2018bayesian} used Bayesian calibration to calibrate the reflood model parameters in TRACE. Furthermore, this work considered multivariate time-dependent outputs and the model bias. However, no results for the quantified model bias were presented and its extrapolation to new domains was not discussed.

Yurko et al. \cite{yurko2014uncertainty} \cite{yurko2015demonstration} used the Function Factorization with Gaussian Process (FFGP) priors model to emulate the behavior of computer code. Calibration of a simple friction-factor example using the Method of Manufactured Solution (MMS) with synthetic data was used to demonstrate the key properties of this method. This approach is better suited for the emulation of complex time series outputs. The FFGP emulator uses pattern recognition techniques to efficiently decompose the training data. The latent or hidden patterns allow more training data to be used which can drastically improve the predictive accuracy of the emulator. The FFGP emulator was shown to outperform the standard GP emulator for the friction factor test problem.

Porter and Mousseau \cite{porter2019bayesian} used Bayesian calibration to quantify the uncertainties in three empirical models in the MELCOR code: single phase friction factor in smooth tubes, single phase HTC for forced convection, and the transfer of mass between two phases. Without considering the model bias term, the authors referred to the relation $\mathbf{y}^{\text{E}} (\mathbf{x}) = \mathbf{y}^{\text{M}} ( \mathbf{x}, \bm{\theta} )  +  \bm{\epsilon}$ as a \textit{fixed effects analysis}, where all experimental data is governed by a single law or physical model without variation between individual experiments. A unique feature is their work is called \textit{mixed effects analysis}, where each individual experiment exhibits variations. Mixed effect Bayesian calibration employs a new relation $\mathbf{y}^{\text{E}} (\mathbf{x}) = \mathbf{y}^{\text{M}} ( \mathbf{x}, \bm{\theta} + \bm{\beta} ) +  \bm{\epsilon}$. The PMPs $\bm{\theta}$ denote \textit{global effects} and $\bm{\beta}$ represents \textit{random effects} that vary for different experiments. A hierarchical Metropolis-within-DRAM method was employed to estimate both the global and random effects.

\section{IUQ Methods Developed/Used in the PREMIUM Benchmark}

In the OECD/NEA PREMIUM benchmark, two methods were offered to the participants and used by multiple organizations, the CIRC{\'E} method from CEA and the IPREM method from the University of Pisa. GRS, IRSN, KAERI, PSI and Tractebel developed their own approaches. In this section, we will provide a brief but self-contained introduction of all these methods, together with discussions on their performance in PREMIUM and applications elsewhere.

\subsection{The CIRC{\'E} Method Developed by CEA}
\label{section:PREMIUM-CIRCE}

CIRC{\'E} \cite{de2001determination} was developed by Commissariat \`{a} l'Energie Atomique (CEA) in France and has been extensively applied to the physical models of the CATHARE 2 code. The name CIRC{\'E} refers to \textit{Calcul des Incertitudes Relatives aux Corr{\'e}lations {\'E}lementaires}, which can be translated into English as \textit{Calculation of the Uncertainties Related to the Elementary Correlations}. CIRC{\'E} is one of the earliest IUQ methods in the nuclear community that are devoted to quantifying the uncertainty of the parameters associated with non-measurable physical models, using intermediate-type experiments. It is a robust statistical approach of data analysis that uses scalar measurement data sensitive to some particular physical models to determine a probabilistic representation of PMPs. Several conference papers have been published \cite{de2001determination} \cite{de2004quantification} \cite{de2012circe}.

The most complete description of the CIRC{\'E} method can be found in Appendix B of PREMIUM's methodology report \cite{reventos2016premium}, including the CIRC{\'E} algorithm, detailed user guidelines, the software and the structure of an input data deck and some recommendations for calculating the derivatives used by CIRC{\'E}. \ref{Appendix-CIRCE} presents a relatively brief but self-contained introduction of the mathematical details of CIRC{\'E}. In brief, CIRC{\'E} starts with a linear assumption of parameter-QoIs relation, then approximates the code with a first order Taylor expansion and calculates the MLE of the mean value (bias) $b$ and standard deviation $\sigma$ of the PMPs using the E-M algorithm. Even though it employs some concepts in Bayesian inference, such as prior, posterior and likelihood, it is still considered as a frequentist/deterministic IUQ method, because MLE is used instead of sampling. CIRC{\'E} is a powerful method that has been successfully applied to various problems especially with the CATHARE code. However, it has a few limitations:
\begin{enumerate}[label=(\arabic*)]
	\setlength{\itemsep}{0.1pt}
	\item The normality and independence assumptions. This will work for many problems, but not always. The principle of MLE used by the E-M algorithm does not apply any more if the PMPs are subject to other type of distributions, such as uniform. In Bayesian IUQ, the prior also assumes independence between the calibration parameters. Nevertheless, the posterior joint distributions can always catch the inter-dependence between the parameters, but CIRC{\'E} cannot.
	
	\item The linearity assumption in the parameter-QoI relation. CIRC{\'E} may need to be used iteratively, if the mean values obtained are far away from where the derivatives were calculated and/or the standard deviations are large. The derivatives should be re-calculated around the new mean values so the linearity assumption can be satisfied. Applications have shown that the number of iterations needed is usually small.
	
	\item CIRC{\'E} cannot be used as a black box because it requires the computation of the sensitivity matrix with the adjoint method. As a result, the adjoint system must be introduced to the code intrusively, which leads to a rather nontrivial task. Furthermore, access to the source code is not always available. As a result, the derivative-based  brute force approach needs to be used to calculate the sensitivity matrix.
	
	\item CIRC{\'E} requires the problem under investigation to be well-posed and identifiable. When calculating the mean (bias) vector, matrix inversion is needed. An invertable matrix can be caused by two vectors $\frac{\partial y^{\text{M}}_j}{\partial \bm{\theta}} = \left[ \frac{\partial y^{\text{M}}_j}{\partial \theta_1}, \frac{\partial y^{\text{M}}_j}{\partial \theta_2}, \ldots, \frac{\partial y^{\text{M}}_j}{\partial \theta_I} \right]^\top$ being collinear. In this case, the left matrix in Equation \ref{equation:PREMIUM-CIRCE-WithBias8-Linear-System} will not be invertible, and both $b_i$ and $\sigma_i$ will be poorly estimated.
	
	\item The selected QoIs need to be significantly different from each other. The precision of CIRC{\'E} results increases with the number of independent QoIs. Besides, each selected QoI must be sensitive to, at least, one of the chosen PMPs. The selection of the QoIs implies a significant amount of user effect.
	
	\item As pointed in \cite{reventos2016premium}, CIRC{\'E} requires the considered QoIs to be sensitive to the studied  parameters and independent, while being numerous enough. When the QoI is time-dependent, oscillatory or discontinuous, a significant deal of engineering judgment should be used to pinpoint numerous QoIs from the corresponding time series.
	
	\item It is recommended that CIRC{\'E} should not be used for more than 3 parameters simultaneously \cite{reventos2016premium}. As pointed in \cite{freixa2016testing}, when more than 1 parameter is considered, there are multiple ways to explain the discrepancies between the model and data, making IUQ an ill-posed problem. In this case, CIRC{\'E} will tend to explain the differences with the most influential parameters and therefore the obtained parameter uncertainty will be dominant. The uncertainties for the less influential parameters might not be representative. Moreover, IUQ results can be quite different as a result of small variations on the parameter or QoI selection.
\end{enumerate}

The CEA results in the PREMIUM benchmark were presented in \cite{nouy2017quantification}. The CEA team used the CATHARE 2 code. Three influential parameters were selected. CIRC{\'E} was used in an iterative way, though only 4 iterations were needed to obtain a converged solution. The quantified parameter uncertainties were not wide, because CIRC{\'E} by design estimates the narrowest uncertainty intervals for the parameters. As a result, the propagated code response uncertainties were also not wide, but sufficient to envelop the experimental data during the verification step using the FEBA tests. However, in the validation step using the PERICLES tests, the calibrated calculation made the clad temperatures worse than the reference calculation. Furthermore, the uncertainty bands did not envelop the experimental data. This was also observed for other participants in PREMIUM.

Eight other teams also used CIRC{\'E} in PREMIUM. Generally, in the verification step of the quantified physical model uncertainties based on the FEBA tests, as well as in the validation step based on the PERICLES tests, CIRC{\'E} users produced the narrowest uncertainty bands. Similar to CEA results, FUQ with CIRC{\'E} users' results can envelop the FEBA data, but not the PERICLES data. Another application of the CIRC{\'E} method can be found in \cite{seynhaeve2015uncertainty}, in which uncertainty in a delayed equilibrium model for choked two-phase flashing flow was successfully quantified.

\subsection{The IPREM Method Developed by UNIPI}

The IPREM method, which means \textit{Input Parameter Range Evaluation Methodology}, was developed at the University of Pisa (UNIPI) \cite{kovtonyuk2012procedure} \cite{kovtonyuk2014development} \cite{kovtonyuk2017development} for IUQ through comparison of sensitivity calculation results of a system TH code with experimental data. IPREM utilizes the mathematical apparatus of FFTBM (Fast Fourier Transform Based Method). A brief introduction to the mathematical foundation of FFTBM and the major aspects of IPREM can be found in \ref{Appendix-IPREM}.

The IPREM method is independent of the system TH code, the type of investigated input parameter and QoIs, because it involves only post-processing of calculation results. No modification of the source code is needed as long as the PMPs can be defined in the code input deck. It is easy to use, cost-efficient, and partially reduces the use of engineering judgment in terms that the proper procedure, mathematical apparatus and corresponding criteria are clearly defined. However, IPREM has a few limitations:
\begin{enumerate}[label=(\arabic*)]
	\setlength{\itemsep}{0.1pt}
	\item The procedures involved do not rely upon substantial statistical basis. It still require engineering considerations and previous experience, especially for the weight factors when calculating AA and the CR limit value when deciding the parameter ranges.
	
	\item It only provides the ranges of variation for the PMPs, rather than the PDFs that contain full uncertainty information. As a result, IPREM often assumes uniform distributions for the parameters.
	
	\item For the sensitivity calculations, the PMPs are perturbed one by one. This inherently implies that IPREM treats the parameters as independent.
	
	\item As illustrated in Figure \ref{figure:PREMIUM-IPREM}, there are scenarios in which IPREM will fail, and the user has to manually adjust the limit value for CR. Examples for such failures can be found in \cite{kovtonyuk2017development}. Different limit values can also result in very distinct ranges of variation, so it requires careful engineering characterization.
	
	\item It can only make use of time-dependent QoIs and data. In real problems, a lot of experimental data are not from transient tests, making them not usable with IPREM. For certain transient tests, the recorded data may not sufficiently cover the transient. For example, some tests only have results recorded at end of transient.
	
	\item IPREM does not consider experimental uncertainty.
	
	\item IPREM can be applied with only one test. For example, the three PREMIUM participants who used IPREM only used FEBA test 216 \cite{skorek2019quantification}. When there are multiple tests available, each test may produce different ranges of variation. Therefore, IPREM needs a robust procedure to combine these different results. It was also pointed out in \cite{kovtonyuk2017development} that the choice of the QoIs may significantly affect the resulting parameter ranges, posing challenges and risks in licensing IPREM for regulatory purpose.
\end{enumerate}

Within the PREMIUM project, IPREM was applied by UNIPI \cite{kovtonyuk2014development} \cite{kovtonyuk2017development} to quantify the input uncertainties of the reflood-influential models in RELAP5/Mod3.3. Within the NURESAFE project, IPREM was also applied \cite{kovtonyuk2017development} to quantify uncertainties in the reflood-related models of the CATHARE2 V2.5 code. The IUQ results were properly verified using data from FEBA and ACHILLES facilities. In the validation step, It was shown that the FUQ results from RELAP5 and CATHARE2 based on the IUQ results envelope the FEBA, PERICLES and ACHILLES data for most measurement data, confirming the validity of using IPREM for IUQ. Three other teams also used IPREM in PREMIUM. VTT developed an approach that combines CIRC{\'E} and IPREM, with CIRC{\'E} for calibration, and IPREM for parameter range determination. Overall IPREM users obtained larger uncertainty ranges despite the fact that only one test (FEBA test 216) was used.

Freixa et al. \cite{freixa2016testing} compared CIRC{\'E} and IPREM within the PREMIUM project. They modeled the FEBA facility using RELAP5/Mod3.3. The comparison of the results of both methodologies is based on the qualitative observation of the PDFs and envelope calculations of the FEBA and extrapolation of the resultant PDFs to the PERICLES tests. CIRC{\'E} used data of all six tests from FEBA Series I, while FFTBM only used one FEBA test. Significant differences were found between these two methods in terms of ranges and distributions of the parameters. The authors performed FUQ studies using the quantified input uncertainties, as well as those from expert judgement. The obtained uncertainty bands with both methodologies enveloped all the selected experimental data. The uncertainty bands generated through expert judgement were wider than the ones obtained by the two IUQ methods. Note that in this work, the authors ignored the experimental uncertainty by assuming it is much smaller than the effect of parameter uncertainties.

Alku \cite{alku2015quantification} used IPREM and CIRC{\'E} to quantify the PMP uncertainties in the APROS code, based on the data from the Finnish VEERA reflooding experiments \cite{puustinen1994veera}. The selected list of PMPs were the same with those identified by VTT in the PREMIUM benchmark. The FFTBM method didn't produce usable results (either the upper or lower bound cannot be determined). Even after using a new set of nominal values obtained with CIRC{\'E}, the IPREM results remain unusable. Furthermore, the quantified parameter uncertainties using CIRC{\'E} were quite different from those obtained based on FEBA data. The distributions using CIRC{\'E} based on FEBA and VEERA either did not overlap or overlapped but have very different variances.

\subsection{CET-based Sample Adjusting Approach Developed by GRS}

In the PREMIUM project, participants from the Gesellschaft f{\"u}r Anlagen- und Reaktorsicherheit (GRS) used a simple trial-and-error method and in parallel a new method base on adjusting samples of input uncertainties combinations \cite{skorek2017input}. The idea behind this approach is very straightforward and quite similar to the trial-and-error method. The most influential PMP for the selected QoIs is identified first. Then this parameter is varied in an initial range simultaneously with other parameters to form $N$ random samples ($N=125$ in \cite{skorek2017input}). The system TH code runs at these samples to produce $N$ set of results. The lower and upper limits for the computed QoIs will be determined and compared to the measurement data. If the data is not fully enveloped by the simulations, then new samples will be generated by expanding the parameters above the upper range, and/or below the lower range, according to the current comparison with data. After several parametric calculations a new range could be determined. It is expected to be the smallest range which enables bounding of measured data by uncertainty limits.

In PREMIUM, GRS used this method to quantify the uncertainties in six PMPs in the system TH code ATHLET. The sample adjusting approach was applied for quantification of only one PMP - correction factor for relative velocity in the bundle geometry, while uncertainties for the other five parameters were based on previous study or the trial-and-error method. In the validation step, two extra input parameters were introduced: the relative velocity in the cross-connections and the bundle total power. These are design variables, and their uncertainties were determined based on sensitivity calculation, expert judgement and experimenter suggestion. Verification with FEBA tests was satisfactory, while validation with PERICLES data was not.

Reasons for validation failure were summarized in a post-PREMIUM analysis \cite{skorek2017input}. The authors believed that ``besides the uncertainties related to the dispersed droplet flow modelling in the system code, the main reason for discrepancy between determined uncertainty limits and experimental data was the weakness of input data for the reference calculation''. Ignoring the experimental uncertainty by this method can also have a negative impact on the IUQ results. Even though only one parameter was considered in the work \cite{skorek2017input}, the authors claimed that simultaneous quantification of multiple parameter is possible. Notwithstanding the claim, the authors believed that ``the number of parameters which have to be quantified by simultaneous variation should be as small as possible''.

\subsection{DIPE Developed by IRSN}

The DIPE method \cite{joucla2008dipe}, which represents \textit{Determination of Input Parameters uncertaintiEs}, was developed by the Institute de Radioprotection et de S{\^u}ret{\'e} Nucl{\'e}aire (IRSN) at France. The motivation was to minimize the expert judgment used to determine the parameters uncertainties in the CATHARE code. The procedures of the DIPE method can be found in \ref{Appendix-DIPE}. DIPE provides ranges on uncertain PMPs allowing the coverage of experimental data by code simulations. It is very easy to understand and implement. However, it has a few limitations:
\begin{enumerate}[label=(\arabic*)]
	\setlength{\itemsep}{0.1pt}
	\item The assumption that the simulation curve is monotonous is problematic in practice which limits the applicability of the DIPE method.
	
	\item Experimental uncertainty is treated as negligible so they are not considered.
	
	\item DIPE results based on different QoIs may be very disparate for the same PMP.
	
	\item DIPE only produces the parameter ranges, not the entire PDF. Expert judgment and engineering considerations are still needed. Also for the case of two parameters, the pair-wise joint distributions cannot be achieved because the converge rate contour lines do not represent the true joint cumulative distribution function (CDF).
	
	\item It is unclear how to calculate the CDF for high-dimensional cases. The number of experimental design may increase exponentially with the dimension. It can also be difficult to determine the input values that produce simulation curves corresponding to the 2.5$^{\text{th}}$ and the 97.5$^{\text{th}}$ percentiles, as several combinations of $\bm{\theta}$ may have the similar simulation curve.
\end{enumerate}

The DIPE method was originally applied on the separate effect tests CANON and MARVIKEN in the qualification of the CATHARE 2 V2.5 parameter uncertainties\cite{joucla2008dipe}. The authors only considered cases with one parameter and two parameters. The resulting range is quite large for a multiplicative factor: $[0.1, 14]$. In PREMIUM, the physical model uncertainties quantified by IRSN using DIPE also were quite wide. Wide uncertainty bands tend to bound the experimental data, giving a good calibration score, but they are not very informative \cite{skorek2019quantification} due to the large uncertainties.

\subsection{MCDA Developed by KAERI}

The MCDA method \cite{heo2014implementation}, which means \textit{Model Calibration through Data Assimilation}, was developed by the Korea Atomic Energy Research Institute (KAERI). MCDA is based upon the data assimilation (DA) methodology to determine the mean values and standard deviations of the PMPs. DA is the process to calibrate the target model by adjusting the parameter values to achieve better agreement between model and data \cite{heo2011optimization}. DA incorporates observations of a system into the numerical model to produce an optimal estimate of the evolving state of the system. By definition, DA is very similar to calibration, but it is usually treated as the calibration of dynamic models, such as weather forecasting. However, such distinction is usually not strictly followed in the open literature, as people usually use calibration and assimilation interchangably.

MCDA consists of both \textit{deterministic} and \textit{probabilistic} DA methods to solve both \textit{linear} and \textit{non-linear} problems based on Bayes' theorem. A detailed introduction of the MCDA method can be found in \ref{Appendix-MCDA}. To determine whether a QoI is linear in the PMPs, a linearity test is required to evaluate the degree of linearity. For a linear system, deterministic approach will be used to obtain the mean value and standard deviation of the parameters. On the other hand, for a non-linear system, probabilistic method based on MCMC sampling will be utilized to estimate the a posteriori distributions of the parameters. In this case, it is very close to Bayesian calibration. Therefore, MCDA is treated as a hybrid frenquentist/Bayesian IUQ method.

The work presented in \cite{heo2014implementation} was an application to the system TH code SPACE, based on post critical heat flux experimental data. Note that the authors selected both PMPs and design variables to have their values adjusted through MCDA. The design variables considered were mass flow rate, pressure, temperature, and power values for different time intervals. Both the deterministic and probabilistic parts of MCDA were applied. Non-Gaussian distributions were obtained for the parameters due to the non-linearity of the system. It was found that the most influential parameters underwent the largest calibration. Generally solution obtained with the probabilistic method was better than that obtained using the deterministic approach since the former does not approximate the QoIs with first-order Taylor series expansion.

During KAERI's participation in PREMIUM, MCDA was used only as an alternative method to CIRC{\'E}. Similar to CIRC{\'E}, MCDA also produce IUQ results that lead to very narrow uncertainty bands in the verification and validation steps. These narrow bands are very informative, but tend to fail to cover the data. The MCDA method was used by Ui et al. \cite{ui2019data} for IUQ of a set of parameters in subchannel TH code CTF based on the BFBT data. The probabilistic method based on MCMC sampling was chosen. It was confirmed that the average value of the bundle-averaged void fraction improved with the IUQ results.

\subsection{Bayesian Inference Method Used by PSI}

The IUQ method used by Paul Scherrer Institute (PSI) is based on Bayesian inference that is very similar to Bayesian calibration. However, it was only at an early stage of development during PSI's participation in PREMIUM. The most recent summary paper of PREMIUM \cite{skorek2019quantification} showed that PSI only participated in the verification and validation steps, with the physical model uncertainties obtained based on expert judgements. In two other works \cite{wicaksono2016bayesian} \cite{wicaksono2018bayesian}, Wicaksono and colleagues at PSI published the details of their IUQ approach based on Bayesian inference. They employed an approach that is similar to MBA \cite{wu2018inversePart1} \cite{wu2017metamodel}, with major differences in the treatment of the model bias term. The novelty of PSI's IUQ method is a convenient treatment of the high-dimensional simulation outputs (e.g., time-and space-dependent rod temperatures). As shown in a few other work \cite{wicaksono2016global} \cite{perret2019global}, PSI researchers used functional data analysis that consists of curve registration and dimensional reduction with PCA, to derive innovative QoIs that capture the variation over the whole course of the FEBA reflood transient. These derived QoIs make global sensitivity analysis and IUQ more convenient to do. Note that the results in \cite{wicaksono2016bayesian} \cite{wicaksono2018bayesian} are from a post-PREMIUM analysis, because PSI used expert judgement for IUQ in PREMIUM. Therefore, the PSI's Bayesian inference method was not included in Table \ref{table:Intro-List-of-IUQ-Methods}. It also won't be evaluated in Section \ref{section:Evaluation-of-IUQ-Methods}, as its performance is considered to be close to MBA.

\subsection{Sampling-based IUQ Developed by Tractebel}

In the PREMIUM project, Tractebel from Belgium has contributed to the development and the proof-of-concept application of a sampling-based IUQ approach \cite{zhang2019development} with the DAKOTA \cite{adams2011dakota} statistical uncertainty/sensitivity analysis tool. Tractebel used this approach to quantify the RELAP5/MOD3.3 reflood-related model input uncertainties. The basic idea of this method is to use DAKOTA's random sampling-based FUQ functionality with iteration to quantify the parameter uncertainties. In sampling-based FUQ, \textit{known} input uncertainties are propagated to \textit{unknown} output uncertainties. In sampling-based IUQ, the input uncertainties are \textit{unknown}, but the output uncertainties are considered to be \textit{known} based on the experiments. In other words, sampling-based IUQ seeks the parameter uncertainties (ranges and distributions) such that the code simulation distributions match the data within the measurement uncertainty \cite{zhang2019development}.

The DAKOTA team had an earlier work \cite{swiler2008model} that employs a similar idea, the so-called calibration/optimization-under-uncertainty. This approach tried to find a statistical characterization of input parameters such that when propagated through the model, they match the statistics on the measurement data. Such uncertainty inversion rely on nesting UQ analysis within a deterministic calibration loop which can incur a considerable computational cost. Different from the DAKOTA team's approach, sampling-based IUQ adopts a more pragmatic trial-and-error process. It consists of four major steps. Firstly, identify the key model input parameters and define their ranges of variation and distributions. Literature review, expert feedback and previous applications are used to provide an initial guess for the ranges of variation and distributions. Secondly, sample the input parameters and run the code to obtain the QoIs with uncertainties. Thirdly, check if the lower and upper bounds of the computed QoIs envelop the experimental data, for each measurement location and each time step. Finally, update the parameter ranges of variation and distributions until a reasonable coverage is found. In case of various output parameters are targeted, a compromise may be needed to ensure the adequate coverage of all output parameters.

Sampling-based IUQ is very easy to understand and implement. It can also consider the experimental uncertainty. However, it also shares the common disadvantages of the trial-and-error approach, including the CET-based sample adjusting method.
\begin{enumerate}[label=(\arabic*)]
	\setlength{\itemsep}{0.1pt}
	\item It assumes no discontinuities in the model that would prevent the optimization process. It considers no bias in the code, that is, the model can predict the reality.
	
	\item It is a relatively conservative IUQ method that leads to large uncertainty ranges for the parameters. The exact distributions cannot be quantified, so uniform/Gaussian are used. Parameters are treated as independent because there is no procedure to quantify their dependence.
	
	\item Optimization of the parameter ranges is based on expert knowledge and visual
	comparison. Therefore, the iterative process cannot stop automatically since there is no quantitative acceptance criterion available. Manual iterations and engineering judgment are needed to optimize the lower/upper bounds in order to envelop the experimental data. The authors mentioned \cite{zhang2019development} that this approach can be improved with an automatic optimization process to minimize the uncertainty bandwidth, which will help to obtain a unique solution of the uncertainty inversion.
	
	\item If the parameter is not influential to the QoI, many iterations may be needed because the parameter range will be large. When multiple QoIs are used, a compromise is needed to ensure the adequate coverage of all the QoIs. There are no quantitative guidelines on how to update the parameter ranges, making it difficult to arrive at an unique solution.
\end{enumerate}

In the PREMIUM project, Tractebel \cite{zhang2019development} used RELAP5/MOD3.3 and assigned Gaussian distributions with specified standard deviation and truncation to well-measured parameters, and uniform distributions to less known parameters. Unlike other participants, Tractebel also considered five design variables for IUQ, in addition to four PMPs. The five design variables are bundle power, power profile, inlet water temperature, system pressure and inlet velocity. It is unclear how the uncertainties in these design variables influence the uncertainties in the PMPs. But it is clearly questionable to apply the input uncertainties in design variables quantified based on FEBA data to PERICLES tests, for the simple reason that they are different experiments.

\subsection{Other Work Relevant to the PREMIUM Benchmark}

Li and and colleagues \cite{li2017investigation} used Bayesian IUQ to quantify the uncertainty in RELAP5 reflood model, without considering the model bias term. The work presented in this paper was a post-PREMIUM analysis, because in the PREMIUM benchmark the authors used the IPREM method but did not obtain satisfactory results (see \cite{skorek2019quantification}). To reduce the computational cost during MCMC sampling, the authors built surrogate models using radial basis function. To improve the surrogate model accuracy, an adaptive approach based on cross-entropy minimization was used to densify training samples in the posterior space. The IUQ results were greatly improved compared to their results reported in the PREMIUM benchmark. In an earlier work \cite{li2016improvement}, the authors tried to improve RELAP5 reflood model based on sensitivity analysis of the influential PMPs. However, only deterministic enhancement factors (multipliers) were identified instead of their uncertainties.

\section{The SAPIUM Project}

The PREMIUM project is by far the most comprehensive international activity that attempt to benchmark the available IUQ methods for nuclear system TH codes. Nevertheless, a strong user effect was observed due to the lack of best practices guidance, as can be seen by the disparities between the participants \cite{skorek2019quantification}. It was summarized in \cite{zhang2019development} that the dramatic distinctions were caused by lack of guidelines in (1) selection of the assessment database, (2) assessment of the applicability of the codes for modeling of the identified important phenomena, (3) modeling of the experiments, and (4) practical IUQ process.

Recently, another OECD/NEA project SAPIUM (2017--2019) has been launched and completed towards the construction of a clear and shared systematic approach for IUQ. Earlier progress of the SAPIUM project can be found in \cite{baccou2018sapium} \cite{baccou2019development-NURETH} \cite{zhang2019role}. Unlike PREMIUM, the underlying idea of the SAPIUM project is not to focus on methods benchmarking, but to provide a methodological document. The main outcome of the SAPIUM project is a systematic approach that consists of five elements to perform a meaningful model IUQ and validation as well as some good-practice for each step. In this section, we briefly summarize the major elements of the SAPIUM approach. A comprehensive discussion on the technical aspects of each element can be found in \cite{baccou2019development-NED} \cite{baccou2020sapium}. The five elements are further decomposed into 17 steps, as listed below:
\begin{enumerate}
	\setlength{\itemsep}{0.1pt}
	\item[-] \textit{Element 1: specification of the IUQ problem and requirements}. It includes three steps: (1) specification of the IUQ purpose, (2) selection of the QoIs and (3) identification of important phenomena.	
	\item[-] \textit{Element 2: development and assessment of the experimental database}. It involes three steps: (4) establishment of a list of available experiments and standardized description of each experiment, (5) assessment of the adequacy of the database, and (6) selection of the experimental database for IUQ and validation.	
	\item[-] \textit{Element 3: selection and assessment of simulation model}. It consists of three steps: (7) selection of code based on capability assessment, (8) assessment of applicability of the simulation model and (9) selection of uncertain input parameters and specification of input validation ranges.	
	\item[-] \textit{Element 4: physical model IUQ}. It contains four steps: (10) aggregation of the information from the experiments and simulations, (11) quantification of input uncertainties by inverse propagation, (12) combination of IUQ results if several quantifications are performed, and (13) confirmation by counterpart tests.	
	\item[-] \textit{Element 5: input uncertainty validation}. It is composed of four steps: (14) determination of numerical approximation and other input data uncertainties for each validation case, (15) forward propagation of all input uncertainties, (16) computation of validation indicators, and (17) analysis of the validation results.	
\end{enumerate}

The SAPIUM project provides a a common and generic framework to facilitate both discussions between participants and applications to several industrial problems. It also tried to address several significant issues that accompany IUQ, such as geometrical and TH phenomena scaling, predictive capability assessment of the computational model, adequacy of the experimental database, acceptability of the validation results, etc. There are a lot more details on the technical aspects of each element/step. Recommendations were also provided for each major element based on the lessons learned from the analysis of the SAPIUM elements \cite{baccou2019development-NED}. Researchers that intend to perform IUQ of their physical models are highly recommended to follow the SAPIUM guidelines.

\section{Other IUQ Approaches}

In this section, more IUQ methods that have been applied for system TH codes will be reviewed. These methods were not used in the PREMIUM project.

\subsection{Bayesian and Non-linear Extensions of CIRC{\'E} by CEA Saclay}

In Section \ref{section:PREMIUM-CIRCE} and \ref{Appendix-CIRCE}, the details and limitations of the CIRC{\'E} method has been introduced. Recently, Damblin and Gaillard at CEA Saclay \cite{damblin2018bayesian} \cite{damblin2020bayesian} in France have extended the classical CIRC{\'E} method \cite{de2001determination} \cite{de2012circe} to Bayesian and non-linear settings. In the paper, this new method is referred to as ``Bayesian CIRC{\'E}'' because it was interpreted as the Bayesian counterpart of the CIRC{\'E} method by the authors.

For the Bayesian extension, it calculates the joint posterior distribution of $(b, \sigma)$ instead of their MLE.  Bayesian CIRC{\'E} treats the mean value (bias) $b$ and standard deviation $\sigma$ of the parameters as random variables. It first assigns prior distributions for $(b, \sigma)$, then obtains the posterior distributions using MCMC sampling. When the system is linear, if  conjugate Gaussian-inverse gamma prior distributions are used, the full conditional posterior distributions are analytically tractable (exact probability distributions are known). Bayesian CIRC{\'E} uses a MCMC sampling procedure called blocked Gibbs sampler, which was originally called substitution sampling \cite{gelfand1990sampling}, to generate posterior samples in an iterative process until reaching convergence.

For the non-linear extension, the linear assumption of classical CIRC{\'E} is dropped. However, one of the full conditional distributions no longer has closed-form and thus becomes intractable. MCMC procedures such as MH has to be used within the Gibbs sampler. Bayesian CIRC{\'E} uses GP to replace the computational model to greatly reduce the computational cost. In the demonstration problem, Bayesian CIRC{\'E} was used for IUQ of two condensation heat transfer physical models in CATHARE 2 code, using the Westinghouse COSI (Condensation on Safety Injection) experimental tests which are SETs.

Bayesian CIRC{\'E} is overall similar to Bayesian calibration. However, it has a few critical differences from the MBA method. Firstly, the model updating equation was acknowledged in \cite{damblin2020bayesian}, but Bayesian CIRC{\'E} omits the model bias term and presumes that the gap between data/model is mostly due to parameter uncertainties. Secondly, MBA deals with the PMPs directly, while Bayesian CIRC{\'E} deals with the parameters (mean, variance) of the assumed normal/log-normal distributions. Finally, Bayesian CIRC{\'E} also has the independence assumption of the PMPs, and it does not quantify the PMPs' mutual dependence in the posterior, because the posteriors are for $(b, \sigma)$.

\subsection{Non-parametric Clustering by PSI}

A non-parametric statistical approach based on multi-dimensional clustering, hereby referred to as the ``non-parametric clustering'' method, was published by PSI in 2007 \cite{vinai2007statistical} for quantification of uncertainty in best estimate code physical models. This is one of the earliest work that was devoted to system TH physical model uncertainty. However, it essentially quantified the uncertainty in code predictions originated from a certain physical model, rather than the PMP uncertainties. This method was initially applied to the drift-flux model of the RETRAN-3D code based on void fraction data \cite{vinai2007statistical}. However, due to a few limitations as discussed below, it was not considered by PSI in the PREMIUM project.

Non-parametric clustering first calculates the code prediction errors $\epsilon_i = \mathbf{y}^{\text{E}} (\mathbf{x}^i) - \mathbf{y}^{\text{M}} ( \mathbf{x}^i, \bm{\theta}_0 )$ for each measurement data point at $\mathbf{x}^i$, using the parameter nominal values $\bm{\theta}_0$. The resulting ``database of errors'' reflects the model's accuracy on the basis of available experiments. The database resides in the so-called ``assessment state space'' which is defined by the design variables $\mathbf{x}$. Multi-dimensional clustering based on an extended Kruskall-Wallis test is then used to identify clusters in $\bm{\epsilon} = \{\epsilon_i\}_{i=1}^{N_{\text{data}}}$ such that the model errors share a common PDF. A non-parametric PDF estimator that combines the universal orthogonal series estimator and the kernel estimator is used to estimate the PDF of each cluster. The result is a collection of multiple PDFs for the code prediction errors in different regions of the assessment state space. When the code is used for FUQ, these PDFs can be sampled and added to the code predictions to represent the physical model uncertainties.

This method does not try to quantify the uncertainties of PMPs. It essentially fits multiple PDFs for the model prediction errors in different regions of the assessment state space. When used for a new experiment, the PDFs are sampled randomly and added to the code prediction as a ``compensation''. Relatively large amount of data is needed to define the assessment state space and fit the corresponding PDFs of the model prediction errors. The resulting PDFs are dependent on $\mathbf{x}$. However, it is questionable to extrapolate them to a new experiment since the test setting can be completely different. Engineering considerations are needed in multiple steps of the method. For example, when choosing $\mathbf{x}$ that define the assessment state space. The authors considered a two-dimensional plane defined by pressure and mass flux, while ignoring power and heat flux \cite{vinai2007statistical}. Clustering in three- or higher-dimensional space will require more data points. Moreover, expert opinion is needed when separating the assessment state space to multiple regions.

\subsection{Data Adjustment and Assimilation by KIT}

Cacuci and Ionescu-Bujor \cite{cacuci2010best} \cite{cacuci2010sensitivity} at the Karlsruhe Institute of Technology (KIT) in Germany proposed a comprehensive predictive modeling methodology for large-scale non-linear time-dependent systems, enabling the reduction of uncertainties in BE predictions by simultaneously calibrating
(adjusting) model parameters and QoIs, through assimilation of experimental data. This method was programmed into a computational module called BEST-EST. This method was referred to as the Data Adjustment and Assimilation (DAA) method by Petruzzi in \cite{petruzzi2019casualidad}. Besides model calibration based on data assimilation, DAA also provides quantitative indicators constructed from response sensitivities to model parameters and covariance matrices (for measurements, model parameters and QoIs) for determining the model/data consistency. Once the inconsistent data is identified and discarded, DAA yields BE values for parameters and predicted QoIs with reduced uncertainties.

The DAA method utilizes the covariance matrices for parameter-parameter, QoI-parameter and QoI-QoI pairs. Based on the maximum entropy principle in conjunction with the Bayes' theorem, it combines the a priori information with the likelihood provided by the simulation model to calculate BE values for the parameters, QoIs, and their reduced uncertainties (covariance matrices). It is based on formulation into a optimization problem that is solved with MLE, therefore, it is considered as a frequentist IUQ method. The DAA method is very mathematically intense compared to the other IUQ methods (see the derivations in \cite{cacuci2010best} \cite{petruzzi2010best}), and it only applies to time-dependent problems. Moreover, the development of the covariance matrices is a very complicated and time consuming process, so it was not used by any participants in the PREMIUM project. KIT used the IPREM method during its participation in PREMIUM.

The DAA method was applied by Petruzzi et al. \cite{petruzzi2010best} to a blow down benchmark experiment. It was demonstrated that the assimilation of consistent experimental data leads to a significant reduction of uncertainties in BE predictions. Badea et al. \cite{badea2012best} used it to calibrate model parameters and boundary conditions for the TH code FLICA4. QoIs from BFBT experiment ``turbine trip without bypass'' including pressure drops, axial and transverse void fraction distributions were used. This work in \cite{badea2012best} was continued in \cite{cacuci2014reducing}, in which Cacuci and Arslan adopted the DAA method to reduce the uncertainties in calibration parameters and time-dependent boundary conditions (power, mass flow rates, and outlet pressure distributions) in FLICA4 based on the BFBT benchmark, yielding best-estimate predictions of axial void fraction distributions.

\subsection{CASUALIDAD by NINE}

The CASUALIDAD method \cite{petruzzi2019casualidad}, which means \textit{Code with the capability of Adjoint Sensitivity and Uncertainty AnaLysis by Internal Data ADjustment and assimilation}, is developed by Petruzzi at Nuclear and INdustrial Engineering (NINE) in Italy. Note that earlier version of CASUALIDAD was developed by the same author at UNIPI \cite{petruzzi2008development} \cite{petruzzi2014uncertainties}. CASUALIDAD is a comprehensive framework that consists of a fully deterministic method based on advanced mathematical tools to internally perform uncertainty and sensitivity analysis in the system TH code. CASUALIDAD includes six key elements. The fifth element is the implementation of IUQ that uses the DAA method. DAA is used to update the a priori PDF of the parameters and QoIs based on available data to get the posterior improved estimation of the the input parameters, QoIs, and covariance matrices. Therefore, the CASUALIDAD method inherits the major characteristics of the DAA method.

\subsection{MLE and MAP by UIUC}

Kozlowski and his colleagues \cite{shrestha2016inverse} \cite{hu2016inverse} at the University of Illinois at Urbana-Champaign (UIUC) in the USA proposed IUQ methods based on Maximum Likelihood Estimation (MLE) and Maximum A Posteriori (MAP) estimation. Based on the Bayes' rule, the only difference betweem MLE and MAP is that MLE treats the prior distribution as a non-informative constant. The MLE/MAP method is very close to CIRC{\'E}. But unlike CIRC{\'E} which uses an iterative process to calculate parameter covariance matrix $\bm{\Sigma}_{\bm{\theta}}$ and mean vector $\bm{b}$, MLE/MAP use the E-M algorithm in a more straightforward manner.

The prototype version of the MLE/MAP method was presented by Shrestha and Kozlowski \cite{shrestha2016inverse}, while a later version was published by Hu and Kozlowski \cite{hu2016inverse}. Similar to CIRC{\'E}, it also relies on the linearity and normality assumptions, and used a change-of-variable in order to deal with dimensionless multipliers instead of the PMPs directly. MLE/MAP also uses the sensitivity matrix which includes the derivative of the QoIs with respect to the PMPs. Based on the normality and linearity assumptions, the model QoIs also follow a multi-dimensional Gaussian distribution $\mathbf{y}^{\text{E}}_i  \sim  \mathcal{N} \left(  \mathbf{y}^{\text{M}}_i  +  \mathbf{S}_i \bm{\mu},  \bm{S}_i  \bm{\Sigma}_{\bm{\theta}}  \bm{S}_i^{\top}  +  \mathbf{\Sigma}_{\bm{\epsilon}, i}  \right)$, where $\mathbf{y}^{\text{E}}_i$, $\mathbf{y}^{\text{M}}_i$, $\mathbf{S}_i$, and $\mathbf{\Sigma}_{\bm{\epsilon}, i}$ are the measurement data, model simulation, sensitivity matrix and data uncertainty for the $i^{\text{th}}$ experiment, respectively. $\bm{\mu}$ and $\bm{\Sigma}_{\bm{\theta}}$ are the mean vector and covariance matrix of $\bm{\theta}$. The corresponding PDF is the likelihood in the Bayesian setting.

The posterior is proportional to the prior-likelihood product, to find the mean and variance of $\bm{\theta}$, optimization using the E-M algorithm is employed by the MLE/MAP method. When conjugate priors are used, such as a normal prior for the mean vector and an inverse gamma prior for the covariance, the posterior is considered as the target function for optimization, hence it is called MAP. When the prior is treated as a non-informative constant, only the likelihood is treated as the target function, hence it is called MLE.
In \cite{hu2016inverse}, the authors briefly discussed the application of MCMC sampling instead of E-M algorithm. Therefore, we classify MLE/MAP as a hybrid frequentist/Bayesian IUQ method in Table \ref{table:Intro-List-of-IUQ-Methods}. MLE represents a set of values for the PMPs at which the probability of observing/reproducing the given data by the model is the highest. The sensitivity matrix is calculated using discrete adjoint methods, which can be complicated. Similar to CIRC{\'E}, the applications of MLE/MAP are limited by several issues: (1) the relation between the QoIs and the input parameters were assumed to be linear. The linearity assumption is valid for parameter values in the vicinity of the nominal value, but becomes non-linear far from the nominal value; (2) the input parameters were assumed to follow normal distributions, and (3) adjoint sensitivity analysis was required to provide the sensitivity matrix.

The MLE/MAP method was applied in \cite{shrestha2016inverse} to calibrate two TRACE PMPs based on the FEBA tests. It was later applied in \cite{hu2016inverse} to quantify the uncertainties in another two TRACE PMPs using BFBT void fraction data. The effect of the chosen conjugate prior distributions was shown to be large. To study the effects of boundary conditions uncertainties on the estimation of PMPs, Abu Saleem and Kozlowski \cite{saleem2019estimation} added an additional bias term to the linear Taylor expansion. Reduction in the average error of the code prediction was found for the case with the bias term. In another work \cite{saleem2019effect}, the authors studied the effect of mesh refinement on IUQ of physical models of the TH code RSTART, based on the BFBT data. It was found that mesh size refinement has a noticeable effect on the estimation of the mean, but negligible effect on the estimation for the variance. The results obtained by MLE and the most refined mesh demonstrated the best agreement of code and data.

\subsection{Other IUQ Methods}

There is some other work that didn't focus on IUQ but involved a calibration step. For example, Phung et al. \cite{phung2015input} developed a procedure for manual input calibration using multiple parameters measured in different test regimes. This procedure was improved to an automated approach algorithm \cite{phung2016automation} to input calibration and RELAP5 code validation against data on two-phase natural circulation flow instability. This procedure consists of a calibration step and a validation step. In calibration, the ranges of PMPs are quantified by minimizing the model-data difference, by optimizing a fitness function that uses normalization and weighting factors to represent contribution from different QoIs. In validation, the parameter ranges are used by the Genetic algorithm (GA) to identify combinations of the uncertain input parameters that provide maximum deviation of code prediction from the experimental data to maintain certain conservatism. GA is a heuristic method that mimics the process of natural selection in order to find a global optimum of the fitness function. However, only the parameter ranges are found, and user judgement is also involved to determine the normalization and weighting factors in the fitness function.

\subsection{IUQ in Other Nuclear M\&S Areas}

This review paper has focused on IUQ for system TH codes. There are closely related IUQ work in other nuclear areas. For example,  Higdon et al. \cite{higdon2013calibration} used FBA to inversely quantify the uncertainties in four tuning parameters of the FRAPCON code based on fission gas release data from 42 experiments. The measurement uncertainty and the model bias were quantified simultaneously. The MBA method was applied to the fission gas release model in the BISON code \cite{wu2018Kriging}, in which the authors proposed a method for cases when time-series data is used. PCA was used to project the original time-series data to the principal component subspace. IUQ was performed on the subspace to achieve better MCMC convergence. This method was also used and compared with variational Bayesian monte carlo \cite{che2021application} to calibrate the BISON fission gas release model for chromia/alumina-doped UO\textsubscript{2} fuel. Both methods showed similar accuracy. Stripling et al. \cite{stripling2013calibration} developed a method for calibration and data assimilation using the Bayesian multivariate adaptive regression splines emulator as a surrogate for the computer code. This method started with sampling of the uncertain input space. The emulator was then used to assign weights to the samples which were applied to produce the posterior distributions of the inputs. This approach was applied to the calibration of a Hyades 2D model of laser energy deposition in beryllium. The major difference of this approach with Bayesian IUQ is that, it generated samples beforehand and the candidate acceptance routine in MCMC sampling was replaced with a weighting scheme. Note that such approach did not include the model bias term.

\section{Evaluation of the Reviewed IUQ Methods}
\label{section:Evaluation-of-IUQ-Methods}

To choose an IUQ method, or a category of methods, the users have to consider the underlying assumptions, the application scenarios, and the merits/limitations of each method. In the step 11 of the SAPIUM project, a few selection criteria were proposed to provide guidance when choosing an IUQ method \cite{baccou2019development-NED}. In this paper, we adopt a few criteria from SAPIUM and added a few others. These criteria are listed in Table \ref{table:Summary-Criteria-of-IUQ-Methods}. Based on the metrics presented in Table \ref{table:Summary-Criteria-of-IUQ-Methods}, the 12 IUQ methods listed in Table \ref{table:Intro-List-of-IUQ-Methods} are evaluated on a scale from 1 to 5: 1 (very poor), 2 (poor), 3 (fair), 4 (good) and 5 (excellent). The evaluation results are presented in Table \ref{table:Summary-Scores-of-IUQ-Methods}.

\begingroup
\setlength{\tabcolsep}{10pt} 
\renewcommand{\arraystretch}{1.3} 
\begin{table}[!ht]
	\footnotesize
	\centering
	\captionsetup{justification=centering}
	\caption{Criteria to evaluate an IUQ method.}
	\label{table:Summary-Criteria-of-IUQ-Methods}
	\begin{tabular}{p{0.18\linewidth}  p{0.7\linewidth}}
		\toprule
		Criteria  &  Explanation   \\ 
		\midrule
		Solidity           &  Does it rely on a rigorous mathematical framework?  \\
		
		Complexity         &  Is it complex and difficult to use?  \\
		
		Accessibility      &  Can it be applied non-intrusively (without modification of the source code)? Is the mathematical theory easy to be programmed?  \\
		
		Independence       &  Is the method built upon important assumptions? Does it require expert judgment or engineering consideration?  \\
		
		Flexibility        &  Does it work for a wide range of problem scenarios, for example, when the data is limited, or when the data is not time-dependent? How many parameters can it deal with together?  \\
		
		Comprehensiveness  &  Can all sources of uncertainties be considered simultaneously (parameter, model, experiment and code)? Does it quantify the parameter distributions or only ranges? Can the parameter joint PDFs also be obtained?  \\
		
		Transparency       &  Is there a clear documentation available, including list of assumptions and user guidelines? Has this method been widely used with success, especially in the nuclear community?  \\
		
		Tractability       &  Is it computationally expensive? Can surrogate modeling be easily applied?  \\
		\bottomrule
	\end{tabular}
\end{table}
\endgroup

\textit{\textbf{Solidity}}: Both Frequentist and Bayesian IUQ methods have rigorous mathematical definitions. Empirical IUQ methods usually do not rely on a robust mathematical framework. An exception is IPREM, as shown in \ref{Appendix-IPREM}. CET-based sample adjusting and sampling-based IUQ have the lowest scores in this criterion. DIPE has a slightly higher score of 2 due to its evaluation of the coverage rate and determination of the pseudo-CDF. Non-parametric clustering, being an empirical method, is considered to have a fair solidity due to the utilization of non-parametric PDF estimation and clustering.

\textit{\textbf{Complexity}}: A method with higher solidity generally has higher complexity, thus a low complexity score. DAA and CASUALIDAD (which uses DAA as an element) are considered to be the most sophisticated IUQ methods, due to the calculation of the covariance matrices for parameter-parameter, QoI-parameter and QoI-QoI pairs. CIRC{\'E} and MLE/MAP use the E-M algorithm, while Bayesian CIRC{\'E} needs the blocked Gibbs sampler, making them also very complex. IPREM and MCDA only use basic linear algebra operations. MBA also have fair complexity because it is relatively easy to build GP emulators and use MCMC sampling. Non-parametric clustering is also assigned a score of 3 because non-parametric PDF estimation and clustering are easy to implement. The empirical methods based on sampling have the highest scores, with DIPE being slightly more complex than CET-based sample adjusting and sampling-based IUQ.

\textit{\textbf{Accessibility}}: Most IUQ methods are non-intrusive because they treat the system TH codes as black-boxes. Only post-processing of the model simulation results is needed for a non-intrusive IUQ method, so that source code modification and re-compilation of the system TH codes are not required. When the sensitivity matrices that contain the derivative information are required, Adjoint Sensitivity Method (ASM) will be needed, which lead to intrusive use of the source code. An example is CIRC{\'E}, DAA and CASUALIDAD. Of course, the sensitivity matrices can be calculated using finite difference for simple problems, but the accuracy is not as good as ASM. Because MCDA and MLE/MAP are both hybrid frequentist/Bayesian methods, they can use MCMC sampling instead of the sensitivity matrix, so their accessibilities are considered to be fair. Another aspect of the accessibility is the level of difficulty to realize the IUQ method's mathematical theory by programming. IPREM, MBA, Bayesian CIRC{\'E} and non-parametric clustering require more efforts to use than CET-based sample adjusting, DIPE and sampling-based IUQ.

\textit{\textbf{Independence}}: Many assumptions have been used in the IUQ methods, such as linearity, normality, parameter independence, etc. Methods that use the linearity assumption include CIRC{\'E}, MLE/MAP, and the deterministic component of MCDA. Bayesian CIRC{\'E} extends the classical CIRC{\'E} method to non-linear settings. Normality assumption is also used by CIRC{\'E}, Bayesian CIRC{\'E}, MCDA and MLE/MAP. CIRC{\'E}, MCDA, MLE/MAP and most empirical methods assume mutual independence of the PMPs. Empirical IUQ methods are generally rooted in less assumptions than the others. Nevertheless, they usually dependent heavily on expert judgment. For example, engineering considerations are needed in multiple steps of non-parametric clustering, such as selection of the design variables and separation of the assessment state space to multiple regions. Optimization of the parameter ranges in other empirical IUQ methods is generally based on expert knowledge and visual comparison. IPREM is based on engineering considerations rather than solely on statistical methods. As pointed out by the method developers \cite{kovtonyuk2017development}, the resulting ranges of input may be heavily dependent on the selected QoIs. There are scenarios in which IPREM will fail, and the user has to manually adjust the limit value for CR. Different limit values can also result in very distinct ranges of variation, so it requires careful engineering characterization. All these features will undermine the independence of the IUQ methods. MBA only assumes the experimental uncertainty to be i.i.d. Gaussian, which is treated as reasonable and has been widely used in many disciplines.

\begingroup
\renewcommand{\arraystretch}{1.3}
\begin{table}[!ht]
	\tiny
	\centering
	\captionsetup{justification=centering}
	\caption{Metric scores of the 12 reviewed IUQ methods according to the 8 evaluation criteria.}
	\label{table:Summary-Scores-of-IUQ-Methods}
	\begin{tabular}{l c c c c c c c c}
		\toprule
		IUQ methods  &  Solidity  &  Complexity  &  Accessibility  &  Independence  &  Flexibility  &  Comprehensiveness  &  Transparency   &  Tractability  \\ 
		\midrule
		CIRC{\'E}                  &  5  &  2  &  1  &  1  &  2  &  3  &  5  &  3   \\		
		IPREM                      &  5  &  3  &  4  &  1  &  2  &  1  &  5  &  4   \\		
		CET-based Sample Adjusting &  1  &  5  &  5  &  3  &  4  &  2  &  3  &  5   \\		
		DIPE                       &  2  &  4  &  5  &  2  &  2  &  1  &  3  &  3   \\		
		MCDA                       &  5  &  3  &  3  &  3  &  4  &  3  &  4  &  4   \\
		Sampling-based IUQ         &  1  &  5  &  5  &  3  &  4  &  2  &  3  &  5   \\		
		MBA                        &  5  &  3  &  4  &  4  &  5  &  5  &  5  &  3   \\		
		Bayesian CIRC{\'E}         &  5  &  2  &  4  &  3  &  4  &  3  &  3  &  3   \\		
		Non-parametric Clustering  &  3  &  3  &  4  &  3  &  3  &  2  &  3  &  3   \\		
		DAA                        &  5  &  1  &  1  &  3  &  2  &  4  &  4  &  3   \\		
		CASUALIDAD                 &  5  &  1  &  1  &  3  &  2  &  4  &  4  &  3   \\		
		MLE and MAP                &  5  &  2  &  3  &  1  &  2  &  3  &  5  &  3   \\
		\bottomrule
	\end{tabular}
\end{table}
\endgroup

\textit{\textbf{Flexibility}}: The flexibility, or applicability of an IUQ method depends on the number of assumptions it relies on, as well as the inverse problem scenarios it will work. Most IUQ methods require the selected QoIs to be sensitive to the targeted parameters. Otherwise, the posterior distributions in frequentist and Bayesian IUQ methods will be wide, and many trial-and-error runs may be needed in empirical IUQ methods. CIRC{\'E} requires the inverse problem to be well-posed and identifiable. Furthermore, the selected QoIs need to be significantly different from each other and independent. Because it uses local sensitivity based on derivatives, it can only deal with inputs that have relatively small variations unless iterative CIRC{\'E} is applied. Finally, CIRC{\'E} should not be used for more than 3 parameters together, so is MLE/MAP. IPREM, DAA and CASUALIDAD only apply to time-dependent IUQ problems. Empirical methods cannot deal with high-dimensional problems, because it gets very complicated to adjust their ranges together to provide a good coverage of the measurement data. The application scenario for DIPE is very restrictive because it requires that the simulation curve is monotonous. Moreover, DIPE application for more than two parameters can be expensive. Also it is unclear how to determine the simulation curves that correspond to the 2.5$^{\text{th}}$ and the 97.5$^{\text{th}}$ percentiles, as several combinations of the parameters may have the similar simulation curve. Non-parametric clustering needs relatively large amount of experimental data. Among the remaining methods, MBA is the most widely applicable method. It works for both time series and scalar QoI. There is no theoretical limit on the number of PMPs they can deal with, yet the GP emulator works best for problems with moderate dimensions (typically less than 50 \cite{wu2018inversePart1}.

\textit{\textbf{Comprehensiveness}}: The IUQ method should be able to incorporate all sources of quantifiable uncertainties in M\&S. Note that there is no unanimous agreement of the classification of uncertainties. As discussed in Section \ref{section:Definition-of-IUQ}, in this paper we adopt the classification in \cite{wu2018inversePart1}, which consists of four sources: parameter, model, data and code. Even though the experimental uncertainty is considered in the formulation of CIRC{\'E}, it is treated to be lower than the physical models uncertainties and therefore they have not been taken into account in some applications, such as \cite{freixa2016testing}. Bayesian CIRC{\'E} calculates the posterior distribution of parameters instead of their MLEs. But it still omits the model bias term and and presumes that the gap between data/model is mostly due to the uncertainties in PMPs. Bayesian CIRC{\'E} also has the independence assumption of the PMPs, and it does not quantify the PMP mutual dependence in the posterior, because the posteriors are sought for $(b, \sigma)$. For the empirical methods, IPREM, DIPE and non-parametric clustering do not consider the experimental uncertainty, while CET-based sample adjusting and sampling-based IUQ only consider it in a visual manner. Furthermore, empirical methods only achieve the parameter ranges, without specifying the type of distributions. Therefore, the user has to choose a distribution for the inputs, for example, uniform or Gaussian. Note that for the case of two or more parameters, DIPE cannot achieve the pair-wise joint distributions because the converge rate contour lines do not represent the true joint CDF. Non-parametric clustering does not quantify the uncertainties of PMPs, but fits multiple PDFs for the model prediction errors, which later serve as statistical compensation to the model prediction. DAA and CASUALIDAD achieve information and parameter joint uncertainties in the parameter-parameter covariance matrix, but the model bias term is not considered.  Only MBA consider the model bias since it is derived based on the model updating equation in Bayesian calibration.

\textit{\textbf{Transparency}}: All of the IUQ methods reviewed in this paper are clearly defined. Their original references have been provided in Table \ref{table:Intro-List-of-IUQ-Methods}. CIRC{\'E}, IPREM, MBA and MLE/MAP are the most widely applied IUQ methods. MCDA, DAA and CASUALIDAD have also been use in multiple applications. The remaining IUQ methods are assigned a score of 3 because they haven't been widely applied, based on the authors' knowledge.

\textit{\textbf{Tractability}}: For a fixed IUQ problem, trial-and-error methods such as CET-based sample adjusting and sampling-based IUQ are considered to be the most tractable. For most of the reviewed IUQ methods, the cost of empirical methods will be high when the number of PMPs is high. With DIPE, the number of experimental design may increase exponentially with the dimension. Bayesian IUQ methods use MCMC samples that requires tens of thousands of model runs. Fortunately, most non-intrusive methods can readily make use of surrogate modeling to significantly reduce the computational cost.

Note that the evaluation scores in Table \ref{table:Summary-Scores-of-IUQ-Methods} are only based the authors' understanding. It should only be used as a guidance to select an IUQ method depending on the problem under investigation. A lower average score does not mean this method should not be used. For example, empirical IUQ methods can be used to find uncertain ranges that can be used as prior information to initiate the Bayesian IUQ methods. Even though the CIRC{\'E} method has several limitations, it has had multiple very successful applications especially with the CATHARE code. It is also worth noting that certain IUQ methods, such as DAA and CASUALIDAD, are part of extensive predictive evaluation frameworks that involve not only IUQ. The evaluation presented in this paper is only for their IUQ component.

\section{Challenges and Research Needs}

\subsection{Mathematical Description of the Model Bias Term in Bayesian IUQ}

The model bias term $\delta(\mathbf{x})$ was first addressed in the seminal work of Kennedy and O'Hagan \cite{kennedy2001bayesian}. It is important to consider $\delta(\mathbf{x})$ as otherwise we would have an unrealistic level of confidence in the computer model predictions \cite{brynjarsdottir2014learning}. Ignoring $\delta(\mathbf{x})$ during IUQ is equivalent to treat the computer model as perfect because Equation (\ref{equation:Bayesian-IUQ1-Bias}) becomes
\begin{equation}	 \label{equation:Bayesian-IUQ1-Bias-Removed}
	\mathbf{y}^{\text{R}} (\mathbf{x}) = \mathbf{y}^{\text{M}} \left( \mathbf{x}, \bm{\theta}^{*} \right)
\end{equation}

In this situation, we are over-confident with the computer model and will have over-fitting during IUQ \cite{brynjarsdottir2014learning}. Over-fitting means that the calibration parameters are so over-tuned to the calibration data that the computer code may perform poorly when applied to other experiments. Without $\delta(\mathbf{x})$, the posterior PDF becomes:
\begin{equation}        \label{equation:Bayesian-IUQ5-Posterior-no-Bias}
	p \left( \bm{\theta}^{*} | \mathbf{y}^{\text{E}}, \mathbf{y}^{\text{M}}\right)  \propto  \frac{ p \left( \bm{\theta}^{*} \right) }{  \sqrt{| \bm{\Sigma}_{\text{code}} + \bm{\Sigma}_{\text{exp}} |} }   \cdot  \text{exp} \left[  - \frac{1}{2} \left[ \mathbf{y}^{\text{E}} - \mathbf{y}^{\text{M}}  \right]^\top \left( \bm{\Sigma}_{\text{code}} + \bm{\Sigma}_{\text{exp}} \right)^{-1} \left[ \mathbf{y}^{\text{E}} - \mathbf{y}^{\text{M}}  \right] \right]
\end{equation}

It was demonstrated in \cite{wu2018inversePart2} that $\delta(\mathbf{x})$ can avoid over-fitting by providing a regularization effect. However, due to the inherent difficulty in the mathematical description of $\delta(\mathbf{x})$ since it is not directly observable, it has been ignored in most of the previous work on Bayesian calibration, see the review in \cite{wu2018inversePart1}.

The greatest challenge in dealing with $\delta(\mathbf{x})$ is that there are no direct observations for $\delta(\mathbf{x})$, making it extremely difficult to learn because there is no training data available. To represent $\delta(\mathbf{x})$ with a GP, or any other statistical models, training data is needed whose inputs are $\mathbf{x}$ and outputs are the differences between the reality and computer code simulations. Such differences cannot be obtained because the reality is never known. Therefore, we need substitutes of such ``observation data''. Three intuitively natural modularization schemes were compared in \cite{liu2009modularization} to estimate the hyperparameters $\bm{\Psi}^{\delta}$ of the GP model for $\delta(\mathbf{x})$.
\begin{enumerate}[label=(\arabic*)]
	\setlength{\itemsep}{0.1pt}
	\item Treat the differences between data and simulation (run at prior means or nominal values of $\bm{\theta}$, while using the same $\mathbf{x}$ with measurement data) as the ``realizations'' of $\delta(\mathbf{x})$. Then $\bm{\Psi}^{\delta}$ can be estimated with MLE. This is the approach used in \cite{wu2018inversePart2}.
	
	\item Sample $\bm{\theta}$ from $\bm{\theta}^{\text{Prior}}$. Run the computer model at every sample of  $\bm{\theta}$, and calculate their differences with experimental data to obtain an ensemble of ``fake observation data'' for $\delta(\mathbf{x})$. This set of data will be used to obtain the posterior distributions for $\bm{\Psi}^{\delta}$. Then sample $\bm{\Psi}^{\delta}$ from their posteriors, which is generated conditioning on $\bm{\theta}^{\text{Prior}}$. The posterior sample mean will be used as the fixed values of $\bm{\Psi}^{\delta}$.
	
	\item Initially assume the model is perfect ($\delta(\mathbf{x}) = \mathbf{0}$). Then solve for $\bm{\theta}^{\text{Posterior}}$. The resulting posterior will be used as a new prior and proceed with approach (1). 
\end{enumerate}

The first approach requires a careful test source allocation process \cite{wu2018inversePart2} to separate the given data for IUQ and training of $\delta(\mathbf{x})$. It is not applicable when there is very limited experimental data. The second approach requires a proper prior for $\bm{\Psi}^{\delta}$, while the third approach is more expensive to apply. It can also be problematic if over-fitting causes the first round of IUQ to be biased, the subsequent IUQ will always be biased since posterior is usually more concentrated than the prior. To sum up, all the available approaches have their own limitations. More research is needed to develop a better mathematical description for the model bias term, especially for cases with limited data.

\subsection{Confounding and Non-identifiability in IUQ}
\label{section:Identifiability}

A second challenge associated with the model bias $\delta(\mathbf{x})$ is the ``non-identifiability'' issue \cite{wu2019demonstration}. The consideration of $\delta(\mathbf{x})$ can avoid over-fitting, but it also poses challenges in the practical applications. One of the mostly concerned and unresolved problem is the lack of identifiability issue \cite{arendt2012quantification} \cite{liu2009modularization}. Identifiability answers the question that whether the true value $\bm{\theta}^{*}$ can theoretically be inferred based on the available measurement data. It is difficult to know how much of the difference between model and data should be attributed to the parameter uncertainty in $\bm{\theta}$, model bias $\delta(\mathbf{x})$ and measurement uncertainty $\bm{\epsilon}$, known as the ``confounding effect''. Different combinations of these uncertainties can account for the same distinction between model and data, making the true value $\bm{\theta}^{*}$ not identifiable. Consequently, IUQ is usually ill-posed due to the fact that there exist multiple solutions.

Previous research to alleviate the non-identifiability issue focused on using informative priors for $\bm{\theta}$ and $\delta(\mathbf{x})$, which is usually not a viable solution because one rarely has such accurate prior knowledge. In a recent work \cite{wu2019demonstration}, it was shown that identifiability is largely related to the sensitivity of $\bm{\theta}$ to the chosen QoIs. In order for a certain PMP to be statistically identifiable, it should be significant to at least one of the QoIs whose data are used for IUQ. It was also demonstrated that ``fake identifiability'' is possible if the QoIs are not appropriately chosen, or if inaccurate but informative prior distributions are specified. However, more future research is necessary to quantitatively address the confounding and non-identifiability issues.

\subsection{Extrapolation of the Model Bias}

A third challenge in $\delta(\mathbf{x})$ is related to the extrapolation of the learned model bias to validation and prediction domains. According to Equation \ref{equation:Bayesian-IUQ1-Bias}, once $\bm{\theta}^{\text{Posterior}}$ is quantified, the realities in the validation and prediction domains are described as:
\begin{equation}       \label{equation:Bayesian-IUQ7-VALidation-PREDiction}
	\begin{aligned}
		\mathbf{y}^{\text{R}} ( \mathbf{x}^{\text{VAL}} )  
		&= \mathbf{y}^{\text{M}} ( \mathbf{x}^{\text{VAL}}, \bm{\theta}^{\text{Posterior}} ) + \delta ( \mathbf{x}^{\text{VAL}} )    \\
		\mathbf{y}^{\text{R}} ( \mathbf{x}^{\text{PRED}} )  
		&= \mathbf{y}^{\text{M}} ( \mathbf{x}^{\text{PRED}}, \bm{\theta}^{\text{Posterior}} ) + \delta ( \mathbf{x}^{\text{PRED}} )
	\end{aligned}
\end{equation}
where $\mathbf{x}^{\text{VAL}}$ and $\mathbf{x}^{\text{PRED}}$ represent the design variables in the validation and prediction domains, respectively. The model bias $\delta ( \mathbf{x}^{\text{VAL}} )$ and $\delta ( \mathbf{x}^{\text{PRED}} )$ are the GP model that is trained based on the data in the IUQ domain, but extrapolated to the validation and prediction domains.

IUQ with FBA/MBA both represent $\delta(\mathbf{x})$ in a fully data-driven manner. Therefore, these methods should be used with great caution. Extrapolation outside the range of the IUQ domain is questionable. As discussed in \cite{higdon2008computer}, the quality of such extrapolation largely depends on the reliability of $\delta(\mathbf{x})$. What we have learned about $\delta(\mathbf{x})$ at the IUQ domain may not be applicable to the validation and prediction domains. See \cite{higdon2004combining} for an example in which $\delta(\mathbf{x})$ is large in magnitude, but $\bm{\theta}^{\text{Posterior}}$ is similar with that obtained when $\delta(\mathbf{x})$ is zero. Besides the reliability of $\delta(\mathbf{x})$, extrapolation using GP emulator is inherently dangerous. GP emulator usually has large mean prediction errors and significant variance outside of its training domain, as shown in Figure \ref{figure:Bayesian-IUQ4-GP}. More research is needed to improve the GP accuracy when extrapolation is used.

In \cite{bachoc2014calibration}, the authors also used a GP to model $\delta(\mathbf{x})$. Outside of the calibration domain, the learned GP is used as a statistical correction to add to the model predictions. The reported results showed that this statistical correction can substantially improve the calibrated computer model for predicting the physical system on new experimental conditions. The estimator for the prediction is a linear combination of the calibrated computer model simulation and the inferred model error. The estimator was derived as a decreasing function of the distance between the new experimental condition $\mathbf{x}^{\text{new}}$ and $\mathbf{x}^{\text{IUQ}}$. However, the derivation in this work depends on a linearity assumption between the QoIs and $\bm{\theta}$. More investigation needs to be done to confirm the effects of extrapolating the GP-based model bias for a non-linear model.

\subsection{Incorporation of Uncertainties from Design Variables}

In Section \ref{section:Definition-of-IUQ}, when classifying the model inputs as design variables $\mathbf{x}$ and calibration parameters $\bm{\theta}$, it was mentioned that both types of inputs can be uncertain, but IUQ only seeks uncertainties in $\bm{\theta}$ because the uncertainties in $\mathbf{x}$ are assumed to be known from the benchmark data. Such a treatment has been used in most IUQ work reviewed in this paper. For example, the CEA team in PREMIUM mentioned \cite{nouy2017quantification} that design variables (called IBP in the paper) should not be considered since they are not part of a physical model of the TH code; their uncertainties are known a priori and should be theoretically given by the experimenters and not estimated. The PREMIUM summary paper \cite{skorek2019quantification} claimed that IUQ of PMPs should not be applied at the same time to design variables that have full physical meaning, unless there is no other source of information about their uncertainty.

In PREMIUM, the considered inputs were mostly PMPs with a few exceptions. For example, the Tractebel team \cite{zhang2019development} considered five parameters which are boundary conditions, including the local heat flux, bundle power, etc. KIT \cite{mendizabal2017post} also considered the rod bundle power for IUQ. In the MCDA paper \cite{heo2014implementation}, seven boundary conditions variables were selected, including mass flow rate, pressure, temperature and four inputs related to power. Even though it is generally agreed that $\mathbf{x}$ should not be the target of IUQ, their uncertainties should still be considered when seeking the parameter uncertainties in $\bm{\theta}$. This is because the uncertainties in $\mathbf{x}$, after being propagated through the computer model, can still contribute to the differences between model and data. In most previous Bayesian IUQ methods, a potentially significant limitation is that the parameter uncertainties in $\mathbf{x}$ were not considered. Therefore, a systematic study is needed to investigate the effects of uncertainties in $\mathbf{x}$ on the uncertainties in $\bm{\theta}$.

\subsection{More Open Issues for Future Development}

IUQ is an emerging area that has many unresolved issues. Besides those discussed above, there are several important problems that need future development. For example, application of the quantified physical model uncertainties at new experiments. In the PREMIUM project, IUQ and verification were based on the FEBA tests, while validation of IUQ results was performed using PERICLES tests. The verification was successful with a good coverage of the experimental data by the FUQ uncertainty bands. However, when the physical model uncertainties from IUQ were extrapolated to the PERICLES validation data, the results were not satisfactory. Even though multiple reasons have been identified \cite{skorek2019quantification} \cite{skorek2017input} for the validation failure, extrapolation of the IUQ results to new experimental settings needs a more systematic study. Note that in Bayesian IUQ methods, extrapolation of $\bm{\theta}^{\text{Posterior}}$ is acceptable since Bayesian IUQ assumes there are true values of  $\bm{\theta}$ that remain unchanged in different experiments.

Another important topic is whether one should use a ``frozen'' version of the TH code during IUQ. Here frozen means the TH code should not be changed during IUQ. Some researchers believe that applying calibration during IUQ is not advantageous \cite{skorek2019quantification}. It was regarded as a not acceptable modification of the code, which makes questionable the best estimate character of the code and its validation. However, based on the discussion in this paper, for Bayesian methods the boundary between IUQ and calibration is not as clear as other methods. Several other important topics during IUQ include scale-up effects of the IUQ results, adequacy of the experimental database, predictive assessment, etc. The readers are highly recommended to refer to \cite{baccou2019development-NED} \cite{baccou2020sapium} for more detailed discussions.

\section{Conclusions}

UQ is the process to quantify the uncertainties in QoIs by propagating the uncertainties in input parameters through the computer model. UQ is an essential step in computational model validation because assessment of the model accuracy requires a concrete, quantifiable measure of uncertainty in the model predictions. The concept of UQ in the nuclear community generally means forward UQ (FUQ), in which the information flow is from the inputs to the outputs. However, there is another equally important component of UQ - inverse UQ (IUQ), that has been significantly underrated until recently. With IUQ, the information flow is from the model outputs and experimental data to the inputs. FUQ requires knowledge in the model input uncertainties, such as the statistical moments, PDFs, upper and lower bounds, which are not always available. Historically, expert opinion or user self-evaluation have been predominantly used to specify such information in VVUQ studies. Such ad-hoc specifications are subjective, lack mathematical rigor, and can sometimes lead to inconsistencies. IUQ is defined as the process to inversely quantify the input uncertainties based on experimental data. It seeks statistical descriptions of the uncertain input parameters that are consistent with the observation data.

This review paper aims to provide a comprehensive and comparative discussion of the major aspects of the IUQ methodologies that have been used in nuclear engineering, with a focus on the physical models in system TH codes. IUQ methods can be categorized by three main groups: \textit{frequentist}, \textit{Bayesian}, and \textit{empirical}. All these three groups of IUQ methods depend on a comparison between code simulations and observation data, though in different manners. Frequentist IUQ tries to identify most likely parameter values, with which the TH model can reproduce the experimental data. Bayesian IUQ targets at finding parameter uncertainties that can explain the disagreement between model and data, typically with MCMC sampling. Empirical IUQ seeks parameter uncertainties with which the model predictions can envelop the measurement data to a desired level. Because of these different mechanisms, these three types of IUQ methods have very different assumptions, application scenarios, treatment of various sources of uncertainties, etc.

We used eight metrics to evaluate an IUQ method, including \textit{solidity}, \textit{complexity}, \textit{accessibility}, \textit{independence}, \textit{flexibility}, \textit{comprehensiveness}, \textit{transparency}, and \textit{tractability}. Twelve IUQ methods are reviewed, compared, and evaluated based on these eight metrics. Such comparative evaluation will provide a good guidance for users to select a proper IUQ method based on their IUQ problem under investigation. We also identified a few open issues and research needs for IUQ in the nuclear area, including three challenges for the model bias term, extrapolation of IUQ results, scale-up effects of the IUQ results, adequacy of the experimental database, predictive assessment of the model after IUQ, etc. IUQ is an emerging area that has many unresolved issues. More research efforts need to be devoted to this area as it has a significant potential to improve the M\&S predictive capability.

\appendix
\section{The CIRC{\'E} Method}    \label{Appendix-CIRCE}

CIRC{\'E} usually deals with a very small number of PMPs, $\bm{\theta} = \{ \theta_i \}_{i=1}^{I}$ where $I = 1, 2, \text{or } 3$, rarely more. Assume there are $J$ QoIs for which the measured data is available, $\mathbf{y} = \{ y_j \}_{j=1}^{J}$ with $J$ typically equals to several tens. The $\theta_i$ parameters are treated as normal random variables. CIRC{\'E} estimates the mean value (also called bias) $b_i$ and the standard deviation $\sigma_i$ of each $\theta_i$. The input information to CIRC{\'E} are: (1) the differences between the experimental data $\mathbf{y}^{\text{E}}$ and the code calculations $\mathbf{y}^{\text{M}}$, denoted as $( y^{\text{E}}_j - y^{\text{M}}_j )$ for the $j^{\text{th}}$ QoI, (2) the derivatives of each QoI with respect to each parameter, denoted as $\frac{\partial y^{\text{M}}_j}{\partial \theta_i}$, (3) the experimental uncertainties (optional), $\bm{\epsilon} = \{ \epsilon_j \}_{j=1}^{J}$ where $\epsilon_j$ is the uncertainty associated with $y^{\text{E}}_j$. CIRC{\'E} uses the Adjoint Sensitivity Method (ASM) \cite{cacuci2003sensitivity} or finite difference to calculate the derivatives. Table \ref{table:PREMIUM-CIRCE-Symbols} lists all the symbols used for CIRC{\'E}.

\begingroup
\renewcommand{\arraystretch}{1.2}
\begin{table}[!ht]
	\footnotesize
	\centering
	\captionsetup{justification=centering}
	\caption{Definitions of symbols used for the CIRC{\'E} method.}
	\label{table:PREMIUM-CIRCE-Symbols}
	\begin{tabular}{p{0.07\linewidth} | p{0.44\linewidth} | p{0.07\linewidth} | p{0.3\linewidth}}
		\toprule
		Symbol  &  Description  &  Symbol  &  Description  \\ 
		\midrule
		$\bm{\theta}$      &  vector of PMPs, $\bm{\theta} = \{ \theta_i \}_{i=1}^{I}$  &  $I$  &  dimension of $\bm{\theta}$    \\
		
		$\mathbf{y}$       &  QoIs, $\mathbf{y} = \{ y_j \}_{j=1}^{J}$  &  $J$  &  dimension of $\mathbf{y}$    \\
		
		$b_i$              &  mean value (bias) of $\theta_i$  &  $\mathbf{b}$  &  mean (bias) vector, $\mathbf{b} = \{ b_i \}_{i=1}^{I}$   \\
				
		$\sigma_i$         &  standard deviation of $\theta_i$  &  $\bm{\Sigma}_{\bm{\theta}}$  &  covariance matrix of $\bm{\theta}$    \\
		
		$p_i$              &  dimensionless multiplier for $\theta_i$  &  $\mathbf{y}^{\text{M}}$  &  QoIs from model simulation    \\
		
		$\mathbf{y}^{\text{E}}$  &  QoIs from experiment  &  $\mathbf{y}^{\text{R}}$  &  QoIs' real values    \\
		
		$\epsilon_j$       &  measurement uncertainty for $y_j^{\text{E}}$  &  $\bm{\epsilon}$  &  $\bm{\epsilon} = \{ \epsilon_j \}_{j=1}^{J}$    \\
		
		$e_j$              &  measurement error, $e_j \sim \mathcal{N} (0, \epsilon_j^2)$   &  $\bm{e}$  &  $\bm{e} = \{ e_j \}_{j=1}^{J}$    \\

		$\bm{\theta}_j$  &  values of $\bm{\theta}$ such that $y^{\text{M}}_j (\bm{\theta}_j) = y^{\text{R}}_j$  &  $\hat{\bm{\Sigma}}_{\bm{\theta}}$  &  estimator of $\bm{\Sigma}_{\bm{\theta}}$  \\
		
		$\bm{\Sigma}_{\bm{\theta}}^{(m-1)}$  &  \textit{a priori} estimation of $\bm{\Sigma}_{\bm{\theta}}$ at the $m^{\text{th}}$ iteration  &  $\bar{\bm{\theta}}_j$  &  mean vector for $\bm{\theta}_j$  \\
		
		$\bm{\Sigma}_{\bm{\theta}}^{(m)}$  &  \textit{a posteriori} estimation of $\bm{\Sigma}_{\bm{\theta}}$ at the $m^{\text{th}}$ iteration  &  $\bm{\Sigma}_{\bm{\theta}_j}$  &  the covariance matrix for $\bm{\theta}_j$  \\
		
		$\bm{\beta}_j$  &  the centralized vector $\bm{\beta}_j = \bm{\theta}_j - \bm{b}$  &  $\bar{\bm{\beta}}_j$  &  mean vector for $\bm{\beta}_j$  \\		
		\bottomrule
	\end{tabular}
\end{table}
\endgroup

\subsection{Change of Variable}

Unlike other IUQ methods, CIRC{\'E} assumes the nominal value of $\theta_i$ to be 0 (this is why the resultant mean value is also called bias because it denotes a shift from 0). However, in system TH codes, the dimensionless multipliers of the physical models have a nominal value of 1. Consequently, a change-of-variable is needed. Define the dimensionless multiplier corresponds to $\theta_i$ as $p_i$. The change-of-variable is performed with a function $f$, $p_i = f (\theta_i)$. There are two properties that this transform function must satisfy. Firstly, when $\theta_i = 0$ we should have $p_i = 1 = f (\theta_i = 0)$. Secondly, the derivative of $y^{\text{M}}_j$ with respect to $p_i$ at $p_i=1$ must equal the derivative of $y^{\text{M}}_j$ with respect to $\theta_i$ at $\theta_i = 0$.
\begin{equation*}
	\frac{\partial y^{\text{M}}_j}{\partial \theta_i} (\theta_i = 0)  =  \frac{\partial y^{\text{M}}_j}{\partial p_i} (p_i = 1) \times f^\prime (\theta_i = 0)
\end{equation*}

In order to  use $\frac{\partial y^{\text{M}}_j}{\partial p_i} (p_i = 1)$ directly for $\frac{\partial y^{\text{M}}_j}{\partial \theta_i} (\theta_i = 0)$, it must hold that $f^\prime (\theta_i = 0) = 1$. Based on $f (0) = 1$ and $f^\prime (0) = 1$, two simple functions can be found for $f$, $p_i = f (\theta_i) = 1 + \theta_i$, or $p_i = f (\theta_i) = \exp(\theta_i)$. Both of these two functions have been used by CIRC{\'E}. The choice is usually based on considerations of linearity and the values found for the bias $b_i$ and standard deviation $\sigma_i$, as explained later.

CIRC{\'E} makes two important assumptions: (1) each $\theta_i$ must obey a \textit{normal} law, (2) the QoI must depend \textit{linearly} on each of the parameters in their domain of uncertainty. These two prerequisites are often referred to as the \textit{normality} and \textit{linearity} assumptions of CIRC{\'E}. CIRC{\'E} uses an iterative Expectation-Maximization (E-M) algorithm \cite{dempster1977maximum} \cite{mclachlan2007algorithm} based on the principle of MLE. Even though CIRC{\'E} employs some ideas from Bayes' theorem, it is still categorized a frequentist IUQ approach, instead of Bayesian, because MCMC sampling is not used to obtain the posterior distributions.

Based on the normality assumption, $\{\theta_i \}_{i=1}^{I}$ follow a Gaussian distribution. Therefore, $p_i = 1 + \theta_i$ also follows a Gaussian distribution, and $p_i = \exp(\theta_i)$ follows a log-normal distribution. These two types of regular distributions make the application of the MLE principle very convenient. The 95\% variation interval for the $\theta_i$ parameter is $[b_i - 2 \sigma_i, b_i + 2 \sigma_i]$, in which the linearity assumption should hold. When the resulting $\sigma_i$ is large, the linearity assumption will likely to be problematic. The 95\% variation interval for $p_i$ depends on the change-of-variable formula. For $p_i = 1 + \theta_i$ it is $[1 + b_i - 2 \sigma_i, 1 + b_i + 2 \sigma_i]$, while for $p_i = \exp(\theta_i)$ it is $[\exp(b_i - 2 \sigma_i), \exp(b_i + 2 \sigma_i)]$. Linearity test is needed inside the final 95\% variation interval found for each $p_i$. The change-of-variable with which the hypothesis of linearity is better verified is selected. Note that the change-of-variable formula is not necessarily the same for all the PMPs.

There are a few more assumptions used by CIRC{\'E}. Firstly, the data uncertainty $\epsilon_j$ is considered to be independent from $( y^{\text{R}}_j - y^{\text{M}}_j )$. $y^{\text{R}}_j$ is the unknown true value, which has to be learned by either experimentation $y^{\text{E}}_j$, or simulation $y^{\text{M}}_j$. Secondly, the model-data difference $( y^{\text{E}}_j - y^{\text{M}}_j )$ must be  higher than $\epsilon_j$. Otherwise, the calculated parameter standard deviation $\sigma_i$ by CIRC{\'E} will be 0. Thirdly, different parameters are assumed to be statistically \textit{independent}. CIRC{\'E} cannot produce the covariance between the parameters. Finally, the selected QoIs must be as independent as possible. CIRC{\'E}'s calculation precision increases with the number of independent QoIs.

CIRC{\'E} seeks the mean (bias) vector $\bm{b}$, as well as the covariance matrix $\bm{\Sigma}_{\bm{\theta}} = \mathrm{diag} (\sigma_1^2, \ldots, \sigma_I^2)$. CIRC{\'E} is designed for two solution cases. In the first case, the mean vector $\bm{b}$ is assumed to be $\bm{0}$ and fixed (no bias), and only $\bm{\Sigma}_{\bm{\theta}}$ is estimated. The second case is more general in which $\bm{b}$ is estimated together with $\bm{\Sigma}_{\bm{\theta}}$.

\subsection{CIRC{\'E} without bias calculation}

We have used $\bm{\theta}$ to represent the vector of PMPs, and $\theta_i$ for the $i^{\text{th}}$ parameter. Define a new vector $\bm{\theta}_j$ as the values of $\bm{\theta}$ such that the corresponding model calculation for the $j^{\text{th}}$ QoI equals the reality:
\begin{equation*}
	y^{\text{M}}_j (\bm{\theta}_j)  =  y^{\text{R}}_j, \quad j = 1, 2, \ldots, J
\end{equation*}
where the dependence on design variables has been left out for notational convenience. $\bm{\theta}_j$ should not be confused with $\theta_i$ because the former is a vector that denotes a realization of all the $I$ parameters. Because the reality $y^{\text{R}}_j$ is unknown, $\bm{\theta}_j$ is also never known. Therefore, it should only be treated as a notion used by CIRC{\'E}. Given $\bm{\theta}_j$ and the fact that $\bm{b} = \bm{0}$, an estimation of the covariance matrix $\bm{\Sigma}_{\bm{\theta}}$ is:
\begin{equation}    \label{equation:PREMIUM-CIRCE-NoBias1-Cov}
	\hat{\bm{\Sigma}}_{\bm{\theta}}  =  \frac{1}{J} \sum_{j=1}^{J} \bm{\theta}_j \bm{\theta}_j^\top
\end{equation}

Based on the linearity assumption, a first-order Taylor expansion holds around the nominal (mean) value of $\bm{\theta}$, which is $\bm{b} = \bm{0}$:
\begin{equation}    \label{equation:PREMIUM-CIRCE-NoBias2-Linearity1}
	y^{\text{R}}_j - y^{\text{M}}_j (\bm{b})  
	=  y^{\text{M}}_j (\bm{\theta}_j) - y^{\text{M}}_j (\bm{b})  
	=  \left( \frac{\partial y^{\text{M}}_j}{\partial \bm{\theta}} \right)^\top  ( \bm{\theta}_j - \bm{b} )  
	=  \left( \frac{\partial y^{\text{M}}_j}{\partial \bm{\theta}} \right)^\top  \bm{\theta}_j
\end{equation}
where $\frac{\partial y^{\text{M}}_j}{\partial \bm{\theta}}$ is the vector of the derivatives of $y^{\text{M}}_j$ with respect to all the $I$ parameters $\bm{\theta}$.
\begin{equation*}
	\frac{\partial y^{\text{M}}_j}{\partial \bm{\theta}}  =  \left[ \frac{\partial y^{\text{M}}_j}{\partial \theta_1}, \frac{\partial y^{\text{M}}_j}{\partial \theta_2}, \ldots, \frac{\partial y^{\text{M}}_j}{\partial \theta_I} \right]^\top
\end{equation*}

For each $y_j$, one can write:
\begin{equation}    \label{equation:PREMIUM-CIRCE-NoBias2-Linearity2}
	y^{\text{E}}_j - y^{\text{M}}_j  =  y^{\text{E}}_j - y^{\text{R}}_j + y^{\text{R}}_j - y^{\text{M}}_j = e_j  +  \left( \frac{\partial y^{\text{M}}_j}{\partial \bm{\theta}} \right)^\top  \bm{\theta}_j
\end{equation}
where $e_j$ is the measurement error for $y_j$ that represents the difference between reality and experimental data. It is not explicitly known but usually treated as a Gaussian noise with zero mean, i.e., $e_j \sim \mathcal{N} (0, \epsilon_j^2)$. By assuming that $(y^{\text{E}}_j - y^{\text{R}}_j)$ and $(y^{\text{R}}_j - y^{\text{M}}_j)$ as independent, one can derive the following sum of variances:
\begin{equation}    \label{equation:PREMIUM-CIRCE-NoBias3-Sum-of-Variance}
	\begin{aligned}
		\mathrm{Var} \left( y^{\text{E}}_j - y^{\text{M}}_j \right)  
		&=  \mathrm{Var} \left( y^{\text{E}}_j - y^{\text{R}}_j \right) + \mathrm{Var} \left( y^{\text{R}}_j - y^{\text{M}}_j \right)    \\
		&=  \mathrm{Var} ( e_j )  +  \mathrm{Var} \left( \left( \frac{\partial y^{\text{M}}_j}{\partial \bm{\theta}} \right)^\top  \bm{\theta}_j \right)  
		=  \epsilon_j^2  +  \left( \frac{\partial y^{\text{M}}_j}{\partial \bm{\theta}} \right)^\top  \bm{\Sigma}_{\bm{\theta}}  \frac{\partial y^{\text{M}}_j}{\partial \bm{\theta}}
	\end{aligned}
\end{equation}

Note that in earlier version of CIRC{\'E} \cite{de2001determination}, Equation \ref{equation:PREMIUM-CIRCE-NoBias2-Linearity1} started from $(y^{\text{E}}_j - y^{\text{M}}_j)$ instead of $(y^{\text{R}}_j - y^{\text{M}}_j)$. As a result, Equation \ref{equation:PREMIUM-CIRCE-NoBias3-Sum-of-Variance} did not consider the experimental uncertainty $\epsilon_j^2$, so were the subsequent derivations.

CIRC{\'E} uses an iterative process to calculate $\bm{\Sigma}_{\bm{\theta}}$ (recall that estimation of $\bm{b}$ is not needed). The iteration is based on the Bayes' theorem and it generally starts with the identity matrix, $\bm{\Sigma}_{\bm{\theta}}^{(0)} = \bm{I}$. Denote the estimation at the $m^{\text{th}}$ iteration as $\bm{\Sigma}_{\bm{\theta}}^{(m)}$. Given a current (also called \textit{a priori}) estimation $\bm{\Sigma}_{\bm{\theta}}^{(m-1)}$ ($m \ge 1$), the next estimation (also called \textit{a posteriori}) $\bm{\Sigma}_{\bm{\theta}}^{(m)}$ is obtained by correcting the current estimation given $\bm{\theta}_j$. Because $\bm{\theta}_j$ is not observable and its exact value is impossible to determine, the correction uses $(y^{\text{E}}_j - y^{\text{M}}_j)$ that can be ``observed''. The process is continued until the covariance matrix converges.

In every iteration, Bayes' theorem does not calculate explicitly $\bm{\theta}_j$, but gives an estimation of the mean vector for $\bm{\theta}_j$, denoted as $\bar{\bm{\theta}}_j$. For notational convenience, define:
\begin{equation*}
	\begin{aligned}
		\mathbf{X}^{(m)}  &=  \epsilon_j^2  +  \left( \frac{\partial y^{\text{M}}_j}{\partial \bm{\theta}} \right)^\top  \bm{\Sigma}_{\bm{\theta}}^{(m)}  \frac{\partial y^{\text{M}}_j}{\partial \bm{\theta}}    \\
		\mathbf{Y}^{(m)}  &=  \bm{\Sigma}_{\bm{\theta}}^{(m)} \frac{\partial y^{\text{M}}_j}{\partial \bm{\theta}} \left( \frac{\partial y^{\text{M}}_j}{\partial \bm{\theta}} \right)^\top  \bm{\Sigma}_{\bm{\theta}}^{(m)}
	\end{aligned}
\end{equation*}

Using $(y^{\text{E}}_j - y^{\text{M}}_j)$ and the \textit{a priori} matrix $\bm{\Sigma}_{\bm{\theta}}^{(m-1)}$, $\bar{\bm{\theta}}_j$ and the covariance matrix $\bm{\Sigma}_{\bm{\theta}_j}$ for each $\bm{\theta}_j$ are calculated as:
\begin{equation}    \label{equation:PREMIUM-CIRCE-NoBias4-Estimate-Theta_j}
	\bar{\bm{\theta}}_j  =  \bm{\Sigma}_{\bm{\theta}}^{(m-1)}  \cdot  \frac{\partial y^{\text{M}}_j}{\partial \bm{\theta}}  \cdot  \frac{ y^{\text{E}}_j - y^{\text{M}}_j }{ \mathbf{X}^{(m-1)} }
\end{equation}
\begin{equation}    \label{equation:PREMIUM-CIRCE-NoBias4-Estimate-Covariance}
	\bm{\Sigma}_{\bm{\theta}_j}  =  \bm{\Sigma}_{\bm{\theta}}^{(m-1)}  -  \frac{ \mathbf{Y}^{(m-1)} }{ \mathbf{X}^{(m-1)} }
\end{equation}

The $\bm{\theta}_j \bm{\theta}_j^\top$ products are need in Equation \ref{equation:PREMIUM-CIRCE-NoBias1-Cov} to calculate the \textit{a posteriori} matrix $\bm{\Sigma}_{\bm{\theta}}^{(m)}$. Replacing them with $\bar{\bm{\theta}}_j \bar{\bm{\theta}}_j^\top$:
\begin{equation}    \label{equation:PREMIUM-CIRCE-NoBias5-Update-Covariance1}
	\mathbb{E} (\bm{\theta}_j \bm{\theta}_j^\top) =  \bar{\bm{\theta}}_j \bar{\bm{\theta}}_j^\top  +  \bm{\Sigma}_{\bm{\theta}_j}
\end{equation}

Finally, $\bm{\Sigma}_{\bm{\theta}}^{(m)}$ can be obtained using the following iterative formula with $\bm{\Sigma}_{\bm{\theta}}^{(m-1)}$, by combining Equations \ref{equation:PREMIUM-CIRCE-NoBias4-Estimate-Theta_j}, \ref{equation:PREMIUM-CIRCE-NoBias4-Estimate-Covariance} and \ref{equation:PREMIUM-CIRCE-NoBias5-Update-Covariance1} in Equation \ref{equation:PREMIUM-CIRCE-NoBias1-Cov}:
\begin{equation}    \label{equation:PREMIUM-CIRCE-NoBias5-Update-Covariance2}
	\begin{aligned}
		\bm{\Sigma}_{\bm{\theta}}^{(m)}  
		&=  \frac{1}{J} \sum_{j=1}^{J} \bm{\theta}_j \bm{\theta}_j^\top  
		 =  \frac{1}{J} \sum_{j=1}^{J}  \left[  \bar{\bm{\theta}}_j \bar{\bm{\theta}}_j^\top  +  \bm{\Sigma}_{\bm{\theta}_j}  \right]    \\
		&=  \frac{1}{J} \sum_{j=1}^{J}  \frac{ \mathbf{Y}^{(m-1)} }{ \mathbf{X}^{(m-1)} }  \cdot  \frac{ ( y^{\text{E}}_j - y^{\text{M}}_j )^2 }{ \mathbf{X}^{(m-1)} }  +  \frac{1}{J} \sum_{j=1}^{J} \bm{\Sigma}_{\bm{\theta}}^{(m-1)}  -  \frac{1}{J} \sum_{j=1}^{J} \frac{ \mathbf{Y}^{(m-1)} }{ \mathbf{X}^{(m-1)} }    \\
		&=  \bm{\Sigma}_{\bm{\theta}}^{(m-1)}  +  \frac{1}{J} \sum_{j=1}^{J} \frac{ \mathbf{Y}^{(m-1)} }{ \mathbf{X}^{(m-1)} }  \cdot  \left( \frac{ ( y^{\text{E}}_j - y^{\text{M}}_j )^2 }{ \mathbf{X}^{(m-1)} } - 1 \right)
	\end{aligned}
\end{equation}

The iterative process consists of a ``Maximization'' step (M-step) and an ``Expectation'' step (E-step). In the M-step, the principle of maximum likelihood is applied. In the E-step, each $\bm{\theta}_j \bm{\theta}_j^\top$ product is replaced by an estimation of its mean value product, $\bar{\bm{\theta}}_j \bar{\bm{\theta}}_j^\top$. E-M algorithm guarantees that the $\bm{\Sigma}_{\bm{\theta}}^{(m)}$ matrix is always defined and positive. Furthermore, the likelihood of observing the data given the inversely quantified parameters increases.

\subsection{CIRC{\'E} with bias calculation}

When the bias vector $\bm{b}$ is considered for updating, CIRC{\'E} no longer treats $\bm{b} = \bm{0}$. At the $m^{\text{th}}$ iteration, $\bm{\Sigma}_{\bm{\theta}}^{(m)}$ is firstly calculated using $\bm{b}^{(m-1)}$ and $\bm{\Sigma}_{\bm{\theta}}^{(m-1)}$. Equation \ref{equation:PREMIUM-CIRCE-NoBias1-Cov} becomes:
\begin{equation}    \label{equation:PREMIUM-CIRCE-WithBias1-Cov}
	\hat{\bm{\Sigma}}_{\bm{\theta}}  =  \frac{1}{J} \sum_{j=1}^{J} \bm{\beta}_j \bm{\beta}_j^\top
\end{equation}
where $\bm{\beta}_j = \bm{\theta}_j - \bm{b}^{(m-1)}$ is the centralized vector. In this case, Bayes' theorem estimates the mean vector $\bar{\bm{\beta}}_j$ for $\bm{\beta}_j$. The difference between the estimation of $\bar{\bm{\theta}}_j$ and $\bar{\bm{\beta}}_j$ is that, the former uses $(y^{\text{E}}_j - y^{\text{M}}_j)$, while the latter uses $\{ y^{\text{E}}_j - y^{\text{M}}_j - ( \frac{\partial y^{\text{M}}_j}{\partial \bm{\theta}} )^\top  \bm{b}^{(m-1)}  \}$ because the linearity relation becomes:
\begin{equation}    \label{equation:PREMIUM-CIRCE-WithBias2-Linearity1}
	y^{\text{E}}_j - y^{\text{M}}_j  
	=  e_j  +  \left( \frac{\partial y^{\text{M}}_j}{\partial \bm{\theta}} \right)^\top  \left( \bm{\theta}_j - \bm{b}^{(m-1)} \right)  
	=  e_j  +  \left( \frac{\partial y^{\text{M}}_j}{\partial \bm{\theta}} \right)^\top  \bm{\beta}_j
\end{equation}

Equation \ref{equation:PREMIUM-CIRCE-NoBias4-Estimate-Theta_j} becomes:
\begin{equation}    \label{equation:PREMIUM-CIRCE-WithBias4-Estimate-Theta_j}
	\bar{\bm{\beta}}_j  =  \bm{\Sigma}_{\bm{\theta}}^{(m-1)}  \cdot  \frac{\partial y^{\text{M}}_j}{\partial \bm{\theta}}  \cdot  \frac{ y^{\text{E}}_j - y^{\text{M}}_j - \left( \frac{\partial y^{\text{M}}_j}{\partial \bm{\theta}} \right)^\top  \bm{b}^{(m-1)} }{ \mathbf{X}^{(m-1)} }
\end{equation}

Similarly, Equation \ref{equation:PREMIUM-CIRCE-NoBias5-Update-Covariance1} becomes: 
\begin{equation}    \label{equation:PREMIUM-CIRCE-WithBias5-Update-Covariance1}
	\mathbb{E} (\bm{\beta}_j \bm{\beta}_j^\top) = \bar{\bm{\beta}}_j \bar{\bm{\beta}}_j^\top  +  \bm{\Sigma}_{\bm{\theta}_j}
\end{equation}

The update formula for $\bm{\Sigma}_{\bm{\theta}}^{(m)}$ is similar to Equation \ref{equation:PREMIUM-CIRCE-NoBias5-Update-Covariance2}, by changing $(y^{\text{E}}_j - y^{\text{M}}_j)$ to $\{ y^{\text{E}}_j - y^{\text{M}}_j - ( \frac{\partial y^{\text{M}}_j}{\partial \bm{\theta}} )^\top  \bm{b}^{(m-1)} \}$:
\begin{equation}    \label{equation:PREMIUM-CIRCE-WithBias5-Update-Covariance2}
	\bm{\Sigma}_{\bm{\theta}}^{(m)}  =  
	\bm{\Sigma}_{\bm{\theta}}^{(m-1)}  +  \frac{1}{J} \sum_{j=1}^{J} \frac{ \mathbf{Y}^{(m-1)} }{ \mathbf{X}^{(m-1)} }  \cdot  \left( \frac{ \left( y^{\text{E}}_j - y^{\text{M}}_j  - \left( \frac{\partial y^{\text{M}}_j}{\partial \bm{\theta}} \right)^\top  \bm{b}^{(m-1)} \right)^2 }{ \mathbf{X}^{(m-1)} } - 1 \right)
\end{equation}

In every iteration, once $\bm{\Sigma}_{\bm{\theta}}^{(m)}$ is calculated using $\bm{b}^{(m-1)}$ and $\bm{\Sigma}_{\bm{\theta}}^{(m-1)}$, $\bm{b}^{(m)}$ can also be updated using MLE, but with $(y^{\text{E}}_j - y^{\text{M}}_j)$. Given $\bm{\theta} \sim \mathcal{N}(\bm{b}, \bm{\Sigma}_{\bm{\theta}})$, $(y^{\text{E}}_j - y^{\text{M}}_j)$ also follows a Gaussian distribution:
\begin{equation*}
	y^{\text{E}}_j - y^{\text{M}}_j  \sim  \mathcal{N} \left(  \left( \frac{\partial y^{\text{M}}_j}{\partial \bm{\theta}} \right)^\top  \bm{b}, \hspace{0.5em} \epsilon_j^2  +  \left( \frac{\partial y^{\text{M}}_j}{\partial \bm{\theta}} \right)^\top  \bm{\Sigma}_{\bm{\theta}}  \frac{\partial y^{\text{M}}_j}{\partial \bm{\theta}}  \right)
\end{equation*}
where the variance is based on Equation \ref{equation:PREMIUM-CIRCE-NoBias3-Sum-of-Variance}. It is easy to write the PDF of such a Gaussian distribution, denote it as $P (y^{\text{E}}_j - y^{\text{M}}_j)$. For different $j$, because CIRC{\'E} assumes the selected QoIs to be mutually independent, the joint distribution is $L = \prod_{j=1}^{J} P (y^{\text{E}}_j - y^{\text{M}}_j)$, which is the likelihood. Taking the logarithm of the likelihood:
\begin{equation}    \label{equation:PREMIUM-CIRCE-WithBias6-LogL}
	\ln(L) = -\frac{J}{2} \ln(2\pi) - \frac{1}{2} \sum_{j=1}^{J} \ln \left( \mathbf{X}^{(m)} \right)  - \frac{1}{2} \sum_{j=1}^{J} \frac{ \left( y^{\text{E}}_j - y^{\text{M}}_j  - \left( \frac{\partial y^{\text{M}}_j}{\partial \bm{\theta}} \right)^\top  \bm{b}^{(m)} \right)^2 }{ \mathbf{X}^{(m)} }
\end{equation}

To maximize $\ln(L)$ with the bias vector $\bm{b}$, one can take the derivative of $\ln(L)$ with respect to each $b_i$, resulting in $I$ equations:
\begin{equation}    \label{equation:PREMIUM-CIRCE-WithBias7-LogL-Derivative}
	\frac{ \partial \ln(L) }{ \partial b_i }  =  
	\sum_{j=1}^{J} \frac{ \left( y^{\text{E}}_j - y^{\text{M}}_j  - \left( \frac{\partial y^{\text{M}}_j}{\partial \bm{\theta}} \right)^\top  \bm{b}^{(m)} \right) \frac{\partial y^{\text{M}}_j}{\partial \theta_i} }{ \mathbf{X}^{(m)} }  
	=  0
\end{equation}

The $I$ equations from Equation \ref{equation:PREMIUM-CIRCE-WithBias7-LogL-Derivative} can form a linear system whose solution is $\bm{b}^{(m)}$:
\begin{equation}    \label{equation:PREMIUM-CIRCE-WithBias8-Linear-System}
	\begin{bmatrix}
		\sum_{j=1}^{J} \frac{1}{\mathbf{X}^{(m)}} \left( \frac{\partial y^{\text{M}}_j}{\partial \theta_1} \right)^2  &  \ldots  &  \sum_{j=1}^{J} \frac{1}{\mathbf{X}^{(m)}}  \frac{\partial y^{\text{M}}_j}{\partial \theta_1} \frac{\partial y^{\text{M}}_j}{\partial \theta_I}   \\
		\vdots  &  \ddots  &  \vdots  \\
		\sum_{j=1}^{J} \frac{1}{\mathbf{X}^{(m)}} \frac{\partial y^{\text{M}}_j}{\partial \theta_I} \frac{\partial y^{\text{M}}_j}{\partial \theta_1}  &  \ldots  &  \sum_{j=1}^{J} \frac{1}{\mathbf{X}^{(m)}} \left( \frac{\partial y^{\text{M}}_j}{\partial \theta_I} \right)^2
	\end{bmatrix}
	\begin{bmatrix}
		b_1^{(m)}  \\  \vdots  \\  b_I^{(m)}
	\end{bmatrix}
	=  
	\begin{bmatrix}
		\sum_{j=1}^{J} \frac{1}{\mathbf{X}^{(m)}}  \frac{\partial y^{\text{M}}_j}{\partial \theta_1} \left( y^{\text{E}}_j - y^{\text{M}}_j \right) \\  
		\vdots  \\  
		\sum_{j=1}^{J} \frac{1}{\mathbf{X}^{(m)}}  \frac{\partial y^{\text{M}}_j}{\partial \theta_I} \left( y^{\text{E}}_j - y^{\text{M}}_j \right)
	\end{bmatrix}
\end{equation}

\subsection{Iterative CIRC{\'E}}

CIRC{\'E} can be applied to cases with high bias with a simple improvement, which is called ``Iterative CIRC{\'E}''. The idea is to apply standard CIRC{\'E} multiple times, with the starting bias vector being the converged one from the previous standard CIRC{\'E} application. The iteration stops when a bias is low in absolute value. Note that here iterations means the number of times for which standard CIRC{\'E} is used, not the number of iterations in E-M algorithm. Experience has shown that usually 3-4 iterations are sufficient \cite{de2001determination}. With iterative CIRC{\'E}, there is still a hypothesis of linearity, but it is much less strong than a standard CIRC{\'E}, because it is only needed around the final mean (bias) vector.

\section{The IPREM Method}    \label{Appendix-IPREM}

\subsection{Accuracy assessment with FFTBM}

FFTBM was originally developed by Ambrosini and Bovalini \cite{ambrosini1990evaluation} at the UNIPI in 1990 to quantify accuracy of TH code calculations. FFTBM has been applied by several researchers for quantitative accuracy assessment \cite{provsek2002review} \cite{petruzzi2010uncertainties} \cite{provsek2011application} \cite{coscarelli2013integrated}. The general idea is to perform Fourier transform of the experimental data and the model-data difference from the time domain to the frequency domain. Then use the resulting amplitudes of Fourier transform to quantify the code accuracy. Comparison in the frequency domain can eliminate the dependence on time duration of the experiments and shape of the time trends.

FFTBM deals with time-dependent QoIs. In other section we have used $\mathbf{y}^{\text{M}} ( \mathbf{x}, \bm{\theta} )$ to denote the computational model that takes design variables $\mathbf{x}$ and calibration parameters $\bm{\theta}$ as inputs, and $\mathbf{y}^{\text{E}} (\mathbf{x})$ to represent physical observations. In this section we use $\mathbf{y}^{\text{M}} (t)$ and $\mathbf{y}^{\text{E}} (t)$ for notational simplicity and to indicate the time-dependence. For an arbitrary time-dependent function $g(t)$, Fourier transform can be applied to transfer $g(t)$ to a corresponding function $\tilde{g} (f)$ in the frequency domain:
\begin{equation}      \label{equation:PREMIUM-IPREM1-Fourier-Transform}
	\tilde{g} (f) = \int_{- \infty}^{+ \infty} g(t) \cdot e^{-2 \pi t f j} dt
\end{equation}

For machined-based numerical computation in practice, Discrete Fourier Transform (DFT) is usually evaluated for this integral problem employing only a finite number of points. For experimental data and computer simulations sampled in digital form, Fast Fourier Transform (FFT) is a way to do DFT with great efficiency. Define the error function as the difference between the model simulation and experimental signal:
\begin{equation}      \label{equation:PREMIUM-IPREM2-Error}
	\mathbf{y}^{\text{M-E}} (t) = \mathbf{y}^{\text{M}} (t) - \mathbf{y}^{\text{E}} (t)
\end{equation}

FFTBM uses the FFT of the experimental signal $\mathbf{y}^{\text{E}} (t)$ and the error function $\mathbf{y}^{\text{M-E}} (t)$:
\begin{equation}      \label{equation:PREMIUM-IPREM3-FFT}
	\begin{aligned}
		\tilde{ \mathbf{y} }^{\text{E}} (f) &= \int_{- \infty}^{+ \infty} \mathbf{y}^{\text{E}} (t) \cdot e^{-2 \pi t f j} dt    \\
		\tilde{ \mathbf{y} }^{\text{M-E}} (f) &= \int_{- \infty}^{+ \infty} \mathbf{y}^{\text{M-E}} (t) \cdot e^{-2 \pi t f j} dt
	\end{aligned}
\end{equation}

FFTBM uses a dimensionless figure-of-merit (FOM), called \textit{Average Amplitude (AA)}:
\begin{equation}      \label{equation:PREMIUM-IPREM4-AA}
	\text{AA}  =  \frac{ \sum_{n=0}^{2^m} | \tilde{ \mathbf{y} }^{\text{M-E}} (f_n) | }{ \sum_{n=0}^{2^m} | \tilde{ \mathbf{y} }^{\text{E}} (f_n) | }
\end{equation}
where $2^m$ is the number of points in the signal as required by FFT. The AA metric represents the relative magnitude of the discrepancy from the comparison between simulation and data. Low AA values mean better agreement between simulation and data. This is the primary motivation to use AA to evaluate the accuracy of the system TH codes. There is another quantify called weighted frequency that supplies different information from AA allowing better identification of the character of accuracy. Details can be found in \cite{reventos2016premium}.

\subsection{Derivation of the IPREM method based on FFTBM}

IPREM quantifies the uncertainty ranges of PMPs based on the FFTBM method. Note that this has to be done with time-dependent QoIs, such as cladding temperature and quench front propagation during core reflood. It makes use of FFT to single-parameter sensitivity calculations of relevant experiments. It then calculates AA as an indicator of the discrepancy between code and data. The parameter ranges are determined by imposing a restrictive criterion of the AA values (related to a single QoI) and global AA (accounting for multiple QoIs). The major steps are described below based on \cite{kovtonyuk2014development} \cite{kovtonyuk2017development}. Table \ref{table:PREMIUM-IPREM-Symbols} lists the mathematical symbols used in this section.

\begingroup
\renewcommand{\arraystretch}{1.3}
\begin{table}[!ht]
	\footnotesize
	\centering
	\captionsetup{justification=centering}
	\caption{Definitions of symbols used for the IPREM method.}
	\label{table:PREMIUM-IPREM-Symbols}
	\begin{tabular}{p{0.07\linewidth} | p{0.36\linewidth} | p{0.07\linewidth} | p{0.38\linewidth}}
		\toprule
		Symbol  &  Description  &  Symbol  &  Description  \\ 
		\midrule
		$\mathbf{x}$                   &  vector of design variables   &  $\theta_{ij}$  &  the $j^{\text{th}}$ perturbation of $\theta_i$  \\
		
		$\bm{\theta}$                  &  vector of PMPs, $\bm{\theta} = \{ \theta_i \}_{i=1}^{I}$  &  $\mathbf{y}^{\text{R}} (t)$  &  the ``Reference'' code run  \\
		
		$\bm{\theta}_0$                &  nominal values of $\bm{\theta}$  &  $\mathbf{y}^{\text{S}}_{ij} (t)$  &  the ``Sensitivity'' code runs at $\theta_{ij}$  \\
		
		$\mathbf{y} (t)$               &  QoIs that are time-dependent  &  $\text{\textbf{AA}}^{\text{R-E}}$  &  AA from comparing $\mathbf{y}^{\text{R}} (t)$ and $\mathbf{y}^{\text{E}} (t)$  \\
		
		$\mathbf{y}^{\text{M}} (t)$    &  QoIs from model simulation  &  $\text{\textbf{AA}}^{\text{S-E}}_{ij}$  &  AA from comparing $\mathbf{y}^{\text{S}}_{ij} (t)$ and $\mathbf{y}^{\text{E}} (t)$  \\
		
		$\mathbf{y}^{\text{E}} (t)$    &  QoIs from experiment  &  $\text{\textbf{AA}}^{\text{S-R}}_{ij}$  &  AA from comparing $\mathbf{y}^{\text{S}}_{ij} (t)$ and $\mathbf{y}^{\text{R}} (t)$  \\
		
		$\mathbf{y}^{\text{M-E}} (t)$  &  Error function  &  $W_z$  &  weight factor for the $z^{\text{th}}$ QoI   \\
		
		$\tilde{ \mathbf{y} }^{\text{E}} (f)$  &  Fourier transform of $\mathbf{y}^{\text{E}} (t)$  &  $w_z$ &  normalized weight factor for the $z^{\text{th}}$ QoI  \\
		
		$\tilde{ \mathbf{y} }^{\text{M-E}} (f)$  &  Fourier transform of $\mathbf{y}^{\text{M-E}} (t)$  &  $\text{AAG}$ &  Global AA  \\
		
		$\text{AA}$                    &  Average Amplitude  &  $\text{CR} (\theta_{ij})$  &  empirical criterion quantity at $\theta_{ij}$ \\
		
		$I$                            &  dimension of $\bm{\theta}$  &  $\eta$  &  threshold value to determine PMP bounds  \\
		
		$J$                            &  number of perturbations for each PMP  &  $\theta_{i}^L$  &  lower bound for $\theta_{i}$  \\
		
		$Z$                            &  dimension of $\mathbf{y} (t)$  &  $\theta_{i}^U$  &  upper bound for $\theta_{i}$  \\
		\bottomrule
	\end{tabular}
\end{table}
\endgroup

Assume there are $I$ PMPs whose ranges need to be quantified, $\bm{\theta} = \{ \theta_i \}_{i=1}^{I}$. There are $Z$ QoIs, $\mathbf{y} (t) = \{ y_z (t) \}_{z=1}^{Z}$. The nominal values of $\bm{\theta}$ are $\bm{\theta}_0$, for which the code calculation is called the ``Reference'' solution, $\mathbf{y}^{\text{R}} (t)$. Note that previously in Bayesian IUQ we have used the superscript ``R'' for ``Reality'', so using ``R'' for ``Reference'' only applies in this section. The next step is to calculate a series of ``Sensitivity'' cases, denoted as $\mathbf{y}^{\text{S}} (t)$, where ``S'' stands for sensitivity. For the $i^{\text{th}}$ PMP $\theta_i$, it is perturbed alone $J$ times around its nominal value (with other parameters fixed at their nominal values). At the $j^{\text{th}}$ perturbation, the code calculation is represented as $\mathbf{y}^{\text{S}}_{ij} (t)$. To sum up:
\begin{itemize}
	\item $\mathbf{y}^{\text{E}} (t)$ is the experimental data, with $y^{\text{E}}_z (t)$ defined for the $z^{\text{th}}$ QoI;
	
	\item $\mathbf{y}^{\text{R}} (t)$ is the ``Reference'' code simulation at $\bm{\theta}_0$, with $y^{\text{R}}_z (t)$ defined for the $z^{\text{th}}$ QoI;
	
	\item $\mathbf{y}^{\text{S}}_{ij} (t)$ is the ``Sensitivity'' code simulation at the $j^{\text{th}}$ perturbation of the $i^{\text{th}}$ PMP $\theta_i$, with $y^{\text{S}}_{ij, z} (t)$ defined for the $z^{\text{th}}$ QoI.
\end{itemize}

FFTBM analyses are then performed for the following cases:
\begin{enumerate}[label=(\alph*)]
	\setlength{\itemsep}{0.1pt}
	\item ``Reference'' case $\mathbf{y}^{\text{R}} (t)$ vs. data $\mathbf{y}^{\text{E}} (t)$: the resulting AA is denoted as $\text{\textbf{AA}}^{\text{R-E}}$, with $\text{AA}^{\text{R-E}}_z$ defined for the $z^{\text{th}}$ QoI. It quantifies the accuracy of the reference simulation based on data.
	
	\item ``Sensitivity'' case $\mathbf{y}^{\text{S}}_{ij} (t)$ vs. data $\mathbf{y}^{\text{E}} (t)$: at the $j^{\text{th}}$ perturbation of $\theta_i$, the resulting AA is denoted as $\text{\textbf{AA}}^{\text{S-E}}_{ij}$, with $\text{AA}^{\text{S-E}}_{ij, z}$ defined for the $z^{\text{th}}$ QoI. It quantifies the accuracy of the sensitivity cases based on data.
	
	\item ``Sensitivity'' case $\mathbf{y}^{\text{S}}_{ij} (t)$ vs. ``Reference'' case $\mathbf{y}^{\text{R}} (t)$: at the $j^{\text{th}}$ perturbation of $\theta_i$, the resulting AA is denoted as $\text{\textbf{AA}}^{\text{S-R}}_{ij}$, with $\text{AA}^{\text{S-R}}_{ij, z}$ defined for the $z^{\text{th}}$ QoI. It quantifies the deviation of the sensitivity cases from the reference simulation.
\end{enumerate}

It is obvious that $\text{\textbf{AA}}^{\text{S-E}}_{ij}$ and $\text{\textbf{AA}}^{\text{S-R}}_{ij}$ given $i$ and $j$, as well as $\text{\textbf{AA}}^{\text{R-E}}$, are all vectors that consist of $Z$ elements for $Z$ QoIs. In order to calculate a Global AA (AAG), IPREM uses a weighted average of the AAs for different QoIs. IPREM relies on engineering judgment to assign a set of weight factors for each QoI, denoted as $W_z$ for the $z^{\text{th}}$ QoI. An example can be found in \cite{kovtonyuk2017development} for ten QoIs in reflood tests. Different weights will be need for different type of experiments. A set of normalized weight factors $\mathbf{w}$ can be calculated as:
\begin{equation}      \label{equation:PREMIUM-IPREM5-Normalized-Weights}
	w_z = \frac{W_z}{\sum_{k=1}^{Z} W_k}, \quad \text{for } z = 1, \ldots, Z
\end{equation}

The AAG can be calculated as:
\begin{equation}      \label{equation:PREMIUM-IPREM6-Global-AA}
	\text{AAG} = \sum_{z=1}^{Z} w_z \text{AA}_z
\end{equation}

The $\text{\textbf{AA}}$ vector in Equation \ref{equation:PREMIUM-IPREM6-Global-AA} can be $\text{\textbf{AA}}^{\text{S-E}}_{ij}$, $\text{\textbf{AA}}^{\text{S-R}}_{ij}$ and $\text{\textbf{AA}}^{\text{R-E}}$. Denote the resulting AAG for the three cases as: $\text{AAG}^{\text{S-E}}_{ij}$, $\text{AAG}^{\text{S-R}}_{ij}$ and $\text{AAG}^{\text{R-E}}$, respectively. With the AAG values for the $j^{\text{th}}$ perturbation of $\theta_i$ calculated, IPREM builds a FOM called the empirical \textit{criterion quantity} $\text{CR} (\theta_{ij})$, where $\theta_{ij}$ is the $j^{\text{th}}$ perturbation for $\theta_i$.
\begin{equation}      \label{equation:PREMIUM-IPREM7-CR}
	\text{CR} (\theta_{ij})  =  \frac{ \text{AAG}^{\text{S-E}}_{ij} + \text{AAG}^{\text{S-R}}_{ij} - \text{AAG}^{\text{R-E}} }{ 1 - \text{AAG}^{\text{S-E}} }
\end{equation}

In Equation \ref{equation:PREMIUM-IPREM7-CR}, $(\text{AAG}^{\text{S-E}}_{ij} + \text{AAG}^{\text{S-R}}_{ij})$ is a measure of the combined global deviation of the ``Sensitivity'' calculations from the ``Reference'' calculation and the experimental data. $\text{AAG}^{\text{R-E}}$ is the global deviation of the ``Reference'' calculation from the data. $(1 - \text{AAG}^{\text{S-E}})$ in the denominator leads to larger range of variation for a PMP in the direction that improves the agreement between code and data.

\begin{figure}[ht]
	\centering
	\includegraphics[width=0.8\textwidth]{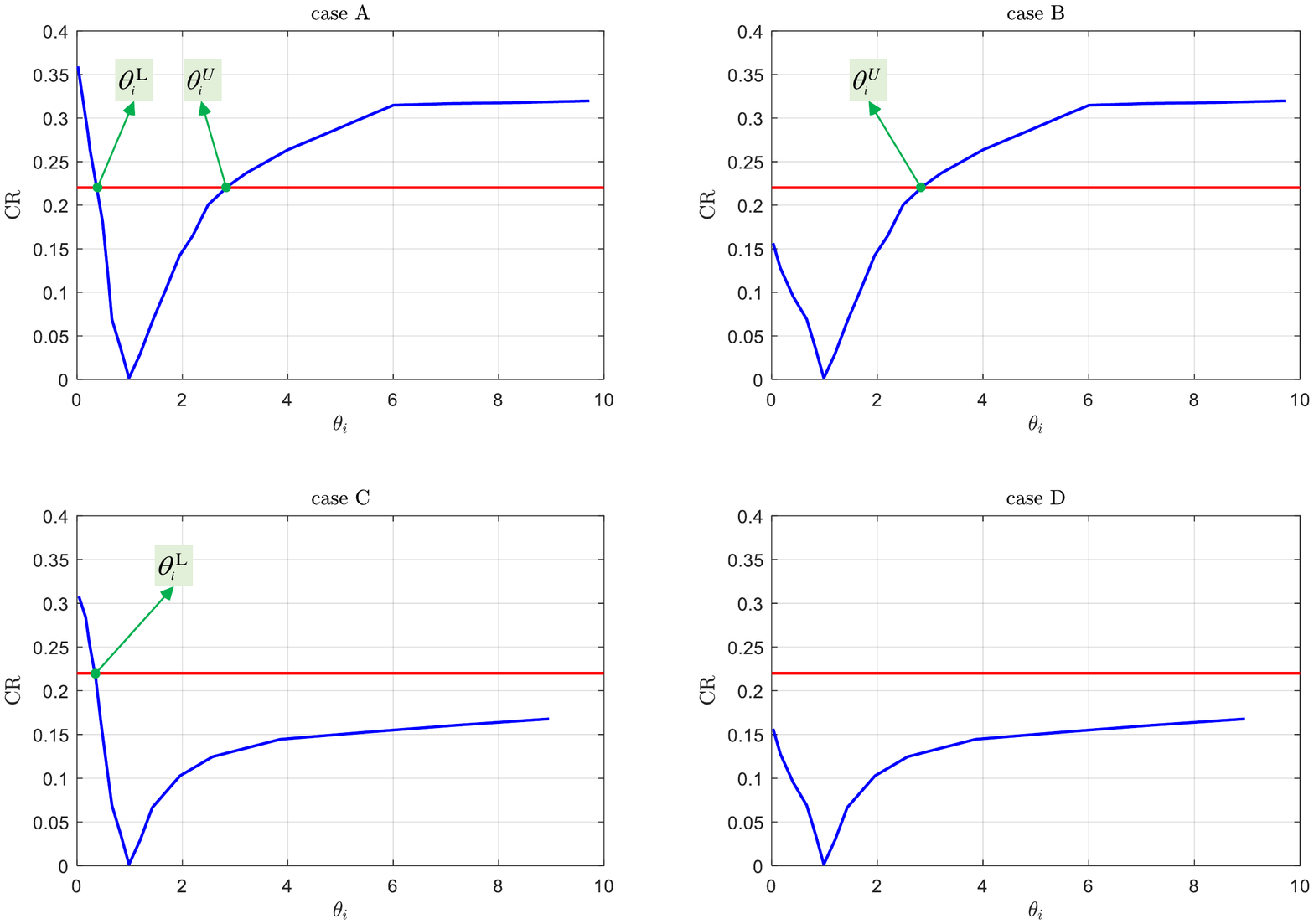}
	\caption[]{Illustration of the IPREM method to quantify input uncertainties (ranges of variation). Case A is adapted from the original DIPE paper \cite{kovtonyuk2017development} with some modifications. Case A shows an ideal situation when both the lower and upper bounds can be identified using the threshold value of 0.22. Case B and C illustrate scenarios when only the upper bound and the lower bound can be found, respectively. Case D shows a situation when IPREM will fail to find the lower/upper bounds.}
	\label{figure:PREMIUM-IPREM}
\end{figure}

The last step of IPREM is to determine the range of variation for each PMP. At the current step, there are $J$ CR values for $\theta_i$ (one at each perturbation), $\text{\textbf{CR}} (\theta_{i}) = \{ \text{CR} (\theta_{ij}) \}_{j=1}^{J}$, for $i = 1, \ldots, I$. The lower and upper bounds of $\theta_i$ are defined by:
\begin{equation}      \label{equation:PREMIUM-IPREM8-Bounds}
	\left[ \theta_{i}^L; \theta_{i}^U \right] = \text{\textbf{CR}} (\theta_{i}) \cap \eta
\end{equation}
where $\eta$ is a limit/threshold value based on engineering judgment. It must be set once and consistently applied for all the analyses performed. For the core reflood scenario in the PREMIUM project, $\eta = 0.22$ was used. The justification of $\eta = 0.22$ is that in the hypothetical case when ``Reference'' calculation exactly matches the experimental data, the maximum allowed deviation of QoIs (at lower/upper bounds of the PMPs) will be 10\%. When $\mathbf{y}^{\text{R}} (t) = \mathbf{y}^{\text{E}} (t)$, $\text{AAG}^{\text{S-E}}_{ij} = \text{AAG}^{\text{S-R}}_{ij}$ and $\text{AAG}^{\text{R-E}} = 0$. Based on Equation \ref{equation:PREMIUM-IPREM7-CR}:
\begin{equation}
	\text{CR}  =  \frac{ 2 \times \text{AAG} }{ 1 - \text{AAG} }  \le  0.22  \quad  \Leftrightarrow  \quad \text{AAG} \le 0.1
\end{equation}

Figure \ref{figure:PREMIUM-IPREM} shows how to use the CR values and the threshold $\eta$ to identify the lower and upper bounds of a PMP. They are selected as the parameter values when CR equals $\eta$. However, there are a few scenarios when IPREM will only be able to find one of the bounds (cases B and C), or none at all (case D). In these cases, the user will have to decrease the value of $\eta$. From case A in Figure \ref{figure:PREMIUM-IPREM}, it can also be noticed that when $\eta$ is large, there is a risk that the upper bound will be very large, such as around ten times the nominal value for multiplicative factors. This explains the fact that the participants who used IPREM in the PREMIUM project tended to identify larger ranges of variation.

\section{The DIPE Method}    \label{Appendix-DIPE}

\begin{figure}[ht]
	\centering
	\includegraphics[width=0.5\textwidth]{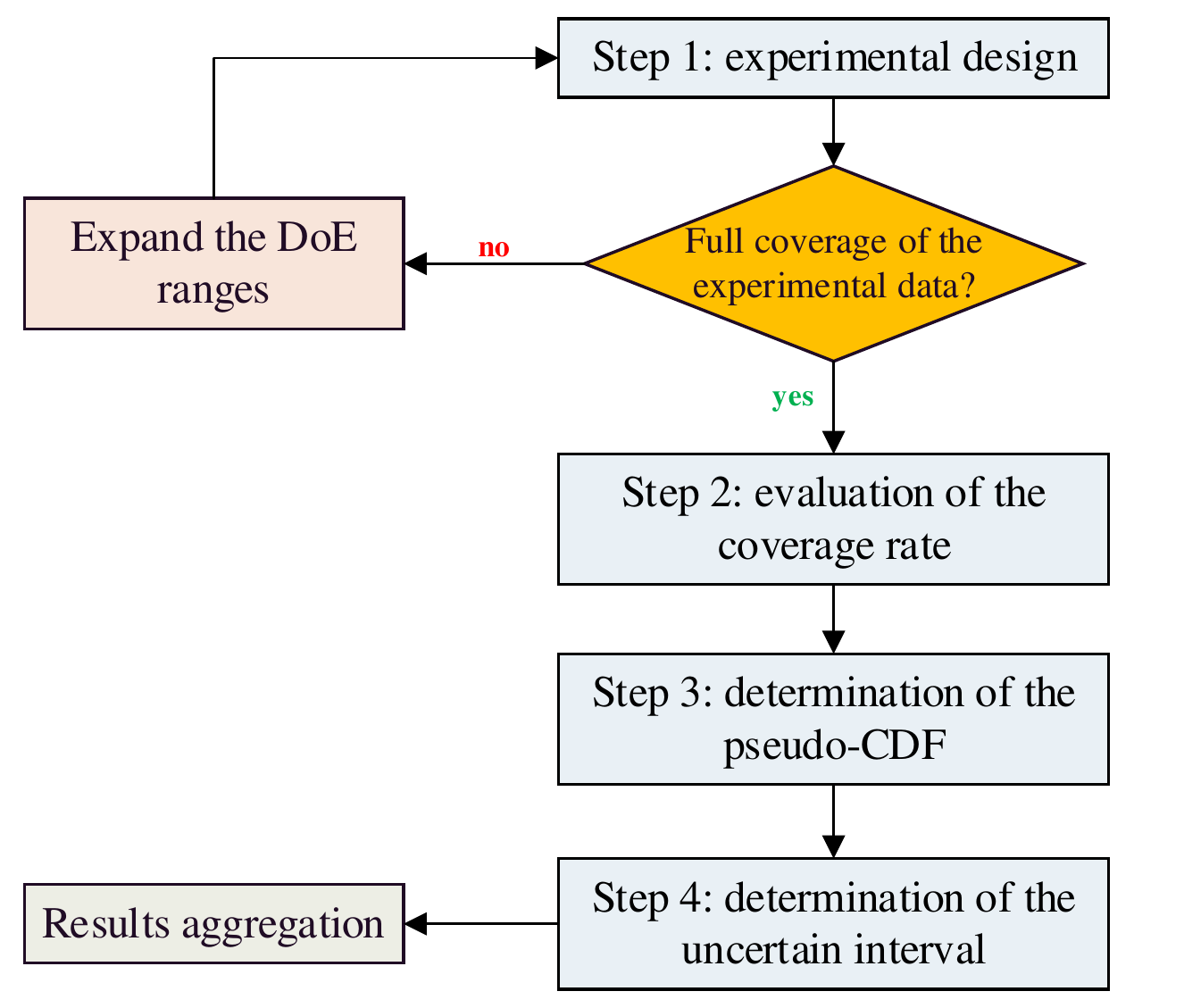}
	\caption[]{Steps of the DIPE method. Figure is adapted from \cite{reventos2016premium}.}
	\label{figure:PREMIUM-DIPE1}
\end{figure}

The DIPE method involves much less math compared to CIRC{\'E}, IPREM and MCDA. In this appendix we briefly explain the DIPE method based on one PMP $\theta$ and one QoI $y$. Figure \ref{figure:PREMIUM-DIPE1} shows the major steps of the DIPE method.

\begin{figure}[htbp]
	\centering
	\includegraphics[width=0.99\textwidth]{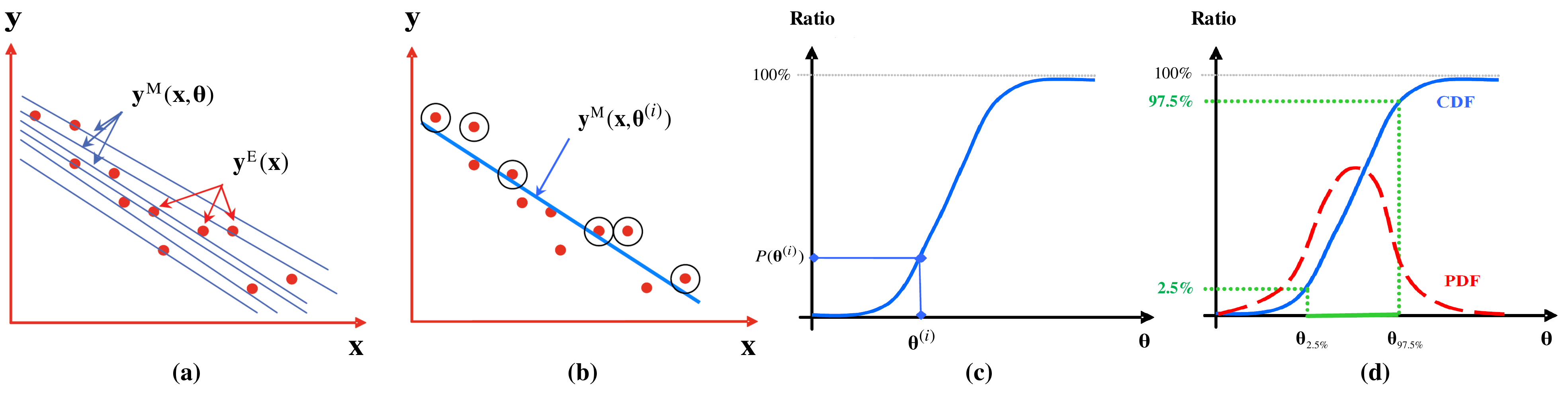}
	\caption[]{Illustration of the DIPE method to quantify input uncertainties. Figures are adapted from the original DIPE paper \cite{joucla2008dipe} with some modifications: (a) experimental data $y^{\text{E}} (\mathbf{x})$ framed by several simulation curves $y^{\text{M}} ( \mathbf{x}, \theta )$ using different values of $\theta$, (b) calculation of the ratio $P(\theta^{\text{(i)}})$ for the simulation $y^{\text{M}} ( \mathbf{x}, \theta^{\text{(i)}} )$, where the circled points means experimental data is larger than the simulated value, (c) CDF built with the ratio $P(\theta^{\text{(i)}})$, and (d) determination of the range of the calibration parameter $\theta$.}
	\label{figure:PREMIUM-DIPE2}
\end{figure}

The first step of DIPE is ``\textit{Experimental design}'', in which a set of values is sampled for $\theta$ based on a prescribed range. Then the TH code runs at these samples to produce an ensemble of simulations, $y^{\text{M}} ( \mathbf{x}, \theta )$. The comparison of the simulations and physical data $y^{\text{E}} (\mathbf{x})$ is shown in Figure \ref{figure:PREMIUM-DIPE2} (a). Note that in this case the design variable $\mathbf{x}$ in Figure \ref{figure:PREMIUM-DIPE2} can be time, and the QoI $y$ can be time-dependent QoIs. If the experimental data cannot be fully bounded by all the simulation curves, the range of $\theta$ needs to be expanded.

The second step of DIPE is ``\textit{Evaluation of the coverage rate}''. For each simulation curve $y^{\text{M}} ( \mathbf{x}, \theta^{\text{(i)}} )$ based on the $i^{\text{th}}$ design value $\theta^{\text{(i)}}$, the number of experimental data points that falls above the simulation is counted as $N^{\text{above}}$, as shown in Figure \ref{figure:PREMIUM-DIPE2} (b). Define the total number of experimental data points as $N^{\text{total}}$. The coverage rate $P(\theta^{\text{(i)}})$ is defined as $P(\theta^{\text{(i)}}) = N^{\text{above}} / N^{\text{total}}$. The ratio $P(\theta^{\text{(i)}})$ denotes the probability that the simulated curve is lower than the experimental data. Thus, $P(\theta^{\text{(i)}})$ obtained for different values of $\theta$ can be combined to form a CDF versus $\theta$, as shown in Figure \ref{figure:PREMIUM-DIPE2} (c).

The third step of DIPE is ``\textit{Determination of the pseudo-CDF}''. The DIPE method relies on several mathematical assumptions. Firstly, the measurement uncertainty is treated as negligible so it is not considered. Secondly, for each experimental data $y^{\text{E}} (\mathbf{x})$, there exists a code simulation $y^{\text{M}} ( \mathbf{x}, \theta )$ that matches this data, while the $\theta$ parameter is a random variable coming from an independent probability law. Thirdly, the simulation curve is monotonous. If all these three assumptions are verified, the obtained function $P(\theta)$ can be considered as an empirical pseudo-CDF of the PMP $\theta$.

The Last step of DIPE is ``\textit{Determination of the uncertainty interval}''. The range of $\theta$ seeks to bound the experimental data with an accuracy of 95\%. To do it, DIPE searches for the simulation curve having 2.5\% of the experimental data \textit{above}, and another simulation curve that has 2.5\% of the experimental data \textit{below}. As shown in Figure \ref{figure:PREMIUM-DIPE2} (d), DIPE seeks the values of $\theta$ for which $P(\theta)$ takes values of 2.5\% and 97.5\% that correspond to the 2.5$^{\text{th}}$ and the 97.5$^{\text{th}}$ percentiles. These values are used to form the uncertain ranges of $\theta$.

If there are multiple experiment tests or QoIs available, the resulting pseudo CDFs for a PMP can be aggregated by averaging. In the case of multiple PMPs, the one-dimensional process is first performed for one parameter while keep the second fixed at a certain value. Repeating this process with different fixing values for the second parameter will produce a set of coverage rates curves and permits the construction of coverage rate contour lines, from which the joint distribution of these two parameters can be obtained. However, these contour lines cannot be considered as a representation of the joint pseudo CDF. Furthermore, this will greatly increase the computational cost. Generally, the number of model simulations increase by the function $d^N$, where $d$ is the dimension of the parameters and $N$ is the number of model runs needed for one-dimensional DIPE.

\section{The MCDA Method}    \label{Appendix-MCDA}

In this section, we provide a self-contained introduction of the MCDA method, using slightly different symbols from the original paper \cite{heo2014implementation}. Table \ref{table:PREMIUM-MCDA-Symbols} lists the symbols used in this appendix.

\begingroup
\renewcommand{\arraystretch}{1.3}
\begin{table}[!ht]
	\footnotesize
	\centering
	\captionsetup{justification=centering}
	\caption{Definitions of symbols used for the MCDA method.}
	\label{table:PREMIUM-MCDA-Symbols}
	\begin{tabular}{p{0.07\linewidth} | p{0.37\linewidth} | p{0.07\linewidth} | p{0.37\linewidth}}
		\toprule
		Symbol  &  Description  &  Symbol  &  Description  \\ 
		\midrule
		$\mathbf{x}$                 &  design variables  &  $\bm{\theta}$  &  calibration parameters  \\
		
		$\bm{\theta}_0$              &  nominal values of $\bm{\theta}$  &  $\mathbf{y}$  &  QoIs  \\
		
		$\mathbf{y}^{\text{M}}$      &  QoIs from model simulation  &  $\mathbf{y}^{\text{E}}$  &  QoIs from experiment  \\
		
		$\bm{\epsilon}$              &  measurement error    &  $c_1$  &  normalization constant for the prior  \\
		
		$c_2$                        &  normalization constant for the likelihood  &  $c$                          &  normalization constant for the posterior  \\
		
		$\bm{\Sigma}_{\bm{\theta}}$  &  covariance for $\bm{\theta}$  &  $\bm{\Sigma}_{\bm{\epsilon}}$&  covariance for $\mathbf{y}^{\text{E}}$  \\
		
		$\bm{\theta}^{\text{prior}}$ &  prior mean vector, $\bm{\theta}^{\text{prior}} = \bm{\theta}_0$    &  $\bm{\theta}^{\text{post}}$ &  posterior mean vector  \\
		
		$\mathbf{S}_{\bm{\theta}}^{\text{prior}}$  &  sensitivity matrix at $\bm{\theta}^{\text{prior}}$  &  $\mathbf{S}_{\bm{\theta}}^{\text{post}}$ &  sensitivity matrix at $\bm{\theta}^{\text{post}}$  \\
		
		$\mathbf{y}^{\text{M}}_{\text{prior}}$ &  code simulation at $\bm{\theta}^{\text{prior}}$  & &  \\
		
		$C (\bm{\theta})$ &  cost function  &  $\alpha$ &  regularization parameter  \\
		
		$\bm{\Sigma}_{\bm{\theta}}^{\text{prior}}$ &  prior covariance matrix of the PMPs  &  $\bm{\Sigma}_{\bm{\theta}}^{\text{post}}$  &  posterior covariance matrix of the PMPs  \\
		
		$\bm{\Sigma}_{\mathbf{y}}^{\text{prior}}$ &  prior covariance matrix of the QoI  &  $\bm{\Sigma}_{\mathbf{y}}^{\text{post}}$  &  posterior covariance matrix of the QoI  \\
		\bottomrule
	\end{tabular}
\end{table}
\endgroup

MCDA assumes the PMPs $\bm{\theta}$ are jointly Gaussian, with mean vector $\bm{\theta}_0$ and covariance matrix $\bm{\Sigma}_{\bm{\theta}}$. The corresponding multivariate Gaussian PDF is given by:
\begin{equation}      \label{equation:PREMIUM-MCDA1-Prior}
	P (\bm{\theta})  =  c_1 \cdot \exp \left[  - \frac{1}{2} \left( \bm{\theta} - \bm{\theta}_0 \right)^\top \bm{\Sigma}_{\bm{\theta}}^{-1} \left( \bm{\theta} - \bm{\theta}_0 \right)  \right]
\end{equation}
where $c_1$ is a normalization constant. Within the Bayesian framework, the prior of $\bm{\theta}$ is determined by expert judgment, and $\bm{\Sigma}_{\bm{\theta}}$ is considered to be diagonal, indicating independence of the PMPs. The observation data $\mathbf{y}^{\text{E}}$ is also assumed to follow Gaussian distributions, while the measurement error $\bm{\epsilon}$ has a covariance matrix $\bm{\Sigma}_{\bm{\epsilon}}$.
\begin{equation}      \label{equation:PREMIUM-MCDA2-Likelihood}
	P (\mathbf{y}^{\text{E}} | \mathbf{y}^{\text{M}})  =  c_2 \cdot \exp \left[  - \frac{1}{2} \left( \mathbf{y}^{\text{E}} - \mathbf{y}^{\text{M}} \right)^\top \bm{\Sigma}_{\bm{\epsilon}}^{-1} \left( \mathbf{y}^{\text{E}} - \mathbf{y}^{\text{M}} \right)  \right]
\end{equation}
where $c_2$ is a normalization constant. The vector $\mathbf{y}^{\text{M}}$ denotes the model prediction. Based on Bayesian inference, it is obvious that Equations \ref{equation:PREMIUM-MCDA1-Prior} and \ref{equation:PREMIUM-MCDA2-Likelihood} represent the prior and likelihood, respectively. The posterior PDF for $\bm{\theta}$ can be written as:
\begin{equation}      \label{equation:PREMIUM-MCDA3-Posterior}
	P (\bm{\theta} | \mathbf{y}^{\text{E}})  =  c \cdot \exp \left[  - \frac{1}{2}  \left(  \left( \bm{\theta} - \bm{\theta}_0 \right)^\top \bm{\Sigma}_{\bm{\theta}}^{-1} \left( \bm{\theta} - \bm{\theta}_0 \right)  +  ( \mathbf{y}^{\text{E}} - \mathbf{y}^{\text{M}} )^\top \bm{\Sigma}_{\bm{\epsilon}}^{-1} ( \mathbf{y}^{\text{E}} - \mathbf{y}^{\text{M}} )  \right)  \right]
\end{equation}
where $c$ is also a normalization constant for the posterior PDF. The difference between Equations \ref{equation:PREMIUM-MCDA3-Posterior} and \ref{equation:Bayesian-IUQ5-Posterior} is that the former uses an explicit format of the Gaussian prior and doesn't consider model bias, while the latter has no restrictions on the type of prior and considers model bias. It has to be pointed out that in by using normalization constant $c_1$, $c_2$ and $c$ in Equations \ref{equation:PREMIUM-MCDA1-Prior}, \ref{equation:PREMIUM-MCDA2-Likelihood} and \ref{equation:PREMIUM-MCDA3-Posterior}, the MCDA method ignores the effect of the ${| \bm{\Sigma} |}^{-1/2}$ term, where $\bm{\Sigma}$ represents the covariance matrices.

\begin{figure}[ht]
	\centering
	\includegraphics[width=0.6\textwidth]{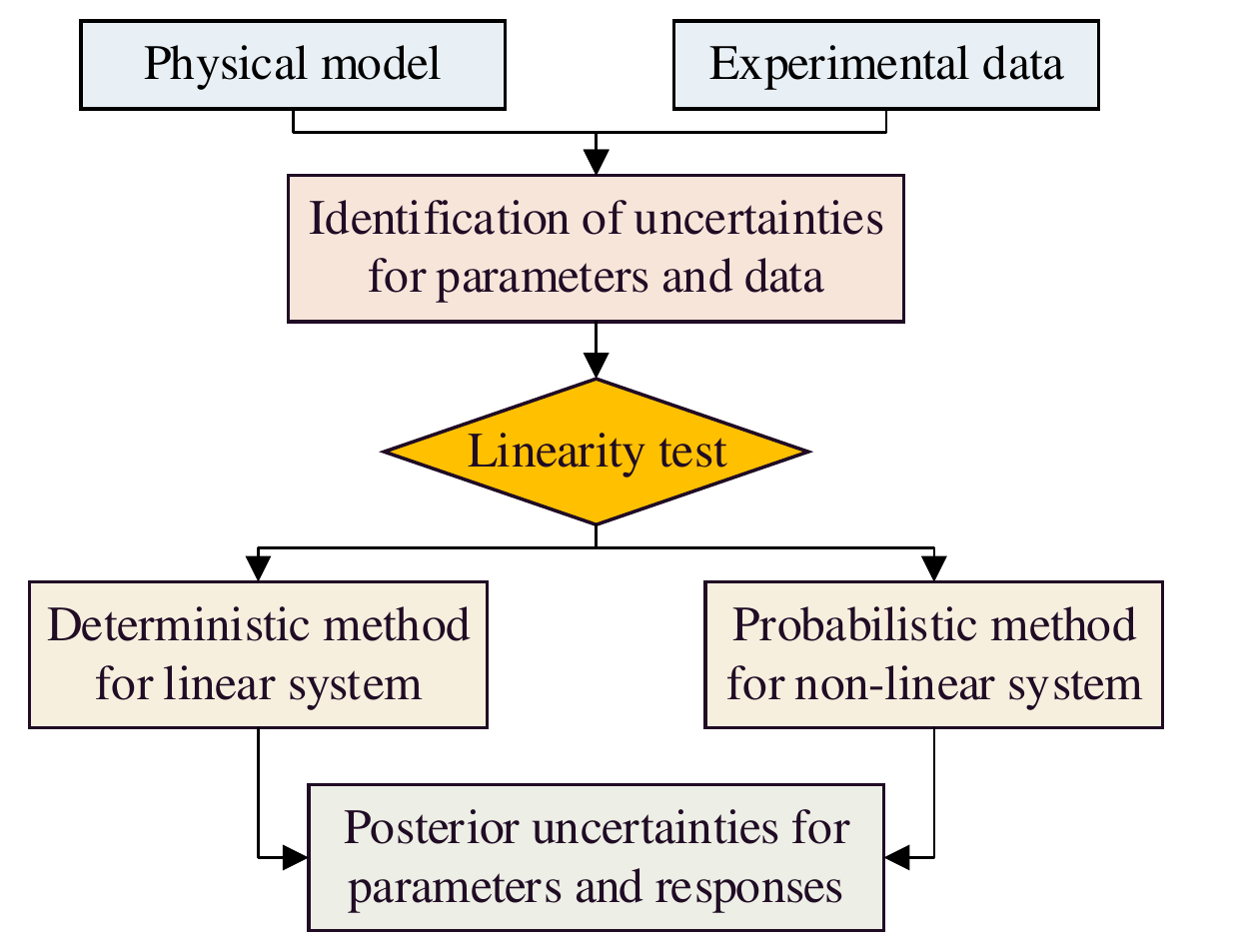}
	\caption[]{The workflow of the MCDA method. Figure is adapted from \cite{heo2014implementation}.}
	\label{figure:PREMIUM-MCDA}
\end{figure}

The mathematical approach used to solve Equation \ref{equation:PREMIUM-MCDA3-Posterior} depends on the linearity of the system, as illustrated in Figure \ref{figure:PREMIUM-MCDA}. Unlike the CIRC{\'E} method that relies soly on the linearity assumption, MCDA also considers non-linear dependencies of uncertain parameters and QoIs. To determine whether a QoI is linear in the PMPs, a linearity test is required to evaluate the degree of linearity. Details for the Chi-square linearity test can be found in Section 3 in \cite{heo2014implementation}. For a \textit{linear} system, MCDA will use a \textit{deterministic} approach to obtain the mean vector and variances of the parameters. For a \textit{non-linear} system, MCDA will use a \textit{probabilistic} method to estimate the posteriori distributions of the parameters.

\subsection{Deterministic MCDA for linear systems}

Given the assumptions that the uncertainties associated with $\bm{\theta}$ and $\mathbf{y}^{\text{E}}$ are Gaussian, for a linear or mildly non-linear system, MCDA uses a deterministic solution process. The system QoI can be approximated by the following equation:
\begin{equation}      \label{equation:PREMIUM-MCDA4-QoI-Linearity}
	\mathbf{y}^{\text{M}}  \cong  \mathbf{y}^{\text{M}}_{\text{prior}}  +  \mathbf{S}_{\bm{\theta}}^{\text{prior}} \left( \bm{\theta} - \bm{\theta}^{\text{prior}} \right)
\end{equation}
where $\bm{\theta}^{\text{prior}}$ denotes the prior mean vector, or nominal values. $\bm{\theta}^{\text{prior}}$ is the same with $\bm{\theta}_0$, here we use a new notation in order to be consistent with $\bm{\theta}^{\text{post}}$ that will be defined later. $\mathbf{S}_{\bm{\theta}}^{\text{prior}}$ is the sensitivity matrix computed at the nominal values $\bm{\theta}^{\text{prior}}$. $\mathbf{y}^{\text{M}}_{\text{prior}}$ is the code simulation at $\bm{\theta}^{\text{prior}}$, $\mathbf{y}^{\text{M}}_{\text{prior}} = \mathbf{y}^{\text{M}} \left( \mathbf{x}, \bm{\theta}^{\text{prior}} \right)$. The prior covariance matrix of the QoI can be calculated by the sandwich rule:
\begin{equation}      \label{equation:PREMIUM-MCDA4-QoI-Prior-Covariance}
	\begin{aligned}
		\bm{\Sigma}_{\mathbf{y}}^{\text{prior}}  
		&=  \mathbb{E} \left(  \left[ \mathbf{y}^{\text{M}} - \mathbf{y}^{\text{M}}_{\text{prior}} \right] \left[ \mathbf{y}^{\text{M}} - \mathbf{y}^{\text{M}}_{\text{prior}} \right]^\top  \right)    \\
		&=  \mathbf{S}_{\bm{\theta}}^{\text{prior}} \cdot  \mathbb{E} \left(  \left[ \bm{\theta} - \bm{\theta}^{\text{prior}} \right] \left[ \bm{\theta} - \bm{\theta}^{\text{prior}} \right]^\top  \right) \cdot  (\mathbf{S}_{\bm{\theta}}^{\text{prior}})^\top    \\
		&=  \mathbf{S}_{\bm{\theta}}^{\text{prior}}  \bm{\Sigma}_{\bm{\theta}}^{\text{prior}}  (\mathbf{S}_{\bm{\theta}}^{\text{prior}})^\top
	\end{aligned}
\end{equation}
where 
\begin{equation}      \label{equation:PREMIUM-MCDA4-Parameter-Prior-Covariance}
	\bm{\Sigma}_{\bm{\theta}}^{\text{prior}} = \mathbb{E} \left(  \left[ \bm{\theta} - \bm{\theta}^{\text{prior}} \right] \left[ \bm{\theta} - \bm{\theta}^{\text{prior}} \right]^\top  \right)
\end{equation}
is the prior covariance matrix of the PMPs. The next step is to determine the mean vector of $\bm{\theta}$ that maximizes the posterior, or equivalently, minimize the negative exponential term of the posterior PDF in Equation \ref{equation:PREMIUM-MCDA2-Likelihood}. Define a cost function $C$:
\begin{equation}      \label{equation:PREMIUM-MCDA5-Cost}
	C (\bm{\theta})  =  \alpha^2 \left( \bm{\theta} - \bm{\theta}^{\text{prior}} \right)^\top (\bm{\Sigma}_{\bm{\theta}}^{\text{prior}})^{-1} \left( \bm{\theta} - \bm{\theta}^{\text{prior}} \right)  +  ( \mathbf{y}^{\text{E}} - \mathbf{y}^{\text{M}} )^\top \bm{\Sigma}_{\bm{\epsilon}}^{-1} ( \mathbf{y}^{\text{E}} - \mathbf{y}^{\text{M}} )
\end{equation}
where $\alpha$ is a regularization parameter that controls the amount of parameter adjustments based on an weighted average of two terms. The first term denotes the mismatch between model and data, $( \mathbf{y}^{\text{E}} - \mathbf{y}^{\text{M}} )^\top \bm{\Sigma}_{\bm{\epsilon}}^{-1} ( \mathbf{y}^{\text{E}} - \mathbf{y}^{\text{M}} )$, while the second term denotes the regularization based on the prior information, $( \bm{\theta} - \bm{\theta}^{\text{prior}} )^\top (\bm{\Sigma}_{\bm{\theta}}^{\text{prior}})^{-1} ( \bm{\theta} - \bm{\theta}^{\text{prior}} )$. According to the MCDA developer \cite{heo2014implementation}, the regularization parameter $\alpha$ is chosen based on the characteristic L-curve \cite{engl1994using}, which is produced by plotting the mismatch term vs. the regularization term by varying $\alpha$. The desired value of $\alpha$ is chosen at a corner where the mismatch term increases rapidly without
any significant change in the regularization term.

The cost $C (\bm{\theta})$ is a function of $\bm{\theta}$. For a linear system, the minimizer can be found by differentiating the cost with respect to $\bm{\theta}$:
\begin{equation}      \label{equation:PREMIUM-MCDA5-Cost-Derivative}
	\begin{aligned}
		\frac{d C (\bm{\theta})}{d \bm{\theta}}  
		&\cong  2 \alpha^2 \left( \bm{\theta} - \bm{\theta}^{\text{prior}} \right)^\top (\bm{\Sigma}_{\bm{\theta}}^{\text{prior}})^{-1}  -  2 ( \mathbf{y}^{\text{E}} - \mathbf{y}^{\text{M}} )^\top \bm{\Sigma}_{\bm{\epsilon}}^{-1}  \frac{d \mathbf{y}^{\text{M}}}{d \bm{\theta}}    \\
		&=  2 \alpha^2 \left( \bm{\theta} - \bm{\theta}^{\text{prior}} \right)^\top (\bm{\Sigma}_{\bm{\theta}}^{\text{prior}})^{-1}  -  2 \left( \mathbf{y}^{\text{E}} - \mathbf{y}^{\text{M}}_{\text{prior}}  -  \mathbf{S}_{\bm{\theta}}^{\text{prior}} \left( \bm{\theta} - \bm{\theta}^{\text{prior}} \right) \right)^\top \bm{\Sigma}_{\bm{\epsilon}}^{-1} \mathbf{S}_{\bm{\theta}}^{\text{prior}}
	\end{aligned}
\end{equation}

By setting $\frac{d C (\bm{\theta})}{d \bm{\theta}} = 0$, the solution to the minimization problem can be found, which is the mean vector of the parameters' posterior distributions:
\begin{equation}      \label{equation:PREMIUM-MCDA6-Parameter-Post-Mean}
	\bm{\theta}^{\text{post}}  =  \bm{\theta}^{\text{prior}}  +  \left( (\mathbf{S}_{\bm{\theta}}^{\text{prior}})^\top \bm{\Sigma}_{\bm{\epsilon}}^{-1} \mathbf{S}_{\bm{\theta}}^{\text{prior}} + \alpha^2 (\bm{\Sigma}_{\bm{\theta}}^{\text{prior}})^{-1}  \right)^{-1}  (\mathbf{S}_{\bm{\theta}}^{\text{prior}})^\top  \bm{\Sigma}_{\bm{\epsilon}}^{-1}  \left( \mathbf{y}^{\text{E}} - \mathbf{y}^{\text{M}}_{\text{prior}} \right)
\end{equation}

For notational simplicity, define:
\begin{equation*}
	\mathbf{K} = \left( (\mathbf{S}_{\bm{\theta}}^{\text{prior}})^\top \bm{\Sigma}_{\bm{\epsilon}}^{-1} \mathbf{S}_{\bm{\theta}}^{\text{prior}} + \alpha^2 (\bm{\Sigma}_{\bm{\theta}}^{\text{prior}})^{-1}  \right)^{-1} (\mathbf{S}_{\bm{\theta}}^{\text{prior}})^\top \bm{\Sigma}_{\bm{\epsilon}}^{-1}
\end{equation*}
which leads to $\bm{\theta}^{\text{post}}  =  \bm{\theta}^{\text{prior}}  +  \mathbf{K} ( \mathbf{y}^{\text{E}} - \mathbf{y}^{\text{M}}_{\text{prior}} )$. For a linear system, the posterior distributions of the parameters are characterized by not only the mean values in Equation \ref{equation:PREMIUM-MCDA6-Parameter-Post-Mean}, but also the covariance matrix.
\begin{equation}      \label{equation:PREMIUM-MCDA6-Parameter-Post-Covariance1}
	\bm{\Sigma}_{\bm{\theta}}^{\text{post}}  
	\equiv  \mathbb{E} \left(  \left[ \bm{\theta} - \bm{\theta}^{\text{post}} \right] \left[ \bm{\theta} - \bm{\theta}^{\text{post}} \right]^\top  \right)
\end{equation}

An explicit expression for the posterior covariance $\bm{\Sigma}_{\bm{\theta}}^{\text{post}}$ can be obtained by substituting Equation \ref{equation:PREMIUM-MCDA6-Parameter-Post-Mean} into Equation \ref{equation:PREMIUM-MCDA6-Parameter-Post-Covariance1}:
\begin{equation}      \label{equation:PREMIUM-MCDA6-Parameter-Post-Covariance2}
	\bm{\Sigma}_{\bm{\theta}}^{\text{post}}  
	\cong  \bm{\Sigma}_{\bm{\theta}}^{\text{prior}}  -  \bm{\Sigma}_{\bm{\theta}}^{\text{prior}} (\mathbf{S}_{\bm{\theta}}^{\text{prior}})^\top \mathbf{K}^\top  -  \mathbf{K} \mathbf{S}_{\bm{\theta}}^{\text{prior}} \bm{\Sigma}_{\bm{\theta}}^{\text{prior}}  +  \mathbf{K} \left( \bm{\Sigma}_{\bm{\epsilon}} + \mathbf{S}_{\bm{\theta}}^{\text{prior}} \bm{\Sigma}_{\bm{\theta}}^{\text{prior}} (\mathbf{S}_{\bm{\theta}}^{\text{prior}})^\top \right) \mathbf{K}^\top
\end{equation}
Note that an approximation has been made in deriving Equation \ref{equation:PREMIUM-MCDA6-Parameter-Post-Covariance2} by ignoring the covariance between the parameters and the QoIs. Besides the mean vector (Equation \ref{equation:PREMIUM-MCDA6-Parameter-Post-Mean}) and covariance matrix (Equation \ref{equation:PREMIUM-MCDA6-Parameter-Post-Covariance2}) of the PMPs, the deterministic component of the MCDA method also directly provides the response uncertainties. For the linear system, the posterior covariance matrix is given by:
\begin{equation}      \label{equation:PREMIUM-MCDA7-QoI-Post-Covariance}
	\bm{\Sigma}_{\mathbf{y}}^{\text{post}}  
	=  \mathbf{S}_{\bm{\theta}}^{\text{post}}  \bm{\Sigma}_{\bm{\theta}}^{\text{post}}  (\mathbf{S}_{\bm{\theta}}^{\text{post}})^\top
\end{equation}
where $\mathbf{S}_{\bm{\theta}}^{\text{post}}$ is the updated sensitivity matrix obtained using the new nominal values of $\bm{\theta}$, i.e., the mean values of the posterior distributions, $\bm{\theta}^{\text{post}}$.

\subsection{Probabilistic MCDA for non-linear systems}

Physical phenomena in nuclear reactors typically exhibit non-linear behavior, especially in fast TH transients, fuel performance and solid mechanics. Linear approximations may be made when the parameter variation is very small, so the parameter-QoI relation can be treated as linear around the nominal value of the parameter. When the QoIs show strong non-linear and discontinuous behavior, the deterministic method derived above cannot be used anymore. Furthermore, the deterministic method assumes the parameter and experimental data uncertainties to be Gaussian, which may not work in certain cases.

For non-linear systems and non-Gaussian uncertainties, MCDA uses a probabilistic method based on MCMC sampling to obtain the posterior distributions. The posterior PDF to be explored by MCMC can be found in Equation \ref{equation:PREMIUM-MCDA3-Posterior}. Note that the prior distributions are not necessarily Gaussian. This probabilistic method is essentially similar to the Bayesian IUQ method we have introduced, however, without the consideration of the model bias term. It also did not consider the computational cost originated from MCMC sampling. In the work presented in \cite{heo2014implementation}, the authors only generated 3,500 MCMC samples, which seems insufficient for a typical Markov chain.

\bibliography{./Journal_Review_IUQ_TH_bib.bib}

\end{document}